\documentclass[twocolumn]{aastex62}

\usepackage{natbib}
\usepackage{amsmath}	
\usepackage{amssymb}	

\received{}
\revised{}
\accepted{}
\submitjournal{ApJ}

\shorttitle{Confirming \textit{Herschel} candidate protoclusters}
\shortauthors{G\'omez-Guijarro et al.}

\begin{document}

\title{Confirming \textit{Herschel} candidate protoclusters from ALMA/VLA CO observations}

\correspondingauthor{C. G\'omez-Guijarro}
\email{carlos.gomez@nbi.ku.dk}

\author{C. G\'omez-Guijarro}
\affil{Cosmic Dawn Center (DAWN)}
\affil{Niels Bohr Institute, University of Copenhagen, Vibenshuset, Lyngbyvej 2, DK-2100 Copenhagen, Denmark}
\affil{DARK, Niels Bohr Institute, University of Copenhagen, Lyngbyvej 2, DK-2100 Copenhagen, Denmark}
\affil{Department of Astronomy, Cornell University, Space Science Building, Ithaca, NY 14853, USA}

\author{D. A. Riechers}
\affil{Department of Astronomy, Cornell University, Space Science Building, Ithaca, NY 14853, USA}

\author{R. Pavesi}
\affil{Department of Astronomy, Cornell University, Space Science Building, Ithaca, NY 14853, USA}

\author{G. E. Magdis}
\affil{Cosmic Dawn Center (DAWN)}
\affil{Niels Bohr Institute, University of Copenhagen, Vibenshuset, Lyngbyvej 2, DK-2100 Copenhagen, Denmark}
\affil{DARK, Niels Bohr Institute, University of Copenhagen, Lyngbyvej 2, DK-2100 Copenhagen, Denmark}
\affil{Institute for Astronomy, Astrophysics, Space Applications and Remote Sensing, National Observatory of Athens, GR-15236 Athens, Greece}

\author{T. K. D. Leung}
\affil{Department of Astronomy, Cornell University, Space Science Building, Ithaca, NY 14853, USA}
\affil{Center for Computational Astrophysics, Flatiron Institute, 162 Fifth Avenue, New York, NY 10010, USA}

\author{F. Valentino}
\affil{Cosmic Dawn Center (DAWN)}
\affil{Niels Bohr Institute, University of Copenhagen, Vibenshuset, Lyngbyvej 2, DK-2100 Copenhagen, Denmark}
\affil{DARK, Niels Bohr Institute, University of Copenhagen, Lyngbyvej 2, DK-2100 Copenhagen, Denmark}

\author{S. Toft}
\affil{Cosmic Dawn Center (DAWN)}
\affil{Niels Bohr Institute, University of Copenhagen, Vibenshuset, Lyngbyvej 2, DK-2100 Copenhagen, Denmark}
\affil{DARK, Niels Bohr Institute, University of Copenhagen, Lyngbyvej 2, DK-2100 Copenhagen, Denmark}

\author{M. Aravena}
\affil{N\'ucleo de Astronom\'ia, Facultad de Ingenier\'ia y Ciencias, Universidad Diego Portales, Av. Ej\'ercito 441, Santiago, Chile}

\author{S. C. Chapman}
\affil{Department of Physics and Atmospheric Science, Dalhousie University, 6310 Coburg Road, Halifax, NS B3H 4R2, Canada}

\author{D. L. Clements}
\affil{Astrophysics Group, Blackett Lab, Physics Department, Imperial College, Prince Consort Road, London SW7 2AZ, UK}

\author{H. Dannerbauer}
\affil{Instituto de Astrof\'isica de Canarias (IAC), Calle V\'ia Lactea s/n, E-38205 La Laguna, Tenerife, Spain}
\affil{Universidad de La Laguna, Dpto. Astrof\'isica, E-38206 La Laguna, Tenerife, Spain}

\author{S. J. Oliver}
\affil{Astronomy Centre, Dept. of Physics \& Astronomy, University of Sussex, Brighton BN1 9QH, UK}

\author{I. P\'erez-Fournon}
\affil{Instituto de Astrof\'isica de Canarias (IAC), Calle V\'ia Lactea s/n, E-38205 La Laguna, Tenerife, Spain}
\affil{Universidad de La Laguna, Dpto. Astrof\'isica, E-38206 La Laguna, Tenerife, Spain}

\author{I. Valtchanov}
\affil{Telespazio Vega UK for ESA, European Space Astronomy Centre, Operations Department, E-28691 Villanueva de la Ca\~nada, Spain}

\begin{abstract}

ALMA 870\,$\mu$m continuum imaging has uncovered a population of blends of multiple dusty star-forming galaxies (DSFGs) in sources originally detected with the \textit{Herschel Space Observatory}. However, their pairwise separations are much smaller that what is found by ALMA follow-up of other single-dish surveys or expected from theoretical simulations. Using ALMA and VLA, we have targeted three of these systems to confirm whether the multiple 870\,$\mu$m continuum sources lie at the same redshift, successfully detecting $^{12}$CO($J = 3$--2) and $^{12}$CO($J = 1$--0) lines and being able to confirm that in the three cases all the multiple DSFGs are likely physically associated within the same structure. Therefore, we report the discovery of two new gas-rich dusty protocluster cores (HELAISS02, $z = 2.171 \pm 0.004$; HXMM20, $z = 2.602 \pm 0.002$). The third target is located in the well known COSMOS overdensity at $z = 2.51$ (named CL J1001+0220 in the literature), for which we do not find any new secure CO(1-0) detection, although some of its members show only tentative detections and require further confirmation. From the gas, dust, and stellar properties of the two new protocluster cores, we find very large molecular gas fractions yet low stellar masses, pushing the sources above the main sequence, while not enhancing their star formation efficiency. We suggest that the sources might be newly formed galaxies migrating to the main sequence. The properties of the three systems compared to each other and to field galaxies may suggest a different evolutionary stage between systems.

\end{abstract}

\keywords{galaxies: clusters: general --- galaxies: evolution --- galaxies: formation --- galaxies: high-redshift --- galaxies: interactions --- galaxies: ISM --- galaxies: starburst --- galaxies: structure --- infrared: galaxies --- radio lines: galaxies --- submillimeter: galaxies}

\section{Introduction} \label{sec:intro}

Galaxies luminous in the far-IR (FIR) and submillimeter (submm) wavelengths comprise the most intense starbursts in the universe, known as dusty star-forming galaxies \citep[DSFGs; see][for a review]{2014PhR...541...45C}. With a redshift distribution that peaks at $z \sim 2$--3 \citep[e.g.,][]{2005ApJ...622..772C}, they constitute an important component of the overall galaxy population at $z \sim 2$ \citep[e.g.,][]{2011A&A...528A..35M}. DSFGs are promising candidates to trace galaxy clusters in formation in formation, the so-called protoclusters \citep[see][]{2016A&ARv..24...14O}. DSFGs have also been proposed as progenitors of the most massive elliptical galaxies in the local universe \citep[e.g.,][]{2008A&A...482...21C,2010MNRAS.406..230R,2013Natur.498..338F,2013ApJ...772..137I,2014ApJ...782...68T,2018ApJ...856..121G}.

At $z \gtrsim 4$ overdensities of galaxies with associated DSFGs have been discovered: GN20 \citep[e.g.,][]{2009ApJ...694.1517D}, HDF850.1 \citep[e.g.,][]{2012Natur.486..233W}, AzTEC-3 \citep[e.g.,][]{2010ApJ...720L.131R,2011Natur.470..233C,2014ApJ...796...84R}, CRLE and HZ10 \citep[e.g.,][]{2015Natur.522..455C,2016ApJ...832..151P,2018ApJ...861...43P}, DRC \citep[e.g.,][]{2018ApJ...856...72O}, SPT2349-56 \citep[e.g.,][]{2018Natur.556..469M}. At $2 \lesssim z \lesssim 3$ several confirmed protoclusters containing dozens of galaxies are known to be DSFGs-rich: GOODS-N $z = 1.99$ protocluster \citep[e.g.,][]{2004ApJ...611..725B,2009ApJ...691..560C}, CL J1449+0856 \citep[e.g.,][]{2011A&A...526A.133G,2015ApJ...801..132V,2016ApJ...829...53V,2018MNRAS.479..703C}, COSMOS $z = 2.10$ protocluster \citep[e.g.,][]{2012ApJ...748L..21S,2014ApJ...795L..20Y}, MRC1138-256 \citep[e.g.,][]{2000A&A...358L...1K,2014A&A...570A..55D}, COSMOS $z = 2.51$ protocluster \citep[e.g.,][]{2007ApJS..172..132B,2010ApJ...708L..36A,2015ApJ...808L..33C,2016ApJ...828...56W,2018A&A...619A..49C,2018ApJ...867L..29W}, SSA22 \citep[e.g.,][]{1998ApJ...492..428S,2015ApJ...815L...8U} \citep[see also][]{2016ApJ...824...36C}.

Large angular scale clusters and cluster candidates have been found by the \textit{Herschel Space Observatory} and \textit{Planck} satellite \citep[e.g.,][]{2014MNRAS.439.1193C,2016MNRAS.461.1719C,2016A&A...596A.100P,2018MNRAS.476.3336G,2018A&A...620A.198M}. In particular, \textit{Herschel} has scanned wide fields at FIR and submm wavelengths with the Spectral and Photometric Imaging Receiver \citep[SPIRE;][]{2010A&A...518L...3G} at 250, 350, and 500\,$\mu$m \citep[e.g.,][]{2010PASP..122..499E,2012MNRAS.424.1614O}. The nature of the \textit{Herschel}/SPIRE wide beam detections is diverse. Among them, gravitationally-lensed \citep[e.g.,][]{2010Sci...330..800N,2012ApJ...756..134B,2013ApJ...762...59W,2013ApJ...779...25B,2015A&A...581A.105C} and $z > 4$ DSFGs \citep[e.g.,][]{2013Natur.496..329R,2014ApJ...780...75D,2018A&A...614A..33D} have been identified in large numbers, with follow-up high spatial resolution observations proven to be very important to uncover their nature. Recently, \textit{Herschel}/SPIRE detections have also found to be blends of multiple DSFGs that could constitute protoclusters \citep[e.g.,][]{2015ApJ...812...43B} \citep[see also][]{2015ApJ...806L..25S}.

\citet{2015ApJ...812...43B} presented ALMA 870\,$\mu$m observations of 29 bright \textit{Herschel}/SPIRE DSFGs from the HerMES survey \citep{2012MNRAS.424.1614O}. The ALMA imaging surprisingly showed that 20/29 objects comprise multiple DSFGs located within a few arcseconds of each other. Such a high fraction with small pairwise physical separations are almost completely unexpected from both a theoretical perspective \citep{2013MNRAS.434.2572H,2015MNRAS.446.1784C,2015MNRAS.446.2291M,2018MNRAS.476.2278H} as well as previous high spatial resolution follow-up of the LArge APEX BOlometer CAmera (LABOCA) and the Submillimetre Common-User Bolometer Array (SCUBA) single-dish observations \citep[][]{2013ApJ...768...91H,2018MNRAS.479.3879W}, suggesting that a portion of the ALMA 870\,$\mu$m counterparts are due to line-of-sight projection effects and are not physically related. In order to investigate whether they are physically associated or not it is necessary to have spectroscopic data with sufficient spatial resolution to distinguish the ALMA counterparts from each other.

In this work we present follow-up CO line observations of three \textit{Herschel} candidate protoclusters from \citet{2015ApJ...812...43B} aiming to confirm whether the multiple ALMA 870\,$\mu$m continuum sources are located at the same redshift or are line-of-sight projections. Note that we will refer to these associations of multiple DSFGs within a few arcseconds of each other as protocluster cores. It is known that the small field of view (FOV) of the ALMA observations is only able to detect the densest peaks of protocluster structures. Confirmation of a larger structure of members located at a similar redshift studying whether the structures will evolve into a cluster at $z = 0$ is required to properly establish the protocluster nature of the candidates, which is beyond the scope of this work \citep[e.g.,][]{2013ApJ...779..127C,2015MNRAS.452.2528M,2017ApJ...844L..23C}.

The layout of the paper is as follows. We introduce the sample and data in Section~\ref{sec:sample_data}. In Section~\ref{sec:analysis} we present the analysis of the observations. Gas, dust and stellar properties of the targets are explored in Section~\ref{sec:results}. We discuss the results, comparing with field galaxies and protoclusters in Section~\ref{sec:discussion}. Summary of the main findings and conclusions are in Section~\ref{sec:summary}.

Throughout this work we adopted a concordance cosmology $[\Omega_\Lambda,\Omega_M,h]=[0.7,0.3,0.7]$ and Chabrier initial mass function (IMF) \citep{2003PASP..115..763C}.

\section{Sample and Data} \label{sec:sample_data}

\begin{figure*}
\begin{center}
\includegraphics[width=\textwidth]{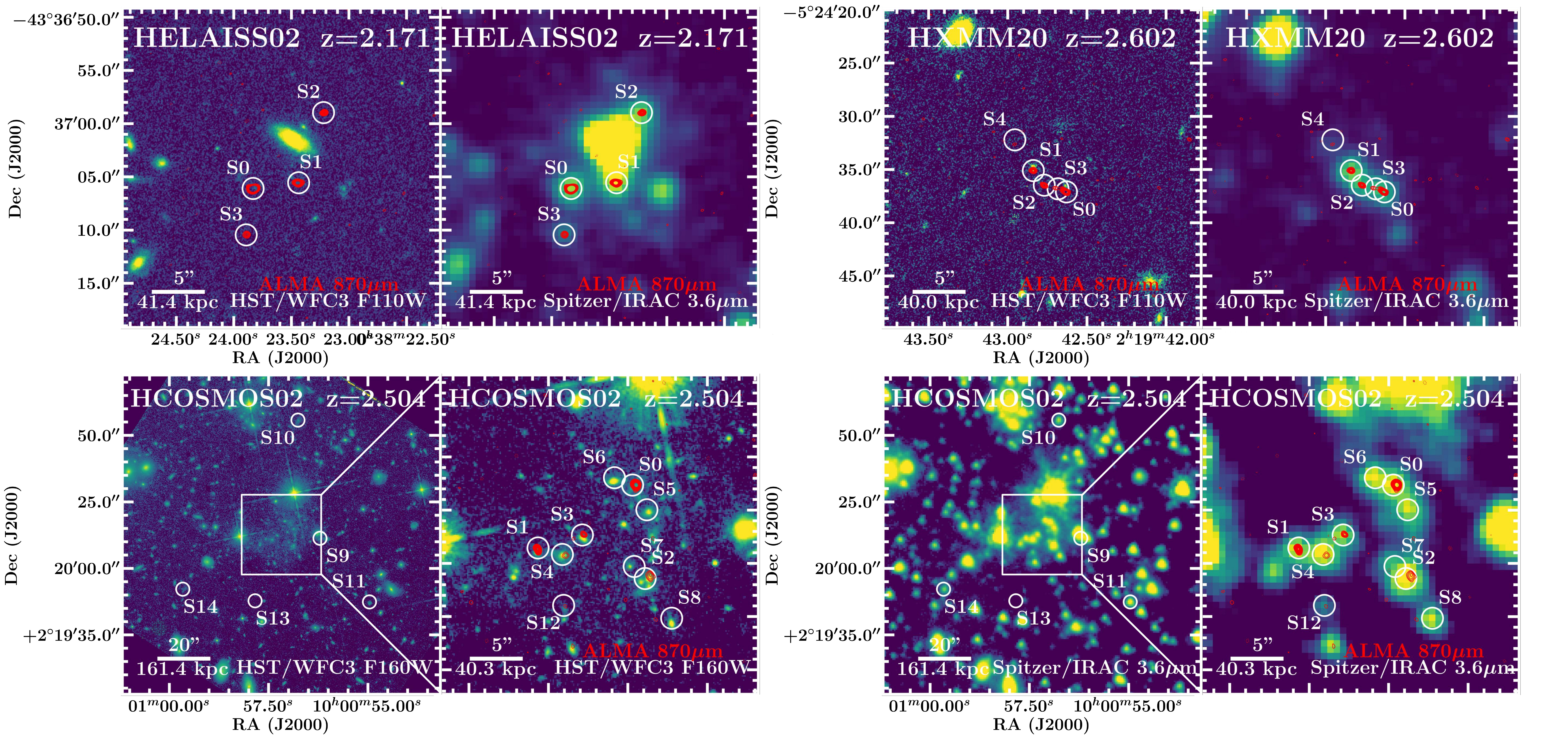}
\caption{Overview of the sample. Top row: HELAISS02 and HXMM20 30\arcsec$\times$30\arcsec \textit{HST}/WFC3 F160W and \textit{Spitzer}/IRAC 3.6\,$\mu$m images. Bottom row: HCOSMOS02 120\arcsec$\times$120\arcsec and 30\arcsec$\times$30\arcsec zoom-in of the central region \textit{HST}/WFC3 F160W and \textit{Spitzer}/IRAC 3.6\,$\mu$m images. ALMA 870\,$\mu$m contours are overlaid in red (starting at $\pm$3\,$\sigma$ and growing in steps of $\pm$1\,$\sigma$, where positive contours are solid and negative contours dotted). CO(1-0) detections presented in Section~\ref{sec:analysis} are encircled and labeled in white.}
\label{fig:overview}
\end{center}
\end{figure*}

\subsection{\textit{Herschel Candidate Protoclusters}} \label{subsec:sample}

We followed up the three sources in the original \textit{Herschel}-ALMA sample from \citet{2015ApJ...812...43B} with the highest multiplicity rate. Each target has at least four ALMA 870\,$\mu$m counterparts (see Figure~\ref{fig:overview} for an overview of the sample). Briefly, the original sample of 29 \textit{Herschel}/SPIRE DSFGs in \citet{2015ApJ...812...43B} was selected to be the brightest set of targets in the ALMA-accessible portion of HerMES \citep{2012MNRAS.424.1614O} available at the time of the Cycle 0 deadline. The intention was to assemble the largest sample of lenses possible, but a comparison of optical imaging with the ALMA imaging surprisingly showed that most of the objects in this subset comprise multiple DSFGs located within a few arcseconds of each other.

The targets HELAISS02 and HXMM20 are new protoclusters candidates. HCOSMOS02 was originally reported in the literature as COSBO-3 by \citet{2007ApJS..172..132B} and shown to be an overdense region with a photometric redshift $z \sim 2.2$--2.4 in \citet{2010ApJ...708L..36A} \citep[see also][]{2012ApJS..200...10S}. Several works have been recently focused on this source. \citet{2015ApJ...808L..33C} spectroscopically confirmed some galaxies in HCOSMOS02 using Keck/MOSFIRE. \citet{2016ApJ...828...56W} (source named CL J1001+0220) reported that there is evidence of virialization and define it as a cluster \citep[see also][]{2017ApJ...846L..31D,2018ApJ...867L..29W}. It also appears to be related with a larger structure composed of several density peaks spanning $2.42 < z < 2.51$ \citep{2013ApJ...765..109D,2015ApJ...802...31D,2015ApJ...808...37C,2015ApJ...808L..33C,2016ApJ...817..160L,2018A&A...619A..49C}. We carried out a redshift search for $^{12}$CO($J = 3$--2) for HELAISS02 and HXMM20 and $^{12}$CO($J = 1$--0) for HXMM20 using prior photometric information that placed our targets at $1.5 < z < 3.5$ with high certainty. In the case of HCOSMOS02, the redshift was established from our Combined Array for Research in Millimeter-wave Astronomy (CARMA) 3\,mm observations targeting $^{12}$CO($J = 3$--2) (see Section~\ref{subsec:other_data}), independently from the Keck/MOSFIRE H$\alpha$ detections in \citet{2015ApJ...808L..33C} and the NOrthern Extended Millimeter Array (NOEMA) $^{12}$CO($J = 5$--4) confirmed with NSF's Karl G. Jansky Very Large Array (VLA) $^{12}$CO($J = 1$--0) observations in \citet{2016ApJ...828...56W}. Knowing the redshift of HCOSMOS02, we performed $^{12}$CO($J = 1$--0) and $^{12}$CO($J = 4$--3) observations.

\subsection{ALMA Observations} \label{subsec:alma_data}

We carried out a spectral scan of the 3\,mm band with ALMA band 3 during Cycle 3 (program 2015.1.00752.S; PI: R. S. Bussmann) targeting $^{12}$CO($J = 3$--2) transition line ($\nu_{\rm{rest}} = 345.79599$\,GHz) for HELAISS02 and HXMM20.

Observations of HELAISS02 were executed between 2016 May 27 and June 17 with 46 usable 12-m antennae. The shortest and longest baselines were 12\,m and 741\,m, respectively. The resulting on-source spectral scan integration time was 25.5\,min. The correlator was set up in five different tunings every one containing four spectral windows of 1.875\,GHz each at 31.25\,MHz (94.95\,km s$^{-1}$ at 98.664\,GHz) resolution in dual polarization, covering the frequency range 84--113.2\,GHz. The radio quasar J2357-5311 was observed as bandpass and secondary flux calibrator and the radio quasar J0030-4224 as amplitude and phase calibrator. Pallas was set to the primary flux calibrator, but it was not observed in the first tuning, so we substituted it for our secondary flux calibrator J2357-5311 in all tunings to be consistent. The flux calibration using J2357-5311 is 15\% lower than using Pallas.

HXMM20 observations were taken on 2016 June 12 with 38 usable 12-m antennae. The shortest and longest baselines were 13\,m and 704\,m, respectively. The on-source spectral scan integration time was 11.6\,min. The correlator configuration was identical to that of HELAISS02. The radio quasars J0006-0623 and J0238+1636 were observed as bandpass and flux calibrators, the first object for the first tuning and the second object for the rest of the tunings. The radio quasar J0209-0438 was observed for amplitude and phase calibration of all the tunings. Pallas was also part of the observations, but the QA assessed a discrepancy of 30\% between the model and the calibrator catalogue; therefore, it was rejected as flux calibrator.

The Common Astronomy Software Applications \citep[CASA;][version 4.6.0 for HELAISS02 and version 4.5.6 for HXMM20]{2007ASPC..376..127M} packages were employed for data reduction and analysis. HELAISS02 and HXMM20 data were mapped using the \texttt{CLEAN} algorithm with natural weighting to get the best point source sensitivity. We used custom masks enclosing the emitting regions in each channel, cleaning down to a 2\,$\sigma$ threshold. For HELAISS02, the resulting synthesized beam size is 1\farcs36$\times$1\farcs14 and the primary beam half power beam width (HPBW) 53\farcs4 at 108.9655\,GHz. For HXMM20, the synthesized beam size is 1\farcs50$\times$1\farcs27 and the primary beam HPBW 60\farcs6 at 96.11968\,GHz. The rms noise per 94.95\,km s$^{-1}$ channel at 108.96550\,GHz is $\sim 0.38$\,mJy beam$^{-1}$ for HELAISS02 and $\sim 0.54$\,mJy beam$^{-1}$ per 94.95\,km s$^{-1}$ channel at 96.11968\,GHz for HXMM20, measured at the phase center.

Line free channels were combined to measure the continuum at $\sim 3$\,mm (see Table~\ref{tab:continuum} and Figure~\ref{fig:continuum}), resulting in a rms noise of $\sim 13$\,$\mu$Jy beam$^{-1}$ for HELAISS02 and $\sim 22$\,$\mu$Jy beam$^{-1}$ for HXMM20, at the phase center. Continuum subtraction is not needed since the continuum level is negligible at the rms noise of the line channels.

\subsection{VLA Observations} \label{subsec:vla_data}

A spectral scan was also carried out with VLA, Ka and Q bands during Cycle 15 semester B (program 15B-065; PI: R. S. Bussmann). We targeted $^{12}$CO($J = 1$--0) transition line ($\nu_{\rm{rest}} = 115.27120$\,GHz) for HXMM20 and HCOSMOS02.

Observations of HXMM20 were taken between 2015 Oct 22 and Nov 14 in D array configuration (shortest baseline 31\,m, longest baseline 997\,m). Total on-source spectral scan integration time was 6.7\,h. We configured three correlator tunings covering Ka and Q band frequencies, each one containing four basebands of 2\,GHz using the 3-bit sampler that provides 2\,MHz channels in dual polarization, covering the frequency range 26.5--48\,GHz. The radio quasars 3C 147 and J0215-0222 acted as flux/bandpass and amplitude/phase calibrators, respectively.

HCOSMOS02 was observed between 2015 Oct 24 and Nov 6 in D array configuration (shortest baseline 34\,m, longest baseline 922\,m). Given the known redshift of this source from our CARMA observations targeting CO(3-2) (see Section~\ref{subsec:other_data}) and independently found by \citet{2016ApJ...828...56W} from CO(1-0), we selected Ka band with the correlator set up covering the frequency range 31.5--33.5\,GHz using the 3-bit sampler providing 2\,MHz channels in dual polarization. The radio quasars 3C 147 and J1018+0530 were used as flux/bandpass and amplitude/phase calibrators, respectively. Additional data is available for HCOSMOS02 from two archival programs (program 15B-210; PI: C. Casey, and program 15B-290; PI: T. Wang. For an upcoming independent analysis of the archival data, see J. Champagne et al., in prep.) We concatenated all three programs, for a total on-source integration time of 33.3\,h. Together the programs cover the frequency range 31.5--34.2\,GHz, but overlap just at 32.2--33.4\,GHz.

CASA (version 4.5.0) was employed for reduction and analysis. We imaged HXMM20 using a $\rm{robust} = 0.5$ Briggs weighting scheme \citep{briggs95} as it gave the best compromise between the spatial resolution required to deblend the different ALMA counterparts and the sensitivity to detect them. For HCOSMOS02 we used a natural weighting scheme to achieve the best point source sensitivity possible. For HXMM20, the resulting synthesized beam size is 2\farcs37$\times$1\farcs92 and the primary beam HPBW 84\farcs3 at 32.04105\,GHz. For HCOSMOS02, the synthesized beam size is 2\farcs88$\times$2\farcs49 and the primary beam HPBW 82\farcs1 at 32.86889\,GHz. The rms noise in a 50\,km s$^{-1}$ channel at 32.04105\,GHz is $\sim 0.12$\,mJy beam$^{-1}$ for HXMM20 and $\sim 31$\,$\mu$Jy beam$^{-1}$ in a 50\,km s$^{-1}$ channel at 32.86889\,GHz for HCOSMOS02, measured at the phase center.

\begin{figure*}
\begin{center}
\includegraphics[width=\textwidth]{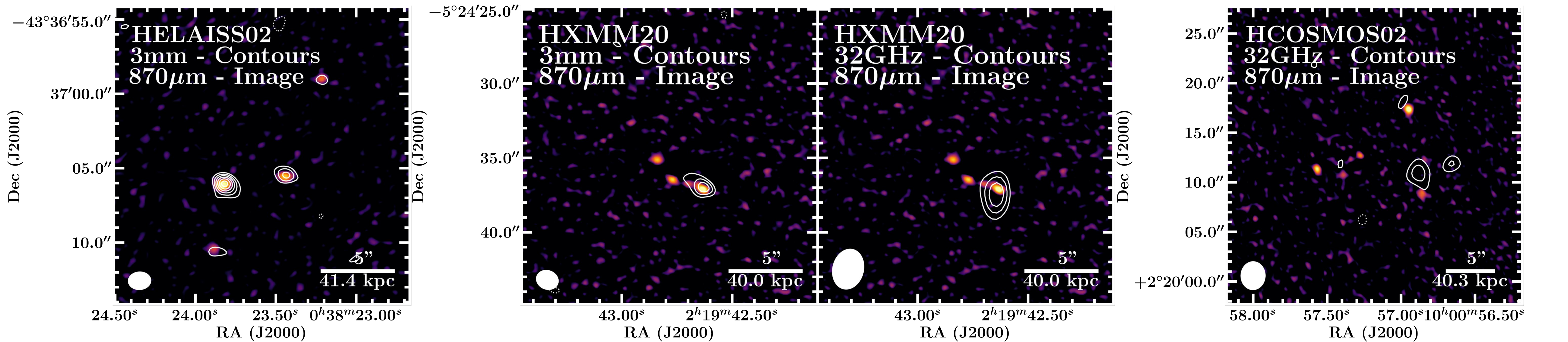}
\caption{From left to right: HELAISS02 3\,mm continuum, HXMM20 3\,mm, 32\,GHz and HCOSMOS02 32\,GHz continuum emission on top of the 870\,$\mu$m continuum image. Contours start at $\pm$3\,$\sigma$ and grow in steps of $\pm$1\,$\sigma$ (HELAISS02 3\,mm $\sigma = 13$\,$\mu$Jy beam$^{-1}$; HXMM20 3\,mm $\sigma = 22$\,$\mu$Jy beam$^{-1}$; HXMM20 32\,GHz $\sigma = 3.5$\,$\mu$Jy beam$^{-1}$; HCOSMOS02 32\,GHz $\sigma = 1.9$\,$\mu$Jy beam$^{-1}$). Positive contours are solid and negative contours dotted.}
\label{fig:continuum}
\end{center}
\end{figure*}

Line free channels were combined to search for continuum emission at $\sim 32$\,GHz (see Table~\ref{tab:continuum} and Figure~\ref{fig:continuum}), resulting in a rms noise of $\sim 3.5$\,$\mu$Jy beam$^{-1}$ for HXMM20 and $\sim 1.9$\,$\mu$Jy beam$^{-1}$ for HCOSMOS02, at the phase center. Continuum subtraction is not needed since the continuum level is negligible at the rms noise of the line channels.

\begin{deluxetable}{lccc}
\tabletypesize{\scriptsize}
\tablecaption{Continuum Measurements \label{tab:continuum}}
\tablehead{\colhead{Name} & \colhead{$S_{\rm{870\mu m}}$}\tablenotemark{a} & \colhead{$S_{\rm{3mm}}$} & \colhead{$S_{\rm{32GHz}}$} \\
\colhead{} & \colhead{[mJy beam$^{-1}$]} & \colhead{[$\mu$Jy beam$^{-1}$]} & \colhead{[$\mu$Jy beam$^{-1}$]}}
\startdata
HELAISS02 \\
  S0 & 9.22 $\pm$ 0.17 & 104 $\pm$ 13 & \nodata \\
  S1 & 4.34 $\pm$ 0.16 & 51 $\pm$ 12 & \nodata \\
  S2 & 4.16 $\pm$ 0.32 & 42 $\pm$ 11 & \nodata \\
  S3 & 2.40 $\pm$ 0.19 & 43 $\pm$ 12 & \nodata \\
HXMM20 \\
  S0 & 7.15 $\pm$ 0.44 & 130 $\pm$ 22 & 21.1 $\pm$ 3.5 \\
  S1 & 3.52 $\pm$ 0.41 & 65 $\pm$ 22 & \nodata \\
  S2 & 3.42 $\pm$ 0.26 & \nodata & \nodata \\
  S3 & 2.46 $\pm$ 0.47 & \nodata & \nodata \\
  S4 & 0.94 $\pm$ 0.18 & \nodata & \nodata \\
HCOSMOS02 \\
  S0 & 5.26 $\pm$ 0.26 & \nodata & 6.4 $\pm$ 1.9 \\
  S1 & 3.77 $\pm$ 0.32 & \nodata & \nodata \\
  S2 & 1.69 $\pm$ 0.25 & \nodata & 8.9 $\pm$ 1.9 \\
  S3 & 1.66 $\pm$ 0.21 & \nodata & \nodata \\
  S4 & 2.23 $\pm$ 0.41 & \nodata & 6.0 $\pm$ 1.9 \\
\enddata
\tablenotetext{a}{From \citet{2015ApJ...812...43B}.}
\end{deluxetable}

\subsection{CARMA and NOEMA Observations} \label{subsec:other_data}

A spectral scan was carried out with CARMA during 2011 (projects cx322 and c0673; PI: D. A. Riechers) targeting $^{12}$CO($J = 3$--2) for HCOSMOS02, since the redshift of this source was unknown at that time. Once the redshift was secured, we targeted $^{12}$CO($J = 4$--3) transition line ($\nu_{\rm{rest}} = 461.04077$\,GHz) with NOEMA, formerly known as the Plateau de Bure Interferometer (PdBI), for HCOSMOS02 (project W0AB; PI: D. A. Riechers).

CARMA observations were executed in seven tracks in E configuration between 2011 January 23 and February 10, plus one track in D configuration in 2011 May 25, using 10--15 antennas. Four regular tunings were set up covering 85.48--111.48GHz at 5.208\,MHz resolution for the E configuration tracks and one custom tuning, within the frequency range of the four regular tunings, for the D configuration track. The resulting on-source spectral scan integration time was 11.9\,h. The radio quasars J0927+390 and 3C273 were observed as bandpass calibrators and the radio quasar J1058+015 as phase calibrator. The radio quasars 3C84 and 3C273 were the flux calibrators. We employed MIRIAD for reduction and imaging. The resulting synthesized beam size at 100\,GHz is 4\farcs47$\times$2\farcs80 (primary beam HPBW 60\farcs6).

NOEMA observations were carried out in two tracks in D configuration observed on 2013 April 10 and 13 using 6 antennas. The tuning frequency was set at 131.139\,GHz. We employed GILDAS for reduction and imaging. The resulting synthesized beam size at the tuning frequency is 3\farcs10$\times$1\farcs77 (primary beam HPBW 38\farcs4).

\section{Confirmation of Protocluster Cores} \label{sec:analysis}

\subsection{HELAISS02} \label{subsec:helaiss02}

Our ALMA spectral scan targeting CO(3-2) for HELAISS02 successfully detected significant emission in all four ALMA 870\,$\mu$m counterparts presented in \citet{2015ApJ...812...43B}. Therefore, we confirmed that they are located at the same redshift at a median value $z = 2.171 \pm 0.004$.

We computed the moment-0 maps for each source, which represent the total intensity integrated over the velocity axis (see Figure~\ref{fig:el02xmm20_mom0}). The velocity channels selected for integration were the line channels that maximize the signal-to-noise ratio (S/N). In Figure~\ref{fig:el02xmm20_spec} we present the spectra extracted at the pixel located at the peak of the 870\,$\mu$m continuum emission. S0, S2 and S3 detections are secure, while S1 appears tentatively detected at $\rm{S/N} < 3$ with its spectrum showing a symmetric negative peak in adjacent channels to the line due to potential sidelobe residuals. We measured centroids, widths and peak fluxes using the CASA task \texttt{specfit} fitting a single Gaussian component. The results are presented in Table~\ref{tab:helaiss02_line}. In order to calculate the integrated line fluxes we performed a 2D Gaussian fit per source and per velocity channel in the spectral cube using the CASA task \texttt{imfit}. For each source we selected the channels used to create their respective moment-0 maps as those to be fitted. No significant emission was detected in the residuals beyond a point source fit; thus, we fixed the Gaussian width and position angle to those of the clean beam and the position to the 870\,$\mu$m peak. The uncertainties in \texttt{imfit} are known to be too small when using fixed parameters, so the quoted uncertainties in Table~\ref{tab:helaiss02_line} are the 1$\sigma$ noise from the moment-0 maps instead. In addition, we calculated the line luminosity expressed in terms of the surface integrated brightness temperature \citep[$L^{'}_{\rm{CO}}$;][]{1992ApJ...398L..29S}.

3\,mm continuum emission was detected at $\rm{S/N} > 3$ for all four ALMA counterparts as well (see Figure~\ref{fig:continuum}). Measurements were also extracted at the pixel located at the peak of the 870\,$\mu$m continuum emission (see Table~\ref{tab:continuum}).

\begin{deluxetable*}{lcccccccc}
\tabletypesize{\scriptsize}
\tablecaption{HELAISS02 CO(3-2) Line Measurements \label{tab:helaiss02_line}}
\tablehead{\colhead{Name}\tablenotemark{a} & \colhead{$\alpha$(J2000)}\tablenotemark{b} & \colhead{$\delta$(J2000)}\tablenotemark{b} & \colhead{$v_{\rm{0,CO(3-2)}}$}\tablenotemark{c} & \colhead{d$v_{\rm{CO(3-2)}}$} & \colhead{$S_{\rm{CO(3-2)}}$} & \colhead{$I_{\rm{CO(3-2)}}$} & \colhead{$\rm{S/N}_{\rm{CO(3-2)}}$} & \colhead{$\log L^{'}_{\rm{CO(3-2)}}$}  \\
\colhead{} & \colhead{[h:m:s]} & \colhead{[$^\circ$:$'$:$''$]} & \colhead{[km s$^{-1}$]} & \colhead{[km s$^{-1}$]} & \colhead{[mJy beam$^{-1}$]} & \colhead{[Jy km s$^{-1}$]} & \colhead{} & \colhead{[K km s$^{-1}$ pc$^{-2}$]}}
\startdata
HELAISS02 & 00 38 23.59 & ${-}$43 37 04.15 & 2.171 $\pm$ 0.004 &  &  &  &  &  \\
  S0 & 00 38 23.76 & ${-}$43 37 06.10 & 236 $\pm$ 25 & 931 $\pm$ 58 & 2.93 $\pm$ 0.24 & 3.13 $\pm$ 0.30 & 11.8 & 10.90 $\pm$ 0.04 \\
  S1\tablenotemark{*} & 00 38 23.48 & ${-}$43 37 05.56 & $-$630 $\pm$ 100 & 910 $\pm$ 230 & 0.54 $\pm$ 0.14 & 0.53 $\pm$ 0.21 & 2.79\tablenotemark{*} & 10.13 $\pm$ 0.17 \\
  S2 & 00 38 23.31 & ${-}$43 36 58.97 & $-$236 $\pm$ 33 & 575 $\pm$ 78 & 1.70 $\pm$ 0.27 & 0.91 $\pm$ 0.16 & 7.38 & 10.36 $\pm$ 0.08 \\
  S3 & 00 38 23.80 & ${-}$43 37 10.46 & 281 $\pm$ 66 & 610 $\pm$ 160 & 0.83 $\pm$ 0.19 & 0.51 $\pm$ 0.19 & 3.48 & 10.11 $\pm$ 0.16 \\
  Total &  &  &  &  &  & 5.08 $\pm$ 0.44 &  & 11.11 $\pm$ 0.04 \\
\enddata
\tablenotetext{a}{Source names correspond to those originally reported in \citet{2015ApJ...812...43B}.}
\tablenotetext{b}{Coordinates correspond to those of the ALMA 870\,$\mu$m continuum sources as originally reported in \citet{2015ApJ...812...43B}.}
\tablenotetext{c}{Velocity offset is centered at the median redshift of the four sources.}
\tablenotetext{*}{Tentative detection.}
\end{deluxetable*}

\subsection{HXMM20} \label{subsec:hxmm20}

The ALMA and VLA spectral scans targeted CO(3-2) and CO(1-0) for HXMM20, respectively. We detected significant emission in all the five ALMA 870\,$\mu$m counterparts in \citet{2015ApJ...812...43B}. Therefore, we also confirmed that they are located at the same redshift at a median value $z = 2.602 \pm 0.002$.

\begin{figure*}
\begin{center}
\includegraphics[width=\textwidth]{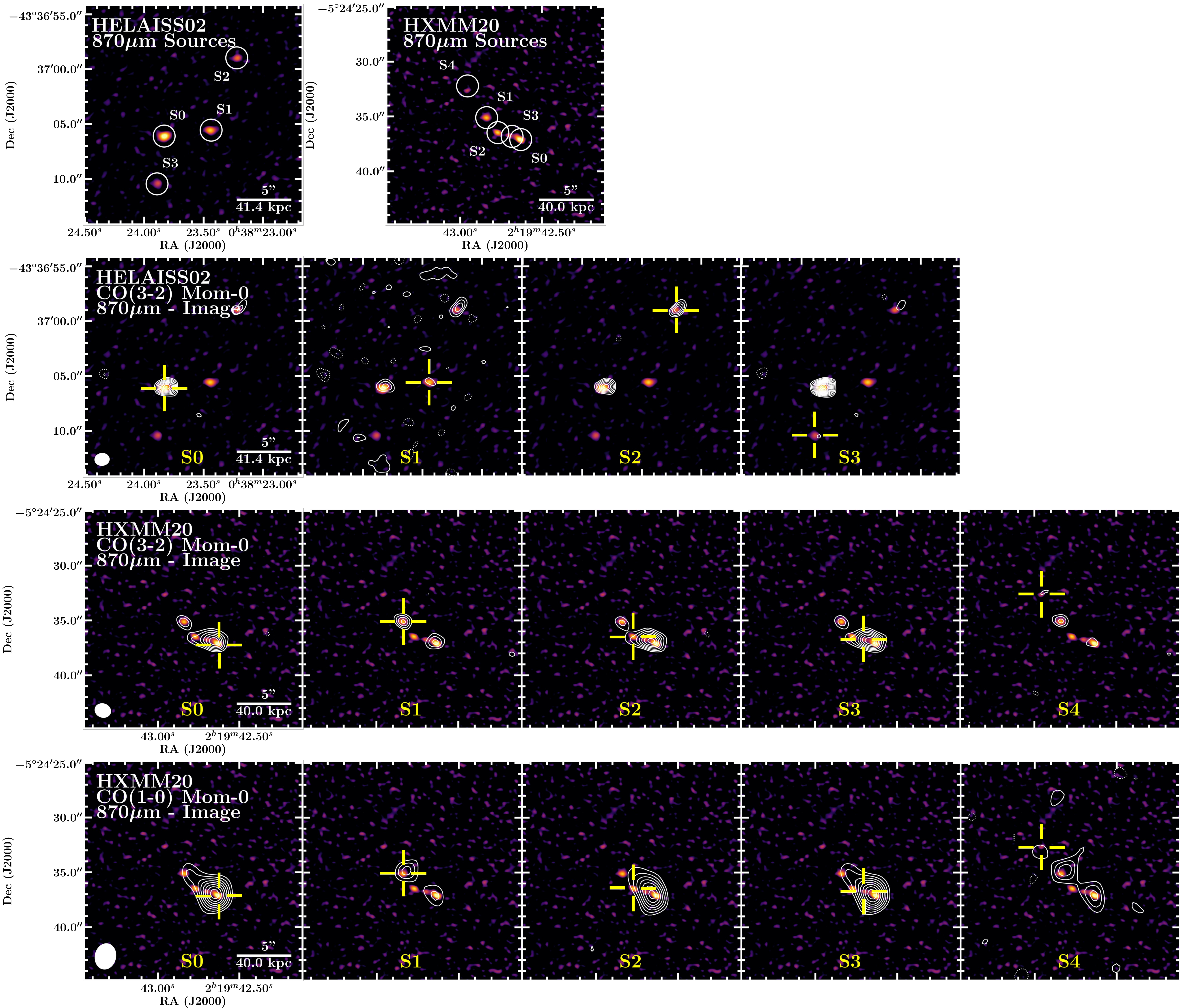}
\caption{HELAISS02 and HXMM20 moment-0 maps. First row: Overview of the ALMA 870\,$\mu$m continuum sources reported in \citet{2015ApJ...812...43B}. Second row: HELAISS02 CO(3-2) moment-0 maps of the 870\,$\mu$m continuum sources in \citet{2015ApJ...812...43B} represented as contours on top of the 870\,$\mu$m continuum image. Third row: HXMM20 CO(3-2) moment-0 maps of the 870\,$\mu$m continuum sources in \citet{2015ApJ...812...43B} on top of the 870\,$\mu$m continuum image. Fourth row: HXMM20 CO(1-0) moment-0 maps of the 870\,$\mu$m continuum sources in \citet{2015ApJ...812...43B} on top of the 870\,$\mu$m continuum image. The source to which each panel refers to is marked with a yellow cross (note that sources spanning a similar velocity range appear also in the panel by construction of a moment-0 map). Contours start at $\pm$3\,$\sigma$ and grow in steps of $\pm$1\,$\sigma$, except for HELAISS02 CO(3-2) S1 and HXMM20 CO(1-0) S4 that start at $\pm$2\,$\sigma$. Positive contours are solid and negative contours dotted.}
\label{fig:el02xmm20_mom0}
\end{center}
\end{figure*}

\begin{figure*}
\begin{center}
\includegraphics[width=\textwidth]{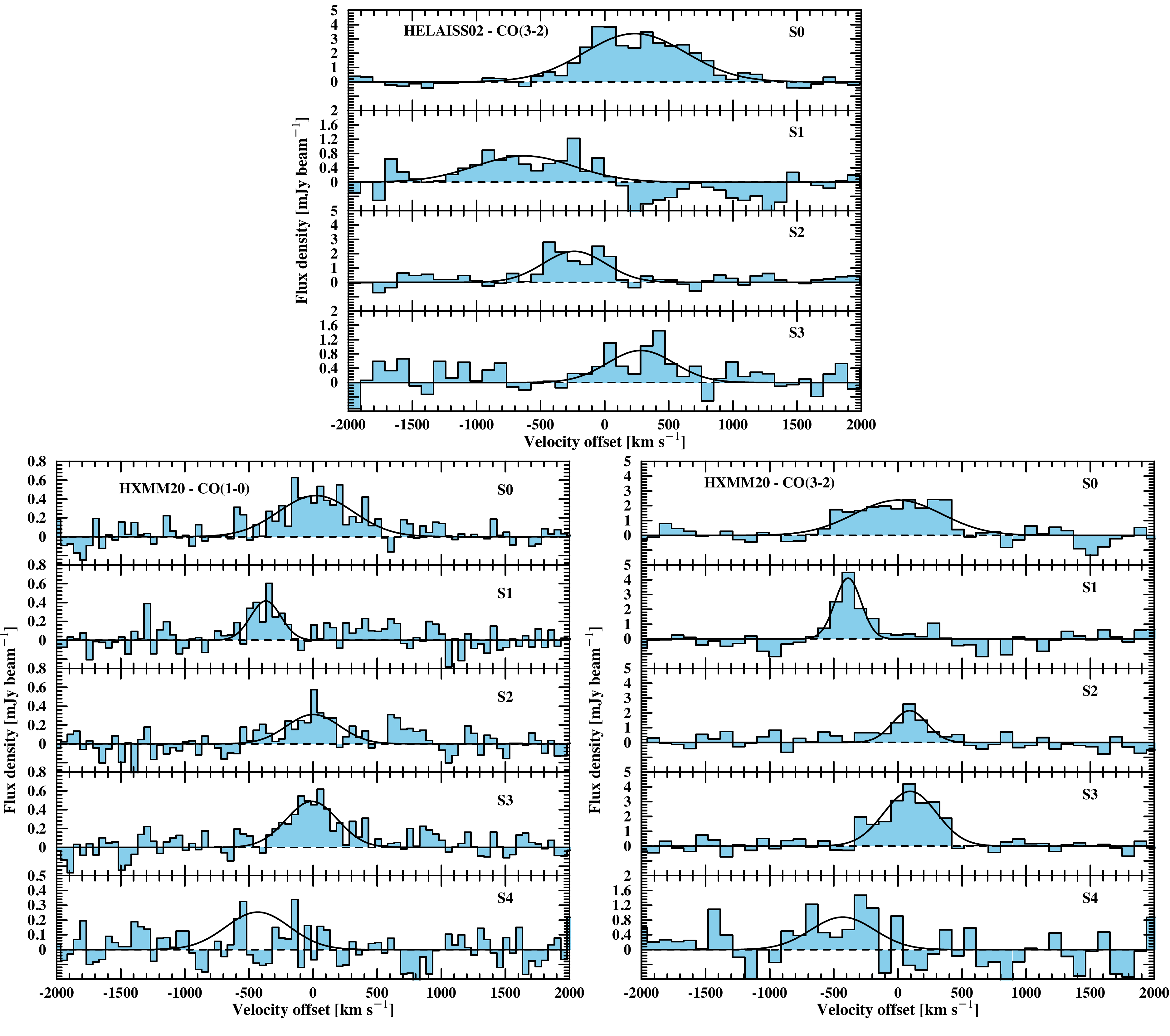}
\caption{HELAISS02 CO(3-2), HXMM20 CO(3-2), and CO(1-0) spectra of the 870\,$\mu$m continuum sources reported in \citet{2015ApJ...812...43B}. The spectra are ordered according to that nomenclature. Velocity offset for each protocluster core is centered at the median redshift of their sources given by the CO(3-2) transition.}
\label{fig:el02xmm20_spec}
\end{center}
\end{figure*}

The moment-0 maps in Figure~\ref{fig:el02xmm20_mom0} show secure detections of S1 and a blend of S0, S2 and S3. S4 is securely detected in the ALMA CO(3-2) observations, although it is only tentatively detected at $\rm{S/N} < 3$ in the VLA CO(1-0) observations. In Figure~\ref{fig:el02xmm20_spec} we present the spectra extracted at the pixel located at the peak of the 870\,$\mu$m continuum emission. We collect the line measurements in Table~\ref{tab:hxmm20_line}, obtained following the same method as in HELAISS02. For HXMM20-S4 we fixed the centroid and width of CO(1-0) line to that of the CO(3-2), since due to the low S/N part of the emission was not properly accounted in a regular Gaussian fit with free parameters. In the case of HXMM20 the 2D Gaussian fit to calculate the integrated line fluxes is particularly important to properly deblend the emission of S0, S2 and S3, since it operates on each channel taking advantage of the variation of the spatial location of the emission that moves across the different sources in velocity space. For consistency, we checked that the recovered fluxes in these blended sources are consistent with that measured in a moment-0 map created by collapsing over the line channels of the three sources in an aperture enclosing all of them. Therefore, we are not double-counting flux in the blended sources. No significant emission was detected in the residuals beyond the point source fit.

We measured the line brightness temperature ratio $r_{\rm{31}} = L^{'}_{\rm{CO(3-2)}}/L^{'}_{\rm{CO(1-0)}}$, resulting in high values as observed in starburst galaxies such as submillimeter galaxies \citep[SMGs; e.g.,][]{2013MNRAS.429.3047B}. In the case of S1 is also consistent with thermalized level populations (see Table~\ref{tab:hxmm20_line}).

3\,mm and 32\,GHz continuum emission was detected for S0 and 3\,mm for S3 (see Figure~\ref{fig:continuum}). Measurements were also extracted at the pixel located at the peak of the 870\,$\mu$m continuum emission (see Table~\ref{tab:continuum}).

\begin{splitdeluxetable*}{lccccccBccccccccc}
\tabletypesize{\scriptsize}
\tablecaption{HXMM20 CO(1-0) and CO(3-2) Line Measurements \label{tab:hxmm20_line}}
\tablehead{\colhead{Name}\tablenotemark{a} & \colhead{$\alpha$(J2000)}\tablenotemark{b} & \colhead{$\delta$(J2000)}\tablenotemark{b} & \colhead{$v_{\rm{0,CO(1-0)}}$}\tablenotemark{c} & \colhead{$v_{\rm{0,CO(3-2)}}$}\tablenotemark{c} & \colhead{d$v_{\rm{CO(1-0)}}$} & \colhead{d$v_{\rm{CO(3-2)}}$} & \colhead{$S_{\rm{CO(1-0)}}$} & \colhead{$S_{\rm{CO(3-2)}}$} & \colhead{$I_{\rm{CO(1-0)}}$} & \colhead{$I_{\rm{CO(3-2)}}$} & \colhead{$\rm{S/N}_{\rm{CO(1-0)}}$} & \colhead{$\rm{S/N}_{\rm{CO(3-2)}}$} & \colhead{$\log L^{'}_{\rm{CO(1-0)}}$} & \colhead{$\log L^{'}_{\rm{CO(3-2)}}$} & \colhead{$r_{\rm{31}}$} \\
\colhead{} & \colhead{[h:m:s]} & \colhead{[$^\circ$:$'$:$''$]} & \colhead{[km s$^{-1}$]} & \colhead{[km s$^{-1}$]} & \colhead{[km s$^{-1}$]} & \colhead{[km s$^{-1}$]} & \colhead{[mJy beam$^{-1}$]} & \colhead{[mJy beam$^{-1}$]} & \colhead{[Jy km s$^{-1}$]} & \colhead{[Jy km s$^{-1}$]} & \colhead{} & \colhead{} & \colhead{[K km s$^{-1}$ pc$^{-2}$]} & \colhead{[K km s$^{-1}$ pc$^{-2}$]} & \colhead{}}
\startdata
HXMM20    & 02 19 42.78 & ${-}$05 24 34.84 &  & 2.602 $\pm$ 0.002 &  &  &  &  &  &  &  &  &  &  &  \\
  S0 & 02 19 42.63 & ${-}$05 24 37.11 & 22 $\pm$ 34 & 0 $\pm$ 42 & 688 $\pm$ 81 & 803 $\pm$ 98 & 0.44 $\pm$ 0.04 & 2.38 $\pm$ 0.25 & 0.32 $\pm$ 0.05 & 2.00 $\pm$ 0.25 & 11.5 & 9.30 & 11.00 $\pm$ 0.07 & 10.84 $\pm$ 0.05 & 0.69 $\pm$ 0.14 \\
  S1 & 02 19 42.84 & ${-}$05 24 35.11 & $-$369 $\pm$ 27 & $-$389 $\pm$ 13 & 278 $\pm$ 63 & 241 $\pm$ 30 & 0.42 $\pm$ 0.08 & 4.11 $\pm$ 0.45 & 0.12 $\pm$ 0.04 & 1.24 $\pm$ 0.17 & 6.15 & 6.74 & 10.57 $\pm$ 0.14 & 10.63 $\pm$ 0.06 & 1.15 $\pm$ 0.41 \\
  S2 & 02 19 42.77 & ${-}$05 24 36.48 & 1 $\pm$ 46 & 91 $\pm$ 24 & 490 $\pm$ 110 & 319 $\pm$ 56 & 0.31 $\pm$ 0.06 & 2.15 $\pm$ 0.32 & 0.16 $\pm$ 0.05 & 0.83 $\pm$ 0.22 & 4.23 & 4.05 & 10.70 $\pm$ 0.14 & 10.46 $\pm$ 0.12 & 0.57 $\pm$ 0.24 \\
  S3 & 02 19 42.68 & ${-}$05 24 36.82 & ${-}$10 $\pm$ 25 & 96 $\pm$ 19 & 473 $\pm$ 58 & 484 $\pm$ 45 & 0.49 $\pm$ 0.05 & 3.70 $\pm$ 0.30 & 0.25 $\pm$ 0.04 & 1.57 $\pm$ 0.21 & 9.04 & 8.86 & 10.89 $\pm$ 0.07 & 10.73 $\pm$ 0.06 & 0.70 $\pm$ 0.14 \\
  S4\tablenotemark{*} & 02 19 42.96 & ${-}$05 24 32.22 & $-$431 $\pm$ 99 & $-$431 $\pm$ 99 & 590 $\pm$ 230 & 590 $\pm$ 230 & 0.25 $\pm$ 0.07 & 0.88 $\pm$ 0.30 & 0.15 $\pm$ 0.05 & 0.55 $\pm$ 0.18 & 2.71\tablenotemark{*} & 3.12 & 10.67 $\pm$ 0.14 & 10.28 $\pm$ 0.14 & 0.41 $\pm$ 0.19 \\
  Total &  &  &  &  &  &  &  &  & 1.00 $\pm$ 0.10 & 6.19 $\pm$ 0.47 &  &  & 11.49 $\pm$ 0.04 & 11.08 $\pm$ 0.03 & 0.69 $\pm$ 0.09 \\
\enddata
\tablenotetext{a}{Source names correspond to those originally reported in \citet{2015ApJ...812...43B}.}
\tablenotetext{b}{Coordinates correspond to those of the ALMA 870\,$\mu$m continuum sources as originally reported in \citet{2015ApJ...812...43B}.}
\tablenotetext{c}{Velocity offset is centered at the median CO(3-2) redshift of the five sources.}
\tablenotetext{*}{Tentative CO(1-0) detection. Centroid and width were fixed to that of CO(3-2).}
\end{splitdeluxetable*}

\subsection{HCOSMOS02} \label{subsec:hcosmos02}

The combined VLA programs for HCOSMOS02 (see Section~\ref{sec:sample_data}\footnote{For an upcoming independent analysis of the archival data, see J. Champagne et al., in prep.}) targeted CO(1-0) at the redshift of the source \citep[$z = 2.506$, found by][and in our CARMA 3\,mm data]{2016ApJ...828...56W}. We analyzed the ALMA 870\,$\mu$m counterparts reported in \citet{2015ApJ...812...43B} (namely HCOSMOS02-S0, S1, S2, S3, and S4).

We carried out a line blind search over the whole frequency range covered by the combination of VLA programs in a FOV as large as the ALMA pipeline allows by default. The sensitivity decays as we move away from the phase center, following the primary beam response, and the pipeline masked regions below 10\% of the phase center sensitivity. This corresponds to a FOV $1.6 \times \rm{HPBW} = 132\farcs0$. The blind search was performed on the final image cube using \texttt{MF3D}\footnote{Code available at https://github.com/pavesiriccardo/MF3D} \citep{2018ApJ...864...49P}. This algorithm implements a Matched Filtering in 3D line search, which is optimized for Gaussian line profiles and either spatially unresolved or slightly resolved emission \citep[see][for details]{2018ApJ...864...49P}. The purities analysis revealed that $\rm{S/N} > 5.8$ is the threshold above which the ratio of spurious negative detections over positive detections is 0. At $5.0 < \rm{S/N} < 5.8$ we found 12 negative detections and 14 positive detections. We checked all the $5.0 < \rm{S/N} < 5.8$ sources. The line extraction showed symmetric negative peaks, or just consistent on spikes of two or three channels. Besides, they did not show an optical/near-IR counterpart. Therefore, we ended up discarding the sources in the range $5.0 < \rm{S/N} < 5.8$ since they were not reliable. We detected eight sources at $\rm{S/N} > 5.8$ (namely S0, S1, S2, S4, S9, S11, S13, and S14).

Additionally, we analyzed the sources in Table~1 from \citet{2016ApJ...828...56W} and in Table~1 from \citet{2018ApJ...867L..29W} that fall within our FOV, comprising all the sources in the tables except for those with the IDs 128484, 129305, 129444, 132636, 132627 that fall outside our FOV and, thus, below 10\% of the primary beam sensitivity.

\begin{figure*}
\begin{center}
\includegraphics[width=0.95\textwidth]{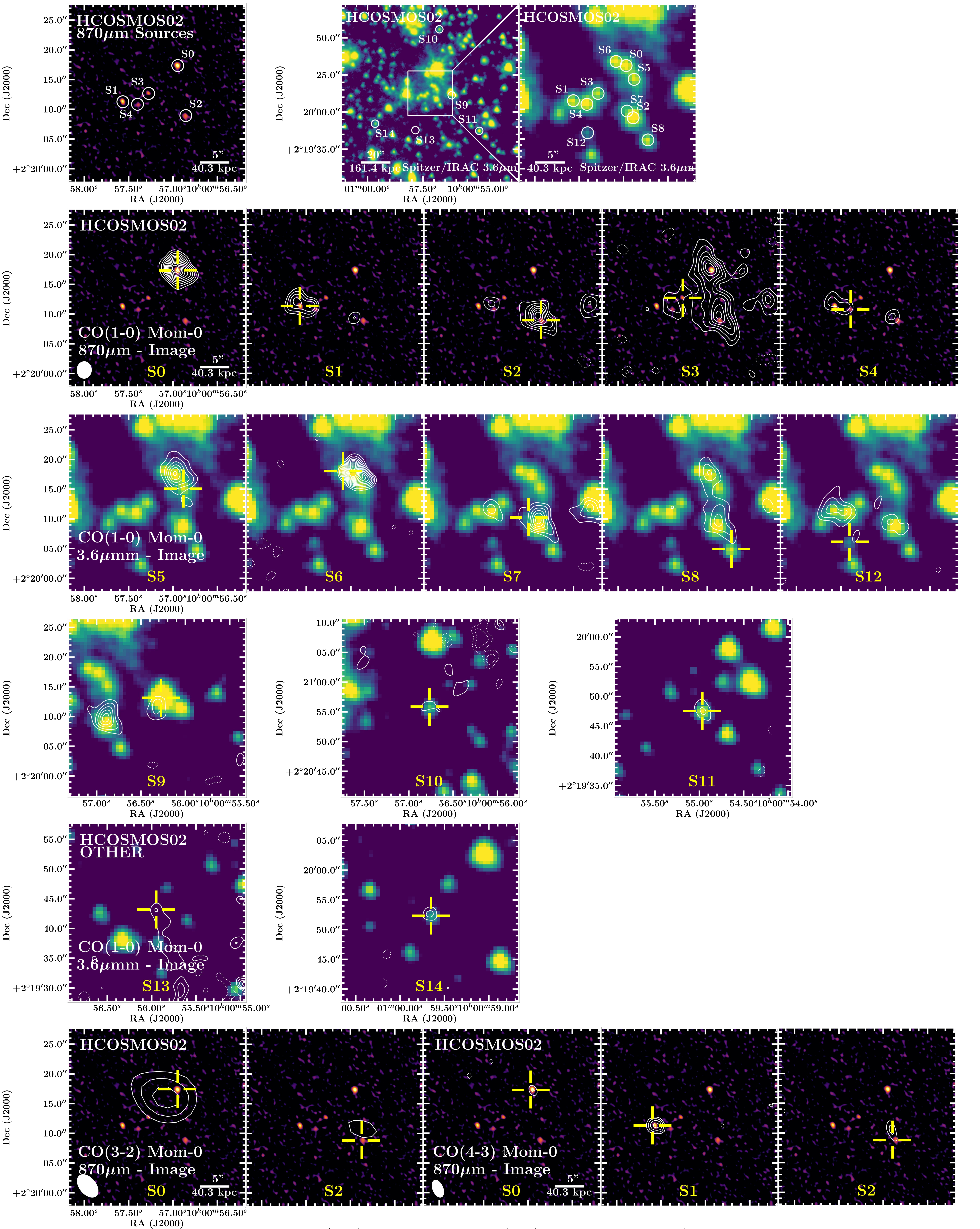}
\caption{HCOSMOS02 moment-0 maps. First row: Overview of the sources. Second row: CO(1-0) moment-0 maps of the 870\,$\mu$m continuum sources in \citet{2015ApJ...812...43B} represented as contours on top of the 870\,$\mu$m continuum image. Third row: CO(1-0) moment-0 maps of the detections in the 30\arcsec$\times$30\arcsec central region on top of the IRAC 3.6\,$\mu$m image. Fourth row: CO(1-0) moment-0 maps of the detections outside the 30\arcsec$\times$30\arcsec central region. Fifth row: Moment-0 maps of the line detections outside the 30\arcsec$\times$30\arcsec central region not part of the HCOSMOS02 structure. Sixth row: CO(3-2) from CARMA and CO(4-3) from NOEMA moment-0 maps of the detected 870\,$\mu$ continuum sources. The source to which each panel refers to is marked with a yellow cross (note that sources spanning a similar velocity range appear also in the panel by construction of a moment-0 map). Contours start at $\pm$3\,$\sigma$ and grow in steps of $\pm$1\,$\sigma$, except for CO(1-0) S3, S5, S8, S10, S12 and S13 that start at $\pm$2\,$\sigma$. Positive contours are solid and negative contours dotted.}
\label{fig:hcosmos02_mom0}
\end{center}
\end{figure*}

\begin{figure*}
\begin{center}
\includegraphics[width=\textwidth]{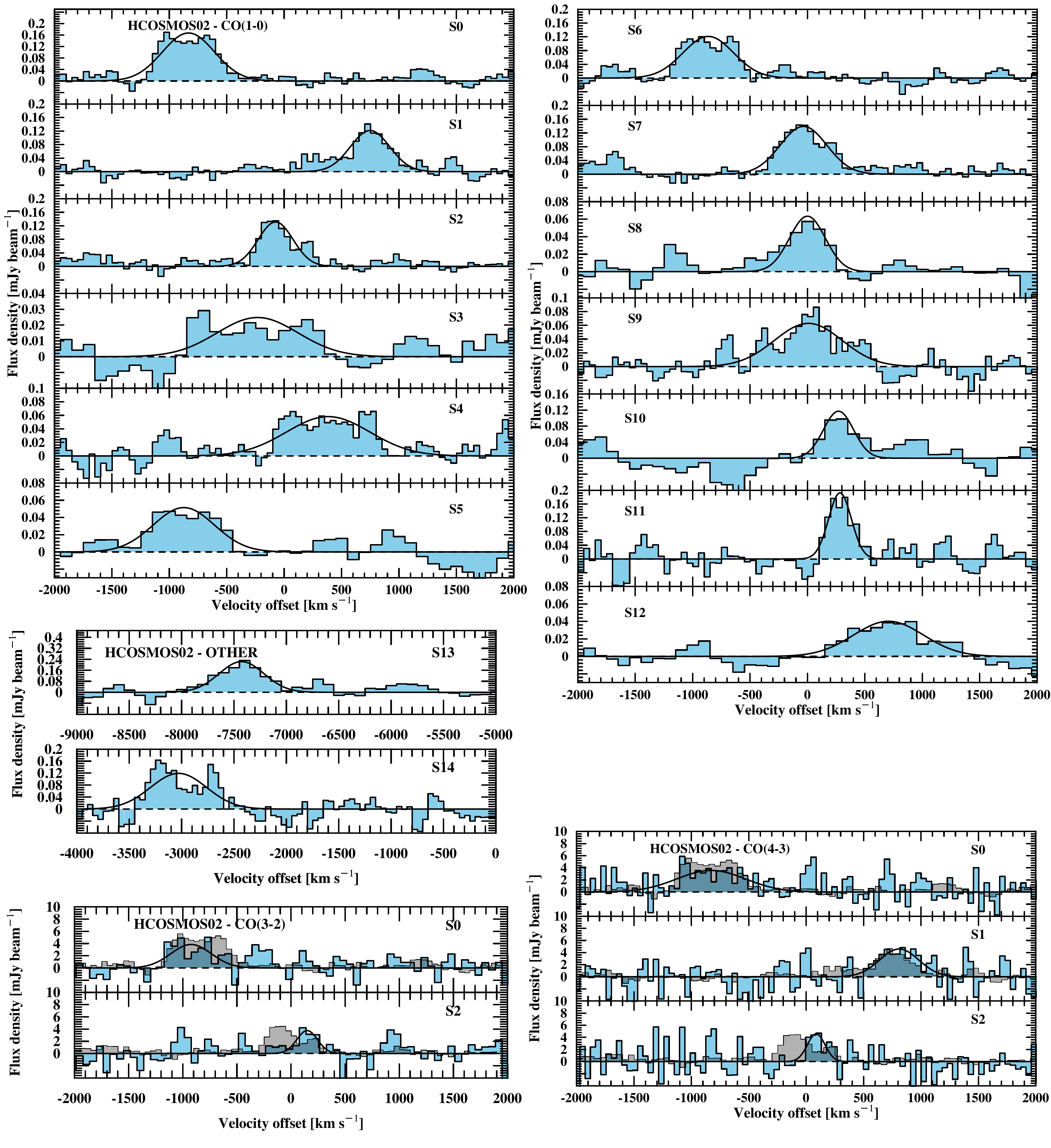}
\caption{HCOSMOS02 CO(1-0), CO(3-2), and CO(4-3) spectra. Source names are those originally reported in \citet{2015ApJ...812...43B} for S0 to S4. The rest are named subsequently with increasing velocity. The spectra are ordered according to the nomenclature. Velocity offset is centered at the median redshift of the sources given by the CO(1-0) transition. Scaled CO(1-0) spectra are overlaid on top of the CO(3-2) and CO(4-3) in gray.}
\label{fig:hcosmos02_spec}
\end{center}
\end{figure*}

We show the moment-0 maps for each source in Figure~\ref{fig:hcosmos02_mom0} and their spectra in Figure~\ref{fig:hcosmos02_spec}. Measurements were performed following the same method as in HELAISS02 and HXMM20. Spectra were extracted at the pixel peak of the 870\,$\mu$m continuum emission for \citet{2015ApJ...812...43B} sources. In the case of the sources from the blind search the spectra were extracted at the position of the detection given by the code, which are consistent with the coordinates in \citet{2016ApJ...828...56W} for the sources that appear in this previous study. The spectra were binned at 100\,km s$^{-1}$ for the sources with $\rm{S/N} < 3$. All measurements are collected in Table~\ref{tab:hcosmos02_line}.

The moment-0 maps show that S0, S1 and S2 look extended. However, checking the \textit{Spitzer}/IRAC 3.6\,$\mu$m (SPLASH; Capak et al., in prep.) image we found that both S0 and S2 are associated with two IRAC counterparts each. In the case of S1 there is no additional IRAC counterpart at the northeast where the excess of CO emission is located, but this excess can be well modeled by an additional component covering a frequency range that is narrower and blueshifted respect to S1. The 2D Gaussian fit for S0, S1 and S2 was performed using an extra component on each source centered at the coordinates of the additional IRAC counterparts for S0 (namely S6), S2 (namely S7), and northeast of S1 with no additional IRAC counterpart (we included its flux contribution in S1). No significant emission was detected beyond the point source fit with the extra components. For consistency, we checked that the recovered fluxes in these blended sources are consistent with that measured in a moment-0 map created by collapsing over the line channels of the blended sources in an aperture enclosing all of them and, thus, we are not double-counting flux (as done for HXMM20 in Section~\ref{subsec:hxmm20}).

All the \citet{2015ApJ...812...43B} sources were securely detected, except for S3, which was only tentatively detected containing potential sidelobe residuals and a symmetric negative peak in the adjacent line channels. S3 is known for displaying prominent stellar, 870\,$\mu$m, and 1.4\,GHz continumm emission, hosting a radio loud AGN \citep{2016ApJ...828...56W,2017ApJ...846L..31D}. In the case of the sources from \citet{2016ApJ...828...56W} and \citet{2018ApJ...867L..29W} we detected the same sources except for those with the IDs 132044 and 131661, at which position we did not retrieve significant emission at $\rm{S/N} > 2$. Note that we also report S6, additional IRAC counterpart next to S0, and also S8 and S12, both of which have IRAC counterparts, but detected at $\rm{S/N} < 3$ and also showing possible sidelobe residuals, which we classify as tentative. The blind search arose a tentative detection for an extra source namely S13 and an extra secure source namely S14 part of the structure encompassing a larger redshift range \citep{2013ApJ...765..109D,2015ApJ...802...31D,2015ApJ...808...37C,2015ApJ...808L..33C,2016ApJ...817..160L,2018A&A...619A..49C}. Note that the sources with IDs 132617 and 129444 in \citet{2016ApJ...828...56W} and \citet{2018ApJ...867L..29W} were not covered by our FOV that stops at 10\% of the sensitivity at the phase center.

All the tentative sources with $\rm{S/N} < 4$ and affected by potential sidelobe contamination need further observations to be securely confirmed.

Additionally, our CARMA program searched for CO(3-2) and our NOEMA program targeted CO(4-3) (see Section~\ref{subsec:other_data}). S0 and S2 are detected in CO(3-2) and S0, S1, and S2 in C0(4-3). Note that the beam size is larger in these observations than in VLA, especially in the case of CARMA. Therefore, CO(3-2) and CO(4-3) could come from several or different neighboring sources. The case of S2 is particularly clear, since CO(3-2) and CO(4-3) line detections are offset in velocity from that of CO(1-0), but also the spatial location of the CO(3-2) and CO(4-3) emissions point towards a contribution from S7, which CO(1-0) is slightly broader and offset from that of S2. The line ratios are unphysical when considering that the CO(3-2) and CO(4-3) are associated to a single source. However, they become physical when adding up the CO(1-0) contribution from S5 and S6 to S0 and the CO(1-0) contribution from S7 to S2.

33\,GHz continuum emission was detected at $\rm{S/N} > 3$ slightly offset form S0, S2 and S4 (see Figure~\ref{fig:continuum}).

\begin{deluxetable*}{lccccccccccccc}
\tabletypesize{\scriptsize}
\tablecaption{HCOSMOS02 CO(1-0) Line Measurements \label{tab:hcosmos02_line}}
\tablehead{\colhead{Name}\tablenotemark{a} & \colhead{Other Name}\tablenotemark{b} & \colhead{$\alpha$(J2000)}\tablenotemark{a} & \colhead{$\delta$(J2000)}\tablenotemark{a} & \colhead{$v_{\rm{0,CO(1-0)}}$}\tablenotemark{c} & \colhead{d$v_{\rm{CO(1-0)}}$} & \colhead{$S_{\rm{CO(1-0)}}$} & \colhead{$I_{\rm{CO(1-0)}}$} & \colhead{$\rm{S/N}_{\rm{CO(1-0)}}$} & \colhead{$\log L^{'}_{\rm{CO(1-0)}}$} \\
\colhead{} & \colhead{} & \colhead{[h:m:s]} & \colhead{[$^\circ$:$'$:$''$]} & \colhead{[km s$^{-1}$]} & \colhead{[km s$^{-1}$]} & \colhead{[mJy beam$^{-1}$]} & \colhead{[Jy km s$^{-1}$]} & \colhead{} & \colhead{[K km s$^{-1}$ pc$^{-2}$]}}
\startdata
HCOSMOS02 & CL J1001+0220 & 10 00 57.18 & $+$02 20 12.70 & 2.504 $\pm$ 0.005 &  &  &  &  &  &  &  &  &  \\
  S0 & 131077 & 10 00 56.95 & $+$02 20 17.35 & $-$833 $\pm$ 26 & 534 $\pm$ 61 & 0.167 $\pm$ 0.017 & 0.105 $\pm$ 0.007 & 15.7 & 10.48 $\pm$ 0.03 \\
  S1 & 130891 & 10 00 57.57 & $+$02 20 11.26 & 748 $\pm$ 27 & 404 $\pm$ 63 & 0.122 $\pm$ 0.016 & 0.129 $\pm$ 0.013 & 6.93 & 10.58 $\pm$ 0.04 \\
  S2 & 130949 & 10 00 56.86 & $+$02 20 08.93 & $-$74 $\pm$ 24 & 358 $\pm$ 57 & 0.131 $\pm$ 0.018 & 0.032 $\pm$ 0.006 & 10.29 & 9.97 $\pm$ 0.08 \\
  S3\tablenotemark{*} & 130933 & 10 00 57.27 & $+$02 20 12.66 & $-$230 $\pm$ 170 & 830 $\pm$ 390 & 0.025 $\pm$ 0.010 & 0.022 $\pm$ 0.014 & 3.85\tablenotemark{*} & 9.81 $\pm$ 0.28 \\
  S4 & 130901 & 10 00 57.40 & $+$02 20 10.83 & 384 $\pm$ 86 & 860 $\pm$ 210 & 0.058 $\pm$ 0.012 & 0.065 $\pm$ 0.014 & 5.13 & 10.28 $\pm$ 0.09 \\
  S5\tablenotemark{*} & 131079 & 10 00 56.88 & $+$02 20 14.93 & $-$874 $\pm$ 64 & 630 $\pm$ 150 & 0.051 $\pm$ 0.011 & 0.034 $\pm$ 0.011 & 2.37\tablenotemark{*} & 9.99 $\pm$ 0.14 \\
  S6 & \nodata & 10 00 57.06 & $+$02 20 18.40 & $-$863 $\pm$ 30 & 519 $\pm$ 71 & 0.121 $\pm$ 0.014 & 0.078 $\pm$ 0.006 & 8.74 & 10.35 $\pm$ 0.03 \\
  S7 & 130842 & 10 00 56.90 & $+$02 20 09.70 & $-$34 $\pm$ 26 & 479 $\pm$ 61 & 0.139 $\pm$ 0.015 & 0.073 $\pm$ 0.007 & 8.98 & 10.33 $\pm$ 0.04 \\
  S8\tablenotemark{*} & \nodata & 10 00 56.70 & $+$02 20 05.20 & 0 $\pm$ 43 & 370 $\pm$ 100 & 0.063 $\pm$ 0.015 & 0.025 $\pm$ 0.009 & 2.22\tablenotemark{*} & 9.86 $\pm$ 0.16 \\
  S9 & no-ID & 10 00 56.32 & $+$02 20 11.50 & 8 $\pm$ 61 & 700 $\pm$ 140 & 0.063 $\pm$ 0.011 & 0.046 $\pm$ 0.013 & 5.27 & 10.13 $\pm$ 0.12 \\
  S10\tablenotemark{*} & 132044 & 10 00 56.76 & $+$02 20 55.72 & 271 $\pm$ 46 & 310 $\pm$ 110 & 0.117 $\pm$ 0.035 & 0.039 $\pm$ 0.018 & 2.59\tablenotemark{*} & 10.06 $\pm$ 0.20 \\
  S11 & 130359 & 10 00 54.96 & $+$02 19 48.10 & 284 $\pm$ 27 & 234 $\pm$ 63 & 0.192 $\pm$ 0.044 & 0.048 $\pm$ 0.017 & 4.26 & 10.15 $\pm$ 0.15 \\
  S12\tablenotemark{*} & \nodata & 10 00 57.38 & $+$02 20 06.40 & 706 $\pm$ 96 & 720 $\pm$ 230 & 0.040 $\pm$ 0.011 & 0.031 $\pm$ 0.013 & 2.57\tablenotemark{*} & 9.96 $\pm$ 0.18 \\
HCOSMOS02 - \\
OTHER \\
  S13\tablenotemark{*} & \nodata & 10 00 57.84 & $+$02 19 47.80 & $-$7428 $\pm$ 12 & 105 $\pm$ 27 & 0.414 $\pm$ 0.094 & 0.046 $\pm$ 0.016 & 3.00\tablenotemark{*} & \nodata \\
  S14 & \nodata & 10 00 59.66 & $+$02 19 52.90 & $-$3029 $\pm$ 74 & 630 $\pm$ 170 & 0.119 $\pm$ 0.029 & 0.080 $\pm$ 0.030 & 4.66 & \nodata \\
\enddata
\tablenotetext{a}{Source names and coordinates correspond to those originally reported in \citet{2015ApJ...812...43B} for S0 to S4. The rest of sources are named subsequently with increasing velocity and their coordinates correspond to the position where the spectrum was extracted as explained in Section~\ref{subsec:hcosmos02}.}
\tablenotetext{b}{From \citet{2016ApJ...828...56W}.}
\tablenotetext{c}{Velocity offset is centered at the median redshift of sources.}
\tablenotetext{*}{Tentative detection.}
\end{deluxetable*}

\begin{splitdeluxetable*}{lccccccBcccccccc}
\tabletypesize{\scriptsize}
\tablecaption{HCOSMOS02 CO(3-2) and CO(4-3) Line Measurements \label{tab:hcosmos02_line2}}
\tablehead{\colhead{Name} & \colhead{$v_{\rm{0,CO(3-2)}}$}\tablenotemark{a} & \colhead{$v_{\rm{0,CO(4-3)}}$}\tablenotemark{a} & \colhead{d$v_{\rm{CO(3-2)}}$} & \colhead{d$v_{\rm{CO(4-3)}}$} & \colhead{$S_{\rm{CO(3-2)}}$} & \colhead{$S_{\rm{CO(4-3)}}$} & \colhead{$I_{\rm{CO(3-2)}}$} & \colhead{$I_{\rm{CO(4-3)}}$} & \colhead{$\rm{S/N}_{\rm{CO(3-2)}}$} & \colhead{$\rm{S/N}_{\rm{CO(4-3)}}$} & \colhead{$\log L^{'}_{\rm{CO(3-2)}}$} & \colhead{$\log L^{'}_{\rm{CO(4-3)}}$} & \colhead{$r_{\rm{31}}$} & \colhead{$r_{\rm{41}}$} \\
\colhead{} & \colhead{[km s$^{-1}$]} & \colhead{[km s$^{-1}$]} & \colhead{[km s$^{-1}$]} & \colhead{[km s$^{-1}$]} & \colhead{[mJy beam$^{-1}$]} & \colhead{[mJy beam$^{-1}$]} & \colhead{[Jy km s$^{-1}$]} & \colhead{[Jy km s$^{-1}$]} & \colhead{} & \colhead{} & \colhead{[K km s$^{-1}$ pc$^{-2}$]} & \colhead{[K km s$^{-1}$ pc$^{-2}$]} & \colhead{} & \colhead{}}
\startdata
HCOSMOS02 \\
  S0 & $-$923 $\pm$ 41 & $-$828 $\pm$ 77 & 442 $\pm$ 98 & 710 $\pm$ 180 & 3.84 $\pm$ 0.72 & 3.56 $\pm$ 0.80 & 1.70 $\pm$ 0.31 & 2.51 $\pm$ 0.57 & 5.48 & 4.40 & 10.74 $\pm$ 0.08 & 10.66 $\pm$ 0.10 & 1.18 $\pm$ 0.24\tablenotemark{*} & 0.97 $\pm$ 0.23\tablenotemark{*} \\
  S1 & \nodata & 810 $\pm$ 50 & \nodata & 400 $\pm$ 120 & \nodata & 4.6 $\pm$ 1.2 & \nodata & 1.84 $\pm$ 0.30 & \nodata & 6.13 & \nodata & 10.53 $\pm$ 0.07 & \nodata & 0.89 $\pm$ 0.17 \\
  S2 & 149 $\pm$ 31 & 87 $\pm$ 31 & 225 $\pm$ 73 & 181 $\pm$ 74 & 3.7 $\pm$ 1.0 & 4.6 $\pm$ 1.6 & 0.84 $\pm$ 0.25 & 0.83 $\pm$ 0.19 & 3.36 & 4.37 & 10.44 $\pm$ 0.13 & 10.18 $\pm$ 0.10 & 0.86 $\pm$ 0.26\tablenotemark{*} & 0.48 $\pm$ 0.12\tablenotemark{*} \\
\enddata
\tablenotetext{a}{Velocity offset is centered at the median redshift as quoted in Table~\ref{tab:hcosmos02_line}.}
\tablenotetext{*}{CO(1-0) contribution from S5 and S6 added up to S0 and CO(1-0) contribution from S7 added up to S2.}
\end{splitdeluxetable*}

\section{Gas, Dust and Stellar Properties} \label{sec:results}

In this section we derive the gas, dust and stellar properties of the confirmed new protoclusters cores HELAISS02 and HXMM20. Particularly, we calculated the molecular gas masses, infrared luminosities, star formation rates, and stellar masses. Note that in the case of HCOSMOS02 these properties have been well studied in \citet{2016ApJ...828...56W} and \citet{2018ApJ...867L..29W}; therefore, we use the values obtained in those works, with updated molecular gas masses based on our CO observations. We compared our CO(1-0) line luminosity values with those in \citet{2018ApJ...867L..29W}. The median of the relative difference between the two estimates is $\sim 7$\% and, thus, we argue that there are no systematics between the two works. Individually, the estimates are in good agreement within a factor of two (for the comparison we added up S6 to S0 to be compared with 131077 in \citet{2018ApJ...867L..29W} and added up S8 to S2 to be compared with 130949 in \citet{2018ApJ...867L..29W}).

\subsection{CO-based Estimates of $M_{\rm{H_2}}$} \label{subsec:mgas_co}

One of the most commonly used methods to derive the molecular gas mass ($M_{\rm{H_2}}$) is by measuring the CO(1-0) line luminosity ($L^{'}_{\rm{CO(1-0)}}$) and assuming an $\alpha_{\rm{CO}}$ conversion factor that relates them through:

\begin{equation}
M_{\rm{H_2}} = \alpha_{\rm{CO}} L^{'}_{\rm{CO(1-0)}} ,
\end{equation}

$\alpha_{\rm{CO}}$ depends on metallicity and likely on the mode of star formation. In the absence of direct gas-phase metallicities and since the majority of our targets are massive ($M_{\rm{*}} > 10^{10}$\,$M_{\odot}$, Table~\ref{tab:gds_prop}), we assumed a solar metallicity for all sources. Then, we adopted $\alpha_{\rm{CO}} = 3.5$ as reported in \citet{2017A&A...603A..93M} for normal SFGs at solar metallicity, calculated as an average value from $\alpha_{\rm{CO}}$--$Z$ relations in the literature \citep{2011ApJ...737...12L,2012ApJ...746...69G,2012ApJ...760....6M}. In the case of HELAISS02 we also converted the CO(3-2) measurements into a CO(1-0) line luminosity. For this conversion we used the line ratio $r_{\rm{31}} = 0.69 \pm 0.09$ derived for HXMM20 from our data (see Table~\ref{tab:hxmm20_line}), assuming that the sample selection criteria are leading to a similar excitation. The $M_{\rm{H_2}}^{\rm{CO}}$ results are collected in Table~\ref{tab:gds_prop}.

\subsection{FIR Properties} \label{subsec:fir}

The available photometry from mid-IR to sub-millimeter can be fitted to derive the dust mass ($M_{\rm{dust}}$) and infrared luminosity ($L_{\rm{IR}}$) estimates of the different ALMA 870\,$\mu$m continuum sources of each protocluster core. We acquired \textit{Spitzer}/MIPS 24\,$\mu$m measurements using the images publicly available from the \textit{Spitzer} Wide-area Infrared Extragalactic Survey \citep[SWIRE;][]{2003PASP..115..897L} in the ELAIS-S1 and XMM-LSS fields, where our protocluster cores are located. Since the sources are blended, following \citet{2018ApJ...856..121G}, we got the fluxes by fitting a PSF model using \texttt{GALFIT} \citep{2002AJ....124..266P}. We required at least a 5$\sigma$ detection to perform the fit. The number of PSFs and the PSFs centroids were set to the number and positions of the 870\,$\mu$m continuum sources, allowing a shift in both the X and Y $< 1$\,pixel from the initial positions. The sources are not detected in \textit{Spitzer}/MIPS 70\,$\mu$m or \textit{Herschel}/PACS 100 and 160\,$\mu$m imaging. In the case of \textit{Herschel}/SPIRE 250, 350, and 500\,$\mu$m we scaled the total fluxes presented in \citet{2015ApJ...812...43B} using the ratio of the 870\,$\mu$m fluxes of each ALMA continuum source by the total 870\,$\mu$m flux also presented in \citet{2015ApJ...812...43B}. Finally, we employed 3\,mm fluxes presented in Table~\ref{tab:continuum}.

We fitted the mid-IR to sub-millimeter spectral energy distribution (SED) with the \citet{2007ApJ...657..810D} models (DL07). The methodology has been presented in detail in various previous studies \citep[e.g.,][]{2012ApJ...760....6M,2016A&A...587A..73B}. In brief, DL07 models describe the mid-IR to sub-millimeter spectrum of a galaxy by a linear combination of two dust components, one arising from dust in the diffuse interstellar medium (ISM), heated by a minimum radiation field $U_{\rm{min}}$ ("diffuse ISM" component) and the other from dust heated by a power-law distribution of starlight, $dM/dU \propto U^{-\alpha}$ extending from $U_{\rm{min}}$ to $U_{\rm{max}}$, associated with the intense photodissociation regions (PDRs, "PDR" component). The relative contribution of the two components is quantified by the parameter $\gamma$ that yields the fraction of the dust exposed to starlight with intensities ranging from $U_{\rm{min}}$ to $U_{\rm{max}}$. Finally, the properties of the grains in the dust models are parameterized by the polycyclic aromatic hydrocarbon (PAH) index, $q_{\rm PAH}$, defined as the fraction of the dust mass in the form of PAH grains. Each observed SED is fitted with a wide range of models generated by combinations of different set of parameters. For our case we considered models with $q_{\rm PAH} = 0.4$--4.6\%, $U_{\rm min} = 0.7$--25, $\gamma = 0.0$--0.8, while following \citet{2007ApJ...663..866D}, we fixed $U_{\rm{max}} = 10^{6}$, and $\alpha = 2$. The best fit were derived through $\chi^{2}$ minimization yielding to $M_{\rm{dust}}$, $U_{\rm{min}}$, $\gamma$ and $q_{\rm PAH}$ estimates. $L_{\rm{IR}}$ was calculated by integrating the best fit to the SED in the range 8--1000\,$\mu$m. To estimate the uncertainties of the parameters we created 1000 realizations of the observed SEDs by perturbing the photometry within the errors and repeating the fit. The corresponding uncertainties are defined by the standard deviation of the distribution of the derived quantities. The $L_{\rm{IR}}$ and $M_{\rm{dust}}$ estimates along with their uncertainties are listed in Table~\ref{tab:gds_prop}, where $\rm{SFR_{IR}}$ estimates were obtained using the $L_{\rm{IR}}$ to $\rm{SFR_{IR}}$ conversion from \citet{kennicutt98} for a Chabrier IMF. In Figure~\ref{fig:el02xmm20_optfir_seds} we present the observed SEDs along with best fit models as derived from our analysis.

\subsection{Dust-based Estimates of $M_{\rm{H_2}}$} \label{subsec:mgas_dust}

A very efficient way to determine the molecular gas reservoir of the galaxies is through their dust emission, either using the metallicity-dependent gas-to-dust mass ratio technique ($\delta_{\rm{GD}}$) \citep[e.g.,][]{2012ApJ...760....6M,2016A&A...587A..73B}, which converts the $M_{\rm{dust}}$ estimates to $M_{\rm{gas}}$ through the well established, almost linear, gas-to-dust mass ratio vs gas-phase metallicity relation ($M_{\rm{gas}}/M_{\rm{dust}}$--$Z$); or through the single band measurement of the dust emission flux on the Rayleigh-Jeans (R-J) side of the SED \citep[e.g.,][]{2014ApJ...783...84S,2015ApJ...799...96G,2016ApJ...833..112S}. Here, and thanks to the detailed coverage of the IR part of the spectrum of our objects, including the R-J tail of the SED, we are in position to use both techniques. We refer to these estimates as $M_{\rm{gas}}^{\rm{GD}}$ and $M_{\rm{gas}}^{\rm{RJ}}$, respectively.

First, we converted the $M_{\rm{dust}}$ estimates, derived as described in the previous section, to $M_{\rm{gas}}$ by adopting the $M_{\rm{gas}}/M_{\rm{dust}}$--$Z$ relation of \citet{2012ApJ...760....6M} ($\log (M_{\rm{dust}}/M_{\rm{gas}}) = (10.54 \pm 1.0) - (0.99 \pm 0.12) \times (12 + \log(O/H))$), where the metallicity is calibrated using the \citet{2004MNRAS.348L..59P} scale. We assumed a solar metallicity for all sources that corresponds to a $M_{\rm{gas}}/M_{\rm{dust}} \sim 90$. The corresponding uncertainties take into quadrature the uncertainties in $M_{\rm{dust}}$ and adopting a 0.2\,dex uncertainty in $Z$. Similarly, we converted the ALMA 3\,mm (rest-frame $\sim 950$\,$\mu$m for HELAISS02 and $\sim 830$\,$\mu$m for HXMM20) flux densities of each source (except for HXMM20-S2, S3 and S4 that are not detected at 3\,mm) to $M_{\rm{gas}}$ through the equation 12 of \citet{2014ApJ...783...84S}. The $M_{\rm{gas}}$ estimates derived by the two approaches are in excellent agreement, compatible within the uncertainties, with an average ratio of 1.24 $\pm$ 0.23. This is not surprising given the implicit assumption of solar gas-phase metallicity in both approaches. The values are summarized in Table~\ref{tab:gds_prop}. Finally, we note that these estimates yield the total gas budget of the galaxies, including contributions from the molecular ($M_{\rm{H_2}}$) and the atomic phase ($M_{\rm{H_I}}$). However, assuming that for high-redshift relatively massive galaxies the molecular gas dominates over the atomic gas within the physical scale probed by the dust continuum observations, $M_{\rm{H_2}} >> M_{\rm{H_I}}$ \citep[e.g.,][]{2006ApJ...650..933B,2008AJ....136.2846B,2009ApJ...698.1467O,2010Natur.463..781T,2010ApJ...713..686D,2011ApJ...730L..19G}, we can then write $M_{\rm{gas}} = M_{\rm{H_2}} + M_{\rm{H_I}} \approx M_{\rm{H_2}}$. 

The CO-independent $M_{\rm{H_2}}$ estimates derived using the two dust-based methods allow us to explore the $\alpha_{\rm{CO}}$ conversion factor of the different sources in each protocluster core \citep[see][for a review]{2013ARA&A..51..207B}. \citet{2012MNRAS.426.2601P,2012ApJ...751...10P} concluded that $\alpha_{\rm{CO}}$ is affected by gas density and temperature, but mostly by the overall dynamical state of the gas. High values are related with self-gravitating gas clouds, such as those found in local star-forming disks like the Milky Way (MW) \citep[e.g., $\alpha_{\rm{CO}} = 4.3$\,K km s$^{-1}$ pc$^{-2}$][]{1996A&A...308L..21S,2001ApJ...547..792D,2010ApJ...710..133A}. Low values are associated to gravitationally unbound gas, such as disturbed gas in local major mergers \citep[e.g., $\alpha_{\rm{CO}} = 0.8$\,K km s$^{-1}$ pc$^{-2}$][]{1997ApJ...478..144S,1998ApJ...507..615D,2008ApJ...680..246T}. We employed $M_{\rm{H_2}}^{\rm{GD}}$, that could be derived for all sources, to calculate $\alpha_{\rm{CO}}$. Our results in Table~\ref{tab:gds_prop} show that HELAISS02 sources have a high $\alpha_{\rm{CO}}$, while HXMM20 sources have a lower $\alpha_{\rm{CO}}$. The integrated measurement for HELAISS02 displays a high $\alpha_{\rm{CO}} = 4.6 \pm 2.4$ consistent with those of MW-like disks, while the lower HXMM20 $\alpha_{\rm{CO}} = 1.8 \pm 0.9$ resembles better those found in mergers. Although the uncertainties are large, it is also worth noting that the lowest $\alpha_{\rm{CO}}$ are associated with the blended sources in HXMM20 (S0, S2 and S3), while the highest $\alpha_{\rm{CO}}$ are related with HELAISS02, where all the sources are well separated from each other, and HXMM20-S1 with a large distance to another neighboring source (with the exception of HXMM-S4, but which CO(1-0) flux is poorly constrained). This agrees with the interpretation of the overall dynamical state of the gas being the major contributor to $\alpha_{\rm{CO}}$, with disturbed gas associated with lower $\alpha_{\rm{CO}}$, which is likely the case of the blended sources of HXMM20, and bound gas linked to higher $\alpha_{\rm{CO}}$, likely the case of the more isolated sources.

\subsection{Stellar Masses} \label{subsec:mstar}

The ELAIS-S1 and XMM-LSS fields, where HELAISS02 and HXMM20 are respectively located, are covered by optical/IR data sets publicly available suitable to determine the stellar masses of the different optical/near-IR counterparts associated to the ALMA 870\,$\mu$m continuum counterparts through SED fitting.

We employed optical/near-IR data from the VISTA Deep Extragalactic Observations \citep[VIDEO;][]{2013MNRAS.428.1281J} survey in the $z$, $y$, $J$, $H$, and $K_{\rm{s}}$ bands; and mid-IR coverage from the \textit{Spitzer} Extragalactic Representative Volume Survey \citep[SERVS;][]{2012PASP..124..714M} at 3.6 and 4.5\,$\mu$m, and from the \textit{Spitzer} Wide-area Infrared Extragalactic Survey \citep[SWIRE;][]{2003PASP..115..897L} at 5.8 and 8.0\,$\mu$m.

\begin{figure*}
\begin{center}
\includegraphics[width=\textwidth]{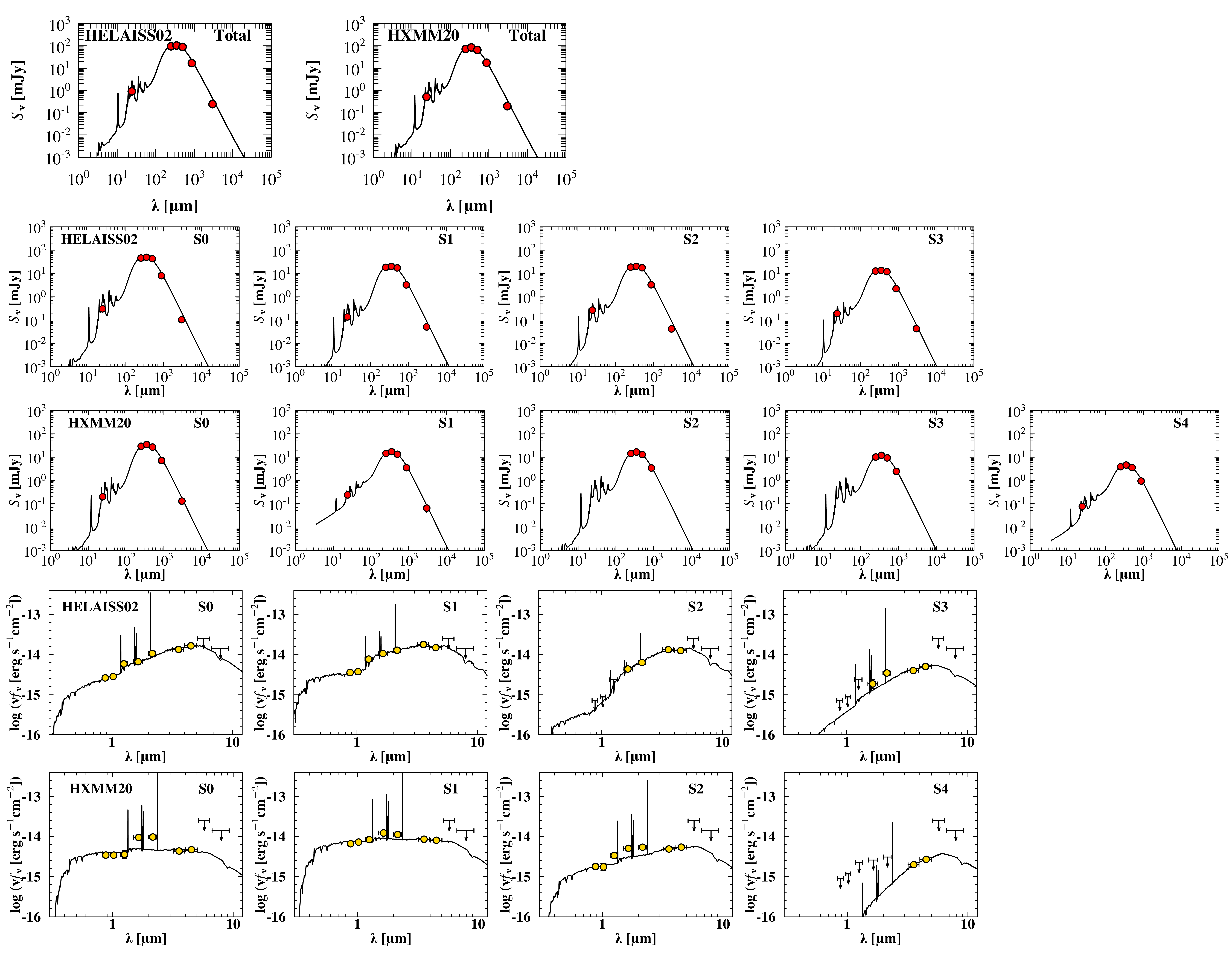}
\caption{HELAISS02 and HXMM20 optical/IR and FIR SEDs. First row: HELAISS02 and HXMM20 FIR SED and best fit for the integrated values over all sources. Second row: HELAISS02 FIR SED and best fit for each 870\,$\mu$m continuum source. Third row: HXMM20 FIR SED and best fit for each 870\,$\mu$m continuum source. Wavelengths are in the observer-frame. Fourth row: HELAISS02 optical/IR SED and best fit for the optical/near-IR counterparts associated to each 870\,$\mu$m continuum source. Fifth row: HXMM20 optical/IR SED and best fit for the optical/near-IR counterparts associated to each 870\,$\mu$m continuum source. Arrows indicate 3$\sigma$ upper limits (5$\sigma$ for the \textit{Spitzer} bands). Wavelengths are in the observer-frame.}
\label{fig:el02xmm20_optfir_seds}
\end{center}
\end{figure*}

The photometry was measured following the procedure described in \citet{2018ApJ...856..121G} for crowded and blended objects. Briefly, from the $z$ to the $K_{\rm{s}}$ bands we performed aperture photometry. The number of apertures is set to the number of 870\,$\mu$m continuum sources. We excluded HXMM20-S3 because it is not clearly detected, being too faint and too close to HXMM20-S0 and HXMM20-S2 to disentangle its individual contribution. Therefore, we did not derive a stellar mass for this source. The apertures were selected in the $K_{\rm{s}}$ band as large as possible (typically 2\arcsec diameter) without overlapping with neighboring apertures. We applied aperture corrections for every band by deriving the growth curve of a PSF in the different bands and computing the correction factor to the fluxes to account for the missing flux outside the aperture. The flux uncertainties were derived from empty aperture measurements. We only use detections above 3$\sigma$ to guarantee a good SED fit (upper limits are included in Figure~\ref{fig:el02xmm20_optfir_seds}). In the case of \textit{Spitzer}/IRAC 3.6 and 4.5\,$\mu$m data the sources appear blended. In this case, the fluxes were calculated from PSF fitting with \texttt{GALFIT} as explained in Section~\ref{subsec:fir} for the \textit{Spitzer}/MIPS 24\,$\mu$m images. The 5$\sigma$ detection criterion to perform the PSF fit was reached for all sources in the 3.6 and 4.5\,$\mu$m bands, but not in the 5.8 and 8.0\,$\mu$m and, thus, these bands were not included in the SED fit (upper limits are shown in Figure~\ref{fig:el02xmm20_optfir_seds}). The number of PSFs was again set to the number of 870\,$\mu$m continuum sources and the PSFs centroids were placed at the positions of $K_{\rm{s}}$ band centroids used as priors, allowing a shift in both the X and Y $< 1$\,pixel from the initial positions. To account the uncertainty in the photometry due to the deblending we performed a number of realizations varying the centroid coordinates randomly within 1 pixel of the best-fit centroid and fixing those coordinates for each realization.

We fitted the resulting SEDs using the code LePHARE \citep{1999MNRAS.310..540A,2006A&A...457..841I} adopting \citet{2003MNRAS.344.1000B} stellar population synthesis models with emission lines to account for nebular line contamination in the broad bands. We assumed a \citet{2003PASP..115..763C} IMF, exponentially declining star formation histories (SFHs) and a \citet{2000ApJ...533..682C} dust law. The parameter grid employed ranges SFH e-folding times 0.1\,Gyr–-30\,Gyr, extinction $0 < A_{V} < 5$, stellar age 1\,Myr--age of the universe at the source redshift and metallicity $Z = 0.004, 0.008$ and 0.02 (i.e., solar). The redshift was fixed to the derived CO(3-2) spectroscopic redshifts for each source (see Tables~\ref{tab:helaiss02_line} and \ref{tab:hxmm20_line}). The derived SEDs are shown in Figure~\ref{fig:el02xmm20_optfir_seds} and the stellar masses in Table~\ref{tab:gds_prop}. Additionally, we explored whether the output stellar extinction $A_{V}$ correlates with $M_{\rm{dust}}$ derived in Section~\ref{subsec:fir}. We found no correlation between them. Some studies have shown that these two quantities could be linked to different stellar populations and depend differently on the viewing angle and on the geometry of the dust distribution \citep[e.g.,][]{2017ApJ...840...15S,2017ApJ...847...21F,2017MNRAS.472.2315P,2018MNRAS.474.1718N,2018ApJ...856..121G}. The plausible different physical origin of the stellar and dust continuum light justifies the use of two different SED fitting techniques, one for the optical/near-IR SED and another one for the FIR SED, as opposed to employing an energy balanced solution that implies a direct relation between stars and dust.

With both the molecular gas and stellar masses we calculated the molecular gas fraction defined as $f_{\rm{H_2}} = M_{\rm{H_2}} / (M_{\rm{*}} + M_{\rm{H_2}})$. The values are also presented in Table~\ref{tab:gds_prop}.

\begin{deluxetable*}{lccccccccc}
\tabletypesize{\scriptsize}
\tablecaption{HELAISS02 and HXMM20 Gas, Dust and Stellar Properties \label{tab:gds_prop}}
\tablehead{\colhead{Name} & \colhead{$\log (M_{\rm{H_2}}^{\rm{CO}}/M_{\odot})$} & \colhead{$\log (M_{\rm{H_2}}^{\rm{GD}}/M_{\odot})$} & \colhead{$\log (M_{\rm{H_2}}^{\rm{RJ}}/M_{\odot})$} & \colhead{$\log (L_{\rm{IR}}/L_{\odot})$} & \colhead{$\log (M_{\rm{dust}}/M_{\odot})$} & \colhead{$\alpha_{\rm{CO}}$} & \colhead{$\rm{SFR_{IR}}$} & \colhead{$\log (M_{\rm{*}}/M_{\odot})$} & \colhead{$f_{\rm{H_2}}$} \\
\colhead{} & \colhead{} & \colhead{} & \colhead{} & \colhead{} & \colhead{} & \colhead{[$M_{\odot}$ K$^{-1}$ km$^{-1}$s pc$^{-2}$]} & \colhead{[$M_{\odot}$ yr$^{-1}$]} & \colhead{} & \colhead{}}
\startdata
HELAISS02 & 11.81 $\pm$ 0.04 & 11.93 $\pm$ 0.22 & 11.82 $\pm$ 0.04 & 13.18 $\pm$ 0.05 & 9.96 $\pm$ 0.09 & 4.6 $\pm$ 2.4 & 1510 $\pm$ 170 & 11.49$_{-0.05}^{+0.06}$ & 0.68 $\pm$ 0.11 \\
  S0 & 11.61 $\pm$ 0.04 & 11.60 $\pm$ 0.23 & 11.45 $\pm$ 0.05 & 12.88 $\pm$ 0.07 & 9.64 $\pm$ 0.12 & 3.5 $\pm$ 1.9 & 760 $\pm$ 120 & 10.48$_{-0.11}^{+0.16}$ & 0.93 $\pm$ 0.02 \\
  S1 & 10.83 $\pm$ 0.09 & 11.27 $\pm$ 0.24 & 11.15 $\pm$ 0.10 & 12.43 $\pm$ 0.06 & 9.30 $\pm$ 0.12 & 9.7 $\pm$ 5.7 & 269 $\pm$ 37 & 11.06$_{-0.11}^{+0.09}$ & 0.37 $\pm$ 0.07 \\
  S2 & 11.07 $\pm$ 0.05 & 11.27 $\pm$ 0.23 & 11.06 $\pm$ 0.11 & 12.48 $\pm$ 0.07 & 9.30 $\pm$ 0.11 & 5.6 $\pm$ 3.0 & 302 $\pm$ 49 & 11.17$_{-0.10}^{+0.08}$ & 0.44 $\pm$ 0.05 \\
  S3 & 10.82 $\pm$ 0.09 & 11.10 $\pm$ 0.23 & 11.07 $\pm$ 0.12 & 12.30 $\pm$ 0.07 & 9.13 $\pm$ 0.11 & 6.7 $\pm$ 3.8 & 200 $\pm$ 32 & 10.13$_{-0.14}^{+0.09}$ & 0.83 $\pm$ 0.04 \\
HXMM20    & 12.04 $\pm$ 0.04 & 11.75 $\pm$ 0.22 & 11.64 $\pm$ 0.07 & 13.23 $\pm$ 0.05 & 9.79 $\pm$ 0.09 & 1.8 $\pm$ 0.9 & 1700 $\pm$ 200 & 10.81$_{-0.15}^{+0.19}$ & 0.94 $\pm$ 0.04 \\
  S0 & 11.54 $\pm$ 0.07 & 11.46 $\pm$ 0.24 & 11.47 $\pm$ 0.07 & 12.82 $\pm$ 0.06 & 9.50 $\pm$ 0.14 & 2.9 $\pm$ 1.7 & 661 $\pm$ 91 & 9.88$_{-0.05}^{+0.10}$ & 0.98 $\pm$ 0.01 \\
  S1 & 11.11 $\pm$ 0.14 & 11.14 $\pm$ 0.23 & 11.17 $\pm$ 0.15 & 12.58 $\pm$ 0.05 & 9.18 $\pm$ 0.11 & 3.7 $\pm$ 2.3 & 380 $\pm$ 44 & 10.12$_{-0.04}^{+0.09}$ & 0.91 $\pm$ 0.03 \\
  S2 & 11.24 $\pm$ 0.14 & 11.08 $\pm$ 0.23 & \nodata & 12.54 $\pm$ 0.06 & 9.12 $\pm$ 0.12 & 2.4 $\pm$ 1.5 & 347 $\pm$ 48 & 10.04$_{-0.07}^{+0.14}$ & 0.94 $\pm$ 0.03 \\
  S3 & 11.44 $\pm$ 0.07 & 10.96 $\pm$ 0.29 & \nodata & 12.64 $\pm$ 0.29 & 8.99 $\pm$ 0.21 & 1.2 $\pm$ 0.8 & 440 $\pm$ 290 & \nodata & \nodata \\
  S4 & 11.21 $\pm$ 0.14 & 10.58 $\pm$ 0.24 & \nodata & 12.00 $\pm$ 0.06 & 8.61 $\pm$ 0.13 & 0.8 $\pm$ 0.5 & 100 $\pm$ 14 & 10.51$_{-0.31}^{+0.29}$ & 0.83 $\pm$ 0.10 \\
\enddata
\end{deluxetable*}

\section{Discussion} \label{sec:discussion}

\subsection{Blends of DSFGs from Single-dish Selected Sources} \label{subsec:blends}

HELAISS02 and HXMM20 are composed of four and five gas-rich DSFGs within a projected diameter of 125\,kpc and 64\,kpc, respectively. The HCOSMOS02 core comprises five gas-rich DSFGs within a projected diameter of 105\,kpc. All the ALMA 870\,$\mu$m continuum sources reported in \citet{2015ApJ...812...43B} for these three candidate protoclusters originally selected as single-dish \textit{Herschel}/SPIRE sources turned out to be located at the same redshift as confirmed by the CO observations presented in our work. Such a high fraction of sources with small pairwise separations located at the same redshift are unexpected from both a theoretical perspective \citep{2013MNRAS.434.2572H,2015MNRAS.446.1784C,2015MNRAS.446.2291M,2018MNRAS.476.2278H} and previous high spatial resolution follow-up of longer wavelength single-dish observations. \citet{2018MNRAS.479.3879W} presented CO observations from six single-dish selected 870\,$\mu$m continuum sources that appeared as blends of at least two individual sources, suggesting that 64\% of these individual sources are unlikely to be physically associated.

Our results are in line with \citet{2013ApJ...772..137I}, that confirmed four ALMA 870\,$\mu$m continuum sources across a $\sim 100$\,kpc region at $z \sim 2.41$ through CO(4-3) and CO(1-0) observations in a \textit{Herschel}/SPIRE-selected hyperluminous infrared galaxy. In addition, recent discoveries of $z > 4$ protoclusters with associated DSFGs resemble the result presented in our work. \citet{2018ApJ...856...72O} discovered a protocluster of at least 10 DSFGs at $z \sim 4.002$, confirmed through [\ion{C}{1}] and high-J CO transitions, located within a 260\,kpc$\times$310\,kpc region. \citet{2018Natur.556..469M} discovered a protocluster at $z \sim 4.31$ of at least 14 gas-rich sources within a projected diameter of 130\,kpc, confirmed from [\ion{C}{2}], with eight of them also detected in CO(4-3) and 12 in 1\,mm continuum.

\subsection{Gas Fractions and Star Formation Efficiencies} \label{subsec:fgas_sfe}

At $z \sim 1.5$--2.5 several works have studied the molecular gas content, efficiency of converting gas into stars and their relation with the specific star formation rate ($\rm{sSFR} = \rm{SFR} / M_{\rm{*}}$) and with field galaxies, those that do not necessarily live in an overdense environment \citep[e.g.,][]{2017ApJ...842L..21N,2017ApJ...842...55L,2017ApJ...849...27R,2017A&A...608A..48D,2018ApJ...856..118H,2018MNRAS.479..703C}. In this section we explore and discuss these matters regarding our sample of protoclusters cores. We employed the properties derived for HELAISS02 and HXMM20 870\,$\mu$m continuum sources in Section~\ref{sec:results} and those derived in \citet{2016ApJ...828...56W} and \citet{2018ApJ...867L..29W} for HCOSMOS02 for its five 870\,$\mu$m continuum sources, with updated molecular gas masses based in our CO observations following the method described in Section~\ref{subsec:mgas_co}.

The well studied correlation between the SFR and the stellar mass of star-forming galaxies (SFGs), so-called main sequence (MS) of star formation \citep[e.g.,][]{2007ApJ...660L..43N,2007A&A...468...33E,2007ApJ...670..156D} permits to distinguish between MS galaxies, as those located within the scatter of the MS, and starburst (SB) galaxies, outliers to the MS exhibiting an elevated sSFR compared to MS galaxies. Another correlation in SFGs arises between the observables $L^{'}_{\rm{CO(1-0)}}$ and $L_{\rm{IR}}$ and, thus, between $M_{\rm{H_2}}$ and $\rm{SFR_{IR}}$ calculated from these observables, commonly referred in the literature as the star formation law or Kennicutt-Schmidt relation \citep[KS relation;][originally defined using star formation rate and gas mass surface densities]{1959ApJ...129..243S,kennicutt98}. There are studies that suggest that MS and SB galaxies follow different relations between these quantities, with SB galaxies having increased star formation efficiency ($\rm{SFE} = \rm{SFR} / M_{\rm{H_2}}$) \citep[e.g.,][]{2010ApJ...714L.118D,2010MNRAS.407.2091G}.

\begin{figure*}
\begin{center}
\includegraphics[width=\textwidth]{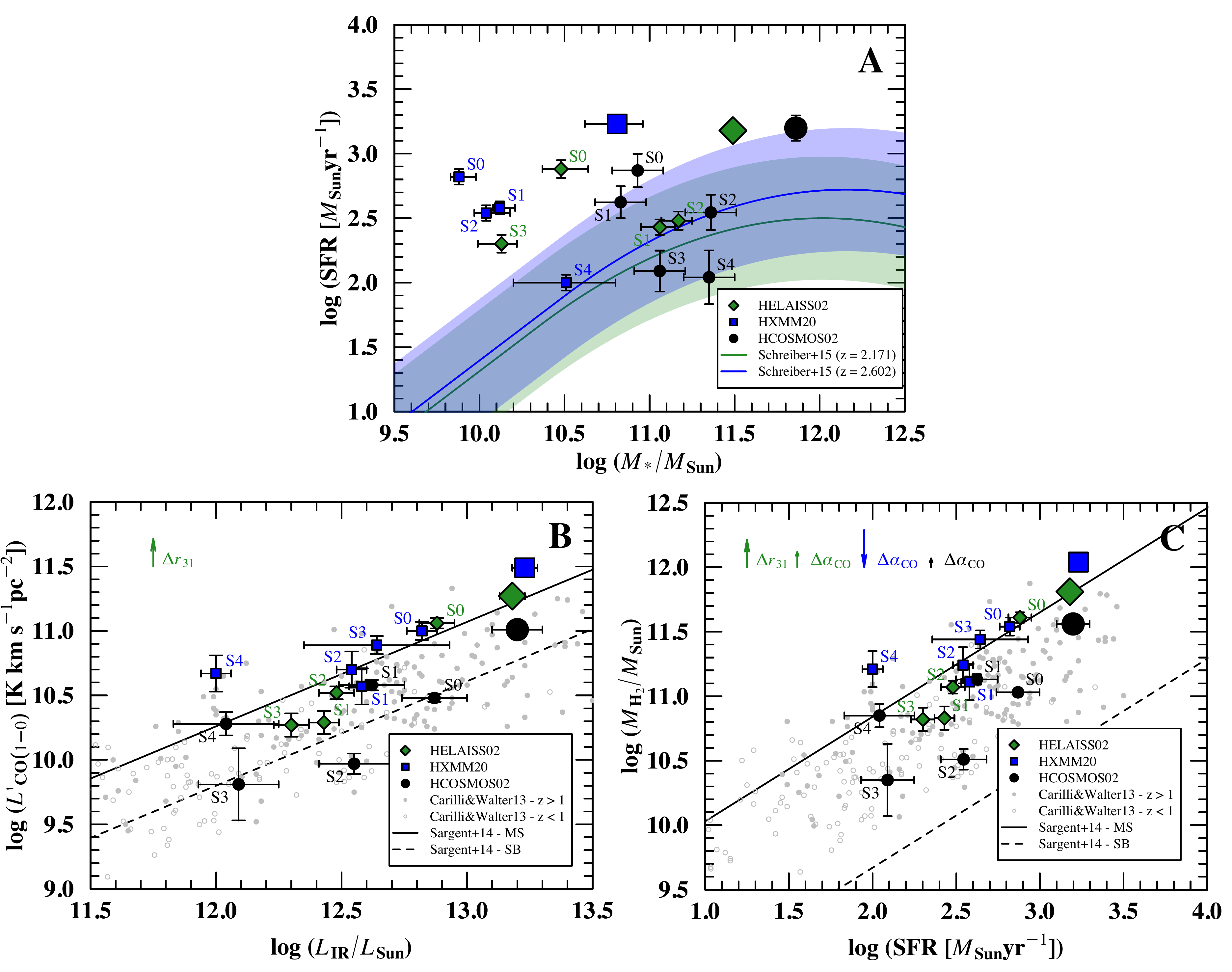}
\caption{Panel A: $\rm{SFR}-M_{\rm{*}}$ plane. The main sequence of star-forming galaxies defined by \citet{2015A&A...575A..74S} converted from Salpeter to Chabrier IMF is plotted at the highest and lowest redshifts of the sample with a 0.5\,dex (3 times) scatter represented by a shadowed region. The big symbols correspond to the total values of the protoclusters. Panel B: $L_{\rm{CO}}-L_{\rm{IR}}$ plane. The arrows indicate the expected displacement of the total values represented with big symbols when using a MS-like excitation conversion ($\Delta r_{\rm{31}}$, affecting HELAISS02). Panel C: $M_{\rm{H_2}}-\rm{SFR_{IR}}$ plane. The arrows indicate the expected displacement of the total values represented with big symbols when using the individual $\alpha_{\rm{CO}}$ values ($\Delta \alpha_{\rm{CO}}$). Trends for main sequence (MS, solid line) and starburst (SB, dashed line) galaxies from \citet{2014ApJ...793...19S} and datapoints from \citet{2013ARA&A..51..105C} are shown as reference in the B and C panels.}
\label{fig:phys_rel}
\end{center}
\end{figure*}

\begin{figure}
\begin{center}
\includegraphics[width=\columnwidth]{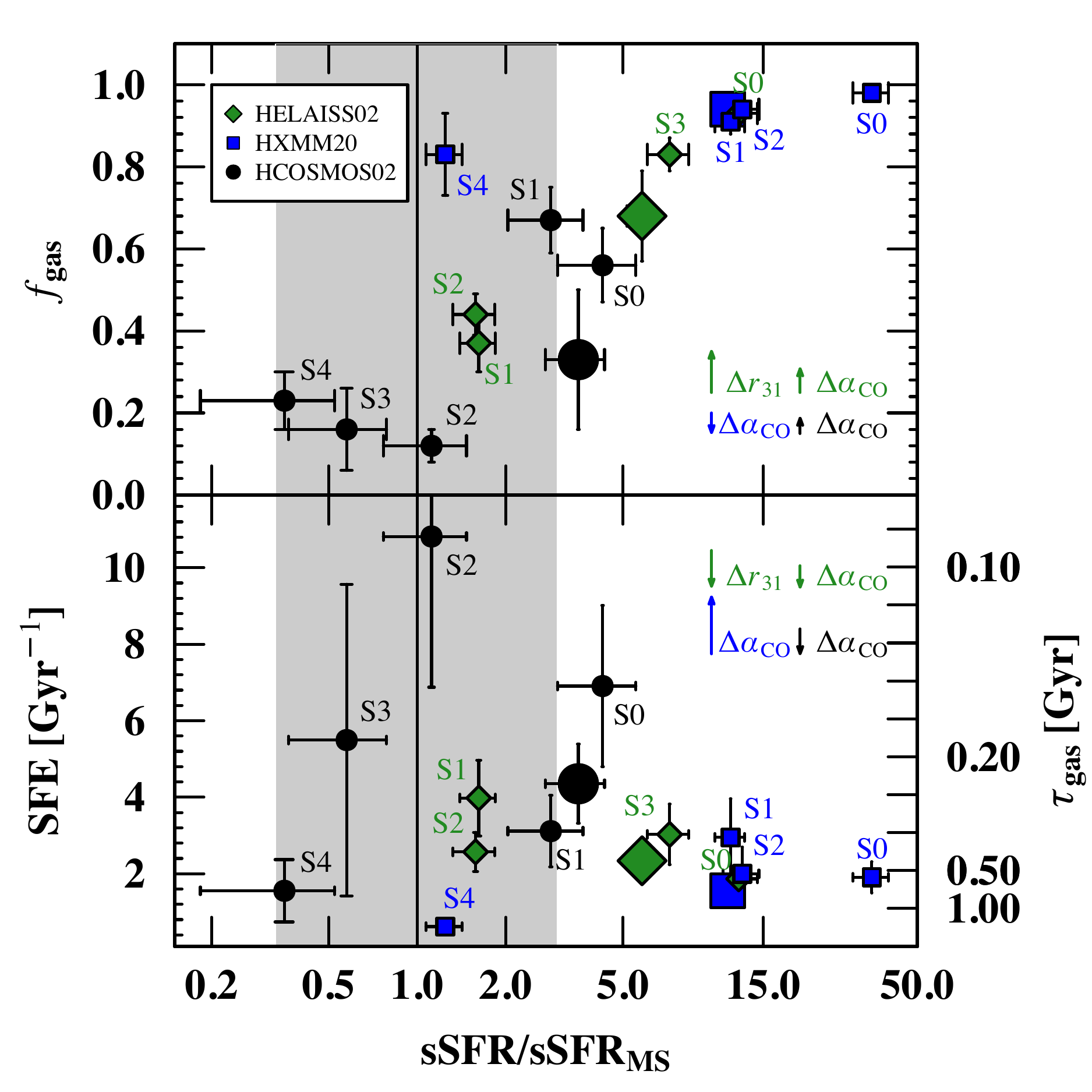}
\caption{$f_{\rm{H_2}}$ (top panel), $\rm{SFE}$ and $\tau_{\rm{H_2}}$ (bottom panel) vs distance to the MS. The shadowed region represents the main sequence of star-forming galaxies defined by \citet{2015A&A...575A..74S} converted from Salpeter to Chabrier IMF with a 0.5\,dex (3 times) scatter. The arrows indicate the expected displacement of the total values when using a MS-like excitation conversion ($\Delta r_{\rm{31}}$, affecting HELAISS02 in both panels) and when using the individual $\alpha_{\rm{CO}}$ ($\Delta \alpha_{\rm{CO}}$, affecting HELAISS02, HXMM20, and HCOSMOS02 in both panels).}
\label{fig:multi_dms}
\end{center}
\end{figure}

In Figure~\ref{fig:phys_rel} we show the location of the protocluster core members in the $\rm{SFR}-M_{\rm{*}}$, $L_{\rm{CO}}-L_{\rm{IR}}$ and $M_{\rm{H_2}}-\rm{SFR_{IR}}$ planes, where $M_{\rm{H_2}}$ comes from the CO-based measurements as derived in Section~\ref{subsec:mgas_co}. We can see that the integrated measurements for HELAISS02 and HXMM20 are consistent with the SB regime in $\rm{SFR}-M_{\rm{*}}$, but with the MS relation in the observables KS plane $L_{\rm{CO}}-L_{\rm{IR}}$. The tension is somewhat smaller in the case of HCOSMOS02, consistent with the MS scatter in $\rm{SFR}-M_{\rm{*}}$ plane. In order to explore the nature of this apparent discrepancies in Figure~\ref{fig:multi_dms} we show how $f_{\rm{H_2}}$ and SFE (or depletion time-scale, $\tau_{\rm{H_2}} = 1/{\rm{SFE}}$) vary as a function of the distance to the MS (DMS), defined as the ratio of the sSFR to the sSFR of the MS at the same stellar mass and redshift ($\rm{sSFR/sSFR_{MS}}$). A number of studies have revealed that the DMS scales with both $f_{\rm{H_2}}$ and SFE, with SFGs having increasing $f_{\rm{H_2}}$ and SFE (lower $\tau_{\rm{H_2}}$) as they move to higher DMS \citep[e.g.,][]{2010ApJ...714L.118D,2010MNRAS.407.2091G,2012ApJ...760....6M,2014ApJ...793...19S,2015ApJ...800...20G,2017ApJ...837..150S,2018ApJ...853..179T}. The integrated measurements for HELAISS02, HXMM20, and HCOSMOS02 follow the expected literature trends in $f_{\rm{H_2}}$. However, the behavior in SFE as a function of DMS is the opposite of what we know from the literature.

It is important to remember the assumptions we made when deriving $M_{\rm{H_2}}$ from CO in Section~\ref{subsec:mgas_co}. The excitation conversion for HELAISS02 ($r_{\rm{31}} = 0.69 \pm 0.09$) and the conversion factor $\alpha_{\rm{CO}} = 3.5$. Adopting a MS-like excitation conversion $r_{\rm{31}} = 0.42 \pm 0.07$ \citep{2015A&A...577A..46D} would increase the $L^{'}_{\rm{CO(1-0)}}$ measurement (and $M_{\rm{H_2}}$) and decrease the SFE (increase $\tau_{\rm{H_2}}$) as represented by the green arrows in Figures~\ref{fig:phys_rel} and \ref{fig:multi_dms}. While the trend in $f_{\rm{H_2}}$ is pretty robust to a change in this assumption, HELAISS02 SFE ($\tau_{\rm{H_2}}$) would move to values similar to HXMM20 within the uncertainties. In the case of $\alpha_{\rm{CO}}$, the values were independently calculated for HELAISS02 and HXMM20 from $M_{\rm{H_2}}$ estimates through the $\delta_{\rm{GD}}$ technique in Section~\ref{subsec:mgas_dust}. \citet{2018ApJ...867L..29W} presented also individual $\alpha_{\rm{CO}}$ values for HCOSMOS02 members. Adopting these values instead, the trend in $f_{\rm{H_2}}$ holds, but that of SFE ($\tau_{\rm{H_2}}$) is less robust (green, blue, and black arrows in Figures~\ref{fig:phys_rel} and \ref{fig:multi_dms}). The excitation assumption also affects the estimates of $\alpha_{\rm{CO}}$ and the mentioned change would lower the values of HELAISS02. Another assumption that affects the $\alpha_{\rm{CO}}$ estimates is the adoption of solar metallicity. If different from solar, we might expect that HXMM20, having lower stellar mass than HELAISS02 and HCOSMOS02, has a lower metallicity and, thus, higher $\alpha_{\rm{CO}}$ \citep[e.g.,][]{2012ApJ...746...69G,2012ApJ...760....6M,2014ApJ...793...19S}.

In addition to the integrated measurements, we explored the behavior of the individual 870\,$\mu$m continuum sources in each protocluster core in the planes mentioned above. A caveat is the scaling assumption we used when deriving the $L_{\rm{IR}}$ and $\rm{SFR_{IR}}$ estimates for HELAISS02 and HXMM20 in Section~\ref{subsec:fir}. We have enough spatial resolution to get individual measurements of most of the sources in the left-hand side of the FIR SED peak through \textit{Spitzer}/MIPS 24\,$\mu$m and in the R-J side of the peak from ALMA 870\,$\mu$m and 3\,mm, but we have no constraints on the actual peak of the SED due the large beam size of \textit{Herschel}/SPIRE compared to the distance between sources. Therefore, we scaled the integrated SPIRE fluxes to the ALMA 870\,$\mu$m measurements for the distinct individual sources. While the R-J side of the FIR SED is enough to constrain $M_{\rm{gas}}$, from $M_{\rm{dust}}$ using the $\delta_{\rm{GD}}$ technique, or through the single band measurement of the dust emission flux (see Section~\ref{subsec:mgas_dust}), the peak and the left-hand side are needed to constrain the overall shape of the SED and, thus, $L_{\rm{IR}}$ and $\rm{SFR_{IR}}$. Consequently, the scaling assumption implies an almost constant SED shape that is dictated almost only based on the region sensitive to $M_{\rm{dust}}$, varying only based on 24\,$\mu$m. This means an almost constant $L_{\rm{IR}}$/$M_{\rm{dust}}$ ratio and, hence, $\rm{SFE} = \rm{SFR} / M_{\rm{H_2}} \propto L_{\rm{IR}}$/$M_{\rm{dust}} \approx \rm{constant}$. The different sources or each protocluster core are by construction bound to have very similar SFE ($\tau_{\rm{H_2}}$ ). HCOSMOS02 870\,$\mu$m continuum sources are less affected by these caveats, since the left-hand side of the FIR SED is better constrained thanks to the \textit{Herschel}/PACS detections \citep{2016ApJ...828...56W}. Additionally, the assumptions affecting the integrated measurements of the excitation conversion for HELAISS02 ($r_{\rm{31}} = 0.69 \pm 0.09$) and the adopted $\alpha_{\rm{CO}} = 3.5$ also applies to the individual sources.

Bearing in mind this caveats, the individual sources reproduce qualitatively the same trends of the integrated measurements. We see that our sources above the MS have higher $f_{\rm{H_2}}$ than those within the MS. Besides SFE ($\tau_{\rm{H_2}}$) seem to decrease (increase) as a function of DMS. Some of the HCOSMOS02 sources display high SFE reaching the SB regime in the observables KS plane $L_{\rm{CO}}-L_{\rm{IR}}$.

In summary, we see that the most massive sources of each protocluster core (HELAISS02-S1 and S2, HXMM20-S4, HCOSMOS02-S2, S3, and S4) are those located within the MS and associated with the lowest gas fraction of each protocluster core. On the other hand, the least massive sources are those located above the MS and are completely dominated by molecular gas. This is also true for the integrated values, being HXMM20 the least massive and the most gas dominated with the highest fraction of galaxies above the MS, HCOSMOS02 the most massive and the least gas dominated with the lowest fraction of galaxies above the MS, while HELAISS02 plays an intermediate role. This points towards a different evolutionary stage of the three protocluster cores. Although there could be a difference in SFE ($\tau_{\rm{H_2}}$) between the different sources as a function of DMS, our current data requires the use of assumptions that are artificially creating any trend in SFE ($\tau_{\rm{H_2}}$). Additional higher spatial resolution at the peak of the FIR SED is paramount to uncover the real SFE of HELAISS02 and HXMM20. HCOSMOS02, with less caveats, points towards a decreasing SFE with DMS.

All this suggests that the molecular gas fraction is pushing the individual sources above the MS, while maintaining a MS-like efficiency as seen for both HELAISS02 and HXMM20 in the observables $L_{\rm{CO}}-L_{\rm{IR}}$ plane. The latter is in agreement with \citet{2017A&A...608A..48D} that concludes that the SFE does not vary in dense environments compared to field galaxies. One possible explanation of why the least massive sources appear above the MS while maintaining a MS-like efficiency in forming stars could be that they are newly formed galaxies migrating to the MS, being the most massive sources already in place probably because they started forming earlier. For example, if the HELAISS02 and HXMM20 sources located above the MS consume half of their available molecular gas at their current SFR by $z \sim 2.00$ and $z \sim 2.37$, respectively, they will be located within the scatter of the MS.

After discussing the overall trends of the integrated and individual measurements, we compare $f_{\rm{H_2}}$ and $\tau_{\rm{H_2}}$ with the gas scaling relations for field galaxies in \citet{2017ApJ...837..150S} at the same redshift, stellar mass and DMS in Table~\ref{tab:field_comp}. The integrated $f_{\rm{H_2}}$ are very similar to field galaxies within the uncertainties, perhaps indicating an overall small excess for HELAISS02 and HXMM20 (1.7 and 3.0$\sigma$, respectively). The individual sources show more discrepancies, with those within the MS (HELAISS02-S1, S2, HCOSMOS02-S2, S3, and S4) having a lack of molecular gas compared to the field, especially in the case of HCOSMOS02. In terms of $\tau_{\rm{H_2}}$, the integrated measurements are larger than field galaxies for HELAISS02 and HXMM20 (2.0 and 5.1$\sigma$) and smaller for HCOSMOS02 (2.1$\sigma$). On a source-by-source basis, given the large caveats affecting the SFE ($\tau_{\rm{H_2}}$) estimates of HELAISS02 and HXMM20, it is difficult to draw conclusions. In summary, our results suggest that two of our protoclusters cores are only slightly more gas rich than field galaxies, but display higher $\tau_{\rm{H_2}}$ due to their MS-like SFE, somewhat unexpected at this redshift, stellar mass and DMS, where galaxies with an enhanced SFE in the field are more common. These two are the ones with the lowest overall stellar mass, while that with the highest overall stellar mass displays lower $\tau_{\rm{H_2}}$ due to some of its members having SB-like SFE and small $f_{\rm{H_2}}$ compared to the field.

\begin{deluxetable}{lcc}
\tabletypesize{\scriptsize}
\tablecaption{HELAISS02 and HXMM20 Molecular Gas Fractions and Depletion Time-scales vs Field Galaxies} \label{tab:field_comp}
\tablehead{\colhead{Name} & \colhead{$f_{\rm{H_2}}/\langle f_{\rm{H_2}} \rangle$}\tablenotemark{a} & \colhead{$\tau_{\rm{gas}}/\langle \tau_{\rm{gas}} \rangle$}\tablenotemark{a} \\
\colhead{} & \colhead{} & \colhead{}}
\startdata
HELAISS02 & 1.40 $\pm$ 0.23 & 1.56 $\pm$ 0.23 \\
  S0 & 1.08 $\pm$ 0.02 & 3.24 $\pm$ 0.59 \\
  S1 & 0.67 $\pm$ 0.13 & 0.37 $\pm$ 0.09 \\
  S2 & 0.86 $\pm$ 0.10 & 0.57 $\pm$ 0.11 \\
  S3 & 0.92 $\pm$ 0.04 & 1.36 $\pm$ 0.36 \\
HXMM20    & 1.15 $\pm$ 0.05 & 4.22 $\pm$ 0.63 \\
  S0 & 1.01 $\pm$ 0.01 & 7.4 $\pm$ 1.6 \\
  S1 & 0.98 $\pm$ 0.03 & 2.22 $\pm$ 0.76 \\
  S2 & 1.00 $\pm$ 0.03 & 3.5 $\pm$ 1.2 \\
  S3 & \nodata & \nodata \\
  S4 & 1.07 $\pm$ 0.13 & 2.24 $\pm$ 0.79 \\
HCOSMOS02 & 0.95 $\pm$ 0.49 & 0.66 $\pm$ 0.16 \\
  S0 & 0.78 $\pm$ 0.13 & 0.46 $\pm$ 0.14 \\
  S1 & 0.93 $\pm$ 0.11 & 0.78 $\pm$ 0.23 \\
  S2 & 0.27 $\pm$ 0.09 & 0.12 $\pm$ 0.04 \\
  S3 & 0.31 $\pm$ 0.19 & 0.14 $\pm$ 0.11 \\
  S4 & 0.63 $\pm$ 0.19 & 0.37 $\pm$ 0.19 \\
\enddata
\tablenotetext{a}{$\langle f_{\rm{H_2}} \rangle$ and $\langle \tau_{\rm{gas}} \rangle$ from Table~2 in \citet{2017ApJ...837..150S} at the redshift, stellar mass and DMS of each source.}
\end{deluxetable}

In the literature the conclusions of studies that tackle gas fractions and efficiencies in protocluster galaxies compared to the field are varied. \citet{2017ApJ...842L..21N} concluded that $f_{\rm{H_2}}$ and $\tau_{\rm{H_2}}$ are higher in $z \sim 1.6$ cluster environments than in the field, from a sample of 11 MS gas-rich sources located in three different targets. \citet{2017ApJ...849...27R} observed two protoclusters members at $z = 1.62$, one of them on the MS and the other below the MS, concluding that both $f_{\rm{H_2}}$ and $\tau_{\rm{H_2}}$ are consistent with the gas scaling relation of field galaxies. \citet{2017ApJ...842...55L} also found consistent $f_{\rm{H_2}}$ with the gas scaling relations in MS protoclusters members at $z \sim 2.49$. \citet{2018ApJ...856..118H} detected 17 member galaxies in CO(2-1) and eight in 870\,$\mu$m dust continuum at $z = 1.46$, arguing that $f_{\rm{H_2}}$ and $\tau_{\rm{H_2}}$ are larger than those from the scaling relations. The sources were located on and below the MS. The authors speculated that the environment of galaxy clusters helps feeding the gas through into the cluster members and reduces the efficiency of star formation. On the other hand, \citet{2018MNRAS.479..703C} found lower $\tau_{\rm{H_2}}$, enhanced SFE and highly excited CO SLEDs in protocluster members at $z = 1.99$, linking such activity to mergers.

The general picture of how dense environments might or not contribute to enhance or suppress the accretion of gas and affect its efficiency to form stars is still debated and unclear. From our observations and based on the literature studies it seems that the evolutionary stage at which each protocluster structure is observed might play an important role in this picture.

\section{Summary and Conclusions} \label{sec:summary}

We selected three \textit{Herschel} candidate protoclusters with multiple ALMA 870\,$\mu$m continuum counterparts with small pairwise separations in order to confirm whether they are or not located at the same redshift by using CO observations. In summary we found:

\begin{itemize}

\item Three out of three candidates are confirmed protocluster core systems, where all the ALMA 870\,$\mu$m continuum sources previously reported are at the same redshift. We confirm the discovery of two new protocluster cores named HELAISS02 ($z = 2.171 \pm 0.004$) and HXMM20 ($z = 2.602 \pm 0.002$).

\item We do not find any new secure CO(1-0) detections in the $z = 2.51$ COSMOS overdensity, in addition to the previously reported ones. Although the system consists on numerous members, some display only tentative CO(1-0) detections and they should be treated with caution requiring further confirmation.

\item The physical conditions of the gas in HELAISS02 and HXMM20 reveal a star formation efficiency consistent with main sequence galaxies, although some of the sources are located in the starburst regime of the $\rm{SFR}-M_{\rm{*}}$ plane due to high gas fractions and yet small stellar masses. We suggest that they could be newly formed galaxies moving into the main sequence.

\item Overall, the three studied protocluster cores display trends when compared to each other and the field. HXMM20 is the least massive system with enhanced gas fraction with respect to the field, while HCOSMOS02 is the most massive system with depleted gas fraction with respect to the field. More precise measurements of star formation efficiencies are needed to confirm a trend in this quantity. We suggest an evolutionary sequence between the three protocluster cores and that the comparison with field galaxies depends on the evolutionary stage of the structure.

\end{itemize}

\acknowledgments

We thank R. S. Bussmann for his contributions to this project and C. M. Casey for useful discussion. We are grateful to the anonymous referee, whose comments have been very useful to improving our work.

C.G.G and S.T. acknowledge support from the European Research Council (ERC) Consolidator Grant funding scheme (project ConTExt, grant number: 648179). D.R. and R.P. acknowledge support from the National Science Foundation under grant number AST-1614213. G.M. and F.V. acknowledge the Villum Fonden research grant 13160 "Gas to stars, stars to dust: tracing star formation across cosmic time". T.K.D.L. acknowledges support by the NSF through award SOSPA4-009 from the NRAO and support by the Simons Foundation. I.P.F. acknowledges support from the Spanish research grants ESP2015-65597-C4-4-R and ESP2017-86852-C4-2-R. H.D. acknowledges financial support from the Spanish Ministry of Economy and Competitiveness (MINECO) under the 2014 Ram\'on y Cajal program MINECO RYC-2014-15686.

The Cosmic Dawn Center is funded by the Danish National Research Foundation.

This paper makes use of the following ALMA data: ADS/JAO.ALMA\#2015.1.00752.S. ALMA is a partnership of ESO (representing its member states), NSF (USA) and NINS (Japan), together with NRC (Canada), MOST and ASIAA (Taiwan), and KASI (Republic of Korea), in cooperation with the Republic of Chile. The Joint ALMA Observatory is operated by ESO, AUI/NRAO and NAOJ.

The National Radio Astronomy Observatory is a facility of the National Science Foundation operated under cooperative agreement by Associated Universities, Inc.

This paper employed \texttt{Astropy}, a community-developed core Python package for Astronomy \citep{2013A&A...558A..33A}; \texttt{APLpy}, an open-source plotting package for Python \citep{2012ascl.soft08017R}; CASA \citep{2007ASPC..376..127M}; \texttt{Matplotlib} \citep{Hunter:2007}; \texttt{Numpy}; \texttt{Photutils} \citep{larry_bradley_2016_164986}; R, a language and environment for statistical computing (R Foundation for Statistical Computing, Vienna, Austria).

\bibliographystyle{aasjournal}{}
\bibliography{pclust.bib}{}

\begin{thebibliography}{}
\expandafter\ifx\csname natexlab\endcsname\relax\def\natexlab#1{#1}\fi
\providecommand{\url}[1]{\href{#1}{#1}}
\providecommand{\dodoi}[1]{doi:~\href{http://doi.org/#1}{\nolinkurl{#1}}}
\providecommand{\doeprint}[1]{\href{http://ascl.net/#1}{\nolinkurl{http://ascl.net/#1}}}
\providecommand{\doarXiv}[1]{\href{https://arxiv.org/abs/#1}{\nolinkurl{https://arxiv.org/abs/#1}}}

\bibitem[{{Abdo} {et~al.}(2010){Abdo}, {Ackermann}, {Ajello}, {Baldini},
  {Ballet}, {Barbiellini}, {Bastieri}, {Baughman}, {Bechtol}, {Bellazzini},
  {Berenji}, {Bloom}, {Bonamente}, {Borgland}, {Bregeon}, {Brez}, {Brigida},
  {Bruel}, {Burnett}, {Buson}, {Caliandro}, {Cameron}, {Caraveo}, {Casandjian},
  {Cecchi}, {{\c{C}}elik}, {Chekhtman}, {Cheung}, {Chiang}, {Ciprini}, {Claus},
  {Cohen- Tanugi}, {Cominsky}, {Conrad}, {Dermer}, {de Palma}, {Digel},
  {Silva}, {Drell}, {Dubois}, {Dumora}, {Farnier}, {Favuzzi}, {Fegan}, {Focke},
  {Fortin}, {Frailis}, {Fukazawa}, {Funk}, {Fusco}, {Gargano}, {Gehrels},
  {Germani}, {Giavitto}, {Giebels}, {Giglietto}, {Giordano}, {Glanzman},
  {Godfrey}, {Grenier}, {Grondin}, {Grove}, {Guillemot}, {Guiriec}, {Harding},
  {Hayashida}, {Horan}, {Hughes}, {Jackson}, {J{\'o}hannesson}, {Johnson},
  {Johnson}, {Kamae}, {Katagiri}, {Kataoka}, {Kawai}, {Kerr}, {Kn{\"o}dlseder},
  {Kuss}, {Lande}, {Latronico}, {Lemoine-Goumard}, {Longo}, {Loparco}, {Lott},
  {Lovellette}, {Lubrano}, {Makeev}, {Mazziotta}, {McEnery}, {Meurer},
  {Michelson}, {Mitthumsiri}, {Mizuno}, {Monte}, {Monzani}, {Morselli},
  {Moskalenko}, {Murgia}, {Nolan}, {Norris}, {Nuss}, {Ohsugi}, {Okumura},
  {Omodei}, {Orlando}, {Ormes}, {Paneque}, {Pelassa}, {Pepe}, {Pesce- Rollins},
  {Piron}, {Porter}, {Rain{\`o}}, {Rando}, {Razzano}, {Reimer}, {Reimer},
  {Reposeur}, {Rodriguez}, {Ryde}, {Sadrozinski}, {Sanchez}, {Sander}, {Saz
  Parkinson}, {Sgr{\`o}}, {Siskind}, {Smith}, {Spandre}, {Spinelli}, {Starck},
  {Strickman}, {Strong}, {Suson}, {Takahashi}, {Tanaka}, {Thayer}, {Thayer},
  {Thompson}, {Tibaldo}, {Torres}, {Tosti}, {Tramacere}, {Uchiyama}, {Usher},
  {Vasileiou}, {Vilchez}, {Vitale}, {Waite}, {Wang}, {Winer}, {Wood}, {Ylinen},
  {Ziegler}, \& {Fermi/LAT Collaboration}}]{2010ApJ...710..133A}
{Abdo}, A.~A., {Ackermann}, M., {Ajello}, M., {et~al.} 2010, \apj, 710, 133,
  \dodoi{10.1088/0004-637X/710/1/133}

\bibitem[{{Aravena} {et~al.}(2010){Aravena}, {Bertoldi}, {Carilli},
  {Schinnerer}, {McCracken}, {Salvato}, {Riechers}, {Sheth}, {Sm{\v
  o}lci{\'c}}, {Capak}, {Koekemoer}, \& {Menten}}]{2010ApJ...708L..36A}
{Aravena}, M., {Bertoldi}, F., {Carilli}, C., {et~al.} 2010, \apjl, 708, L36,
  \dodoi{10.1088/2041-8205/708/1/L36}

\bibitem[{{Arnouts} {et~al.}(1999){Arnouts}, {Cristiani}, {Moscardini},
  {Matarrese}, {Lucchin}, {Fontana}, \& {Giallongo}}]{1999MNRAS.310..540A}
{Arnouts}, S., {Cristiani}, S., {Moscardini}, L., {et~al.} 1999, \mnras, 310,
  540, \dodoi{10.1046/j.1365-8711.1999.02978.x}

\bibitem[{{Astropy Collaboration} {et~al.}(2013){Astropy Collaboration},
  {Robitaille}, {Tollerud}, {Greenfield}, {Droettboom}, {Bray}, {Aldcroft},
  {Davis}, {Ginsburg}, {Price-Whelan}, {Kerzendorf}, {Conley}, {Crighton},
  {Barbary}, {Muna}, {Ferguson}, {Grollier}, {Parikh}, {Nair}, {Unther},
  {Deil}, {Woillez}, {Conseil}, {Kramer}, {Turner}, {Singer}, {Fox}, {Weaver},
  {Zabalza}, {Edwards}, {Azalee Bostroem}, {Burke}, {Casey}, {Crawford},
  {Dencheva}, {Ely}, {Jenness}, {Labrie}, {Lim}, {Pierfederici}, {Pontzen},
  {Ptak}, {Refsdal}, {Servillat}, \& {Streicher}}]{2013A&A...558A..33A}
{Astropy Collaboration}, {Robitaille}, T.~P., {Tollerud}, E.~J., {et~al.} 2013,
  \aap, 558, A33, \dodoi{10.1051/0004-6361/201322068}

\bibitem[{{Berta} {et~al.}(2016){Berta}, {Lutz}, {Genzel},
  {F{\"o}rster-Schreiber}, \& {Tacconi}}]{2016A&A...587A..73B}
{Berta}, S., {Lutz}, D., {Genzel}, R., {F{\"o}rster-Schreiber}, N.~M., \&
  {Tacconi}, L.~J. 2016, \aap, 587, A73, \dodoi{10.1051/0004-6361/201527746}

\bibitem[{{Bertoldi} {et~al.}(2007){Bertoldi}, {Carilli}, {Aravena},
  {Schinnerer}, {Voss}, {Smolcic}, {Jahnke}, {Scoville}, {Blain}, {Menten},
  {Lutz}, {Brusa}, {Taniguchi}, {Capak}, {Mobasher}, {Lilly}, {Thompson},
  {Aussel}, {Kreysa}, {Hasinger}, {Aguirre}, {Schlaerth}, \&
  {Koekemoer}}]{2007ApJS..172..132B}
{Bertoldi}, F., {Carilli}, C., {Aravena}, M., {et~al.} 2007, \apjs, 172, 132,
  \dodoi{10.1086/520511}

\bibitem[{{Bigiel} {et~al.}(2008){Bigiel}, {Leroy}, {Walter}, {Brinks}, {de
  Blok}, {Madore}, \& {Thornley}}]{2008AJ....136.2846B}
{Bigiel}, F., {Leroy}, A., {Walter}, F., {et~al.} 2008, \aj, 136, 2846,
  \dodoi{10.1088/0004-6256/136/6/2846}

\bibitem[{{Blain} {et~al.}(2004){Blain}, {Chapman}, {Smail}, \&
  {Ivison}}]{2004ApJ...611..725B}
{Blain}, A.~W., {Chapman}, S.~C., {Smail}, I., \& {Ivison}, R. 2004, \apj, 611,
  725, \dodoi{10.1086/422353}

\bibitem[{{Blitz} \& {Rosolowsky}(2006)}]{2006ApJ...650..933B}
{Blitz}, L., \& {Rosolowsky}, E. 2006, \apj, 650, 933, \dodoi{10.1086/505417}

\bibitem[{{Bolatto} {et~al.}(2013){Bolatto}, {Wolfire}, \&
  {Leroy}}]{2013ARA&A..51..207B}
{Bolatto}, A.~D., {Wolfire}, M., \& {Leroy}, A.~K. 2013, \araa, 51, 207,
  \dodoi{10.1146/annurev-astro-082812-140944}

\bibitem[{{Bothwell} {et~al.}(2013){Bothwell}, {Smail}, {Chapman}, {Genzel},
  {Ivison}, {Tacconi}, {Alaghband-Zadeh}, {Bertoldi}, {Blain}, {Casey}, {Cox},
  {Greve}, {Lutz}, {Neri}, {Omont}, \& {Swinbank}}]{2013MNRAS.429.3047B}
{Bothwell}, M.~S., {Smail}, I., {Chapman}, S.~C., {et~al.} 2013, \mnras, 429,
  3047, \dodoi{10.1093/mnras/sts562}

\bibitem[{Bradley {et~al.}(2016)Bradley, Sipocz, Robitaille, Tollerud,
  Vin'cius, Deil, Barbary, GŸnther, Cara, Droettboom, Bostroem, Bray, Bratholm,
  Pickering, Craig, Barentsen, Pascual, adonath, Greco, Kerzendorf,
  StuartLittlefair, Ferreira, D'Eugenio, \& Weaver}]{larry_bradley_2016_164986}
Bradley, L., Sipocz, B., Robitaille, T., {et~al.} 2016, astropy/photutils:
  v0.3, \dodoi{10.5281/zenodo.164986}.
\newblock \url{https://doi.org/10.5281/zenodo.164986}

\bibitem[{{Briggs}(1995)}]{briggs95}
{Briggs}, D.~S. 1995, PhD thesis, {New Mexico Institute of Mining and
  Technology}

\bibitem[{{Bruzual} \& {Charlot}(2003)}]{2003MNRAS.344.1000B}
{Bruzual}, G., \& {Charlot}, S. 2003, \mnras, 344, 1000,
  \dodoi{10.1046/j.1365-8711.2003.06897.x}

\bibitem[{{Bussmann} {et~al.}(2012){Bussmann}, {Gurwell}, {Fu}, {Smith}, {Dye},
  {Auld}, {Baes}, {Baker}, {Bonfield}, {Cava}, {Clements}, {Cooray}, {Coppin},
  {Dannerbauer}, {Dariush}, {De Zotti}, {Dunne}, {Eales}, {Fritz}, {Hopwood},
  {Ibar}, {Ivison}, {Jarvis}, {Kim}, {Leeuw}, {Maddox}, {Micha{\l}owski},
  {Negrello}, {Pascale}, {Pohlen}, {Riechers}, {Rigby}, {Scott}, {Temi}, {Van
  der Werf}, {Wardlow}, {Wilner}, \& {Verma}}]{2012ApJ...756..134B}
{Bussmann}, R.~S., {Gurwell}, M.~A., {Fu}, H., {et~al.} 2012, \apj, 756, 134,
  \dodoi{10.1088/0004-637X/756/2/134}

\bibitem[{{Bussmann} {et~al.}(2013){Bussmann}, {P{\'e}rez-Fournon}, {Amber},
  {Calanog}, {Gurwell}, {Dannerbauer}, {De Bernardis}, {Fu}, {Harris}, {Krips},
  {Lapi}, {Maiolino}, {Omont}, {Riechers}, {Wardlow}, {Baker}, {Birkinshaw},
  {Bock}, {Bourne}, {Clements}, {Cooray}, {De Zotti}, {Dunne}, {Dye}, {Eales},
  {Farrah}, {Gavazzi}, {Gonz{\'a}lez Nuevo}, {Hopwood}, {Ibar}, {Ivison},
  {Laporte}, {Maddox}, {Mart{\'{\i}}nez-Navajas}, {Michalowski}, {Negrello},
  {Oliver}, {Roseboom}, {Scott}, {Serjeant}, {Smith}, {Smith}, {Streblyanska},
  {Valiante}, {van der Werf}, {Verma}, {Vieira}, {Wang}, \&
  {Wilner}}]{2013ApJ...779...25B}
{Bussmann}, R.~S., {P{\'e}rez-Fournon}, I., {Amber}, S., {et~al.} 2013, \apj,
  779, 25, \dodoi{10.1088/0004-637X/779/1/25}

\bibitem[{{Bussmann} {et~al.}(2015){Bussmann}, {Riechers}, {Fialkov},
  {Scudder}, {Hayward}, {Cowley}, {Bock}, {Calanog}, {Chapman}, {Cooray}, {De
  Bernardis}, {Farrah}, {Fu}, {Gavazzi}, {Hopwood}, {Ivison}, {Jarvis},
  {Lacey}, {Loeb}, {Oliver}, {P{\'e}rez-Fournon}, {Rigopoulou}, {Roseboom},
  {Scott}, {Smith}, {Vieira}, {Wang}, \& {Wardlow}}]{2015ApJ...812...43B}
{Bussmann}, R.~S., {Riechers}, D., {Fialkov}, A., {et~al.} 2015, \apj, 812, 43,
  \dodoi{10.1088/0004-637X/812/1/43}

\bibitem[{{Ca{\~n}ameras} {et~al.}(2015){Ca{\~n}ameras}, {Nesvadba}, {Guery},
  {McKenzie}, {K{\"o}nig}, {Petitpas}, {Dole}, {Frye}, {Flores-Cacho},
  {Montier}, {Negrello}, {Beelen}, {Boone}, {Dicken}, {Lagache}, {Le Floc'h},
  {Altieri}, {B{\'e}thermin}, {Chary}, {de Zotti}, {Giard}, {Kneissl}, {Krips},
  {Malhotra}, {Martinache}, {Omont}, {Pointecouteau}, {Puget}, {Scott},
  {Soucail}, {Valtchanov}, {Welikala}, \& {Yan}}]{2015A&A...581A.105C}
{Ca{\~n}ameras}, R., {Nesvadba}, N.~P.~H., {Guery}, D., {et~al.} 2015, \aap,
  581, A105, \dodoi{10.1051/0004-6361/201425128}

\bibitem[{{Calzetti} {et~al.}(2000){Calzetti}, {Armus}, {Bohlin}, {Kinney},
  {Koornneef}, \& {Storchi-Bergmann}}]{2000ApJ...533..682C}
{Calzetti}, D., {Armus}, L., {Bohlin}, R.~C., {et~al.} 2000, \apj, 533, 682,
  \dodoi{10.1086/308692}

\bibitem[{{Capak} {et~al.}(2011){Capak}, {Riechers}, {Scoville}, {Carilli},
  {Cox}, {Neri}, {Robertson}, {Salvato}, {Schinnerer}, {Yan}, {Wilson}, {Yun},
  {Civano}, {Elvis}, {Karim}, {Mobasher}, \& {Staguhn}}]{2011Natur.470..233C}
{Capak}, P.~L., {Riechers}, D., {Scoville}, N.~Z., {et~al.} 2011, \nat, 470,
  233, \dodoi{10.1038/nature09681}

\bibitem[{{Capak} {et~al.}(2015){Capak}, {Carilli}, {Jones}, {Casey},
  {Riechers}, {Sheth}, {Carollo}, {Ilbert}, {Karim}, {Lefevre}, {Lilly},
  {Scoville}, {Smolcic}, \& {Yan}}]{2015Natur.522..455C}
{Capak}, P.~L., {Carilli}, C., {Jones}, G., {et~al.} 2015, \nat, 522, 455,
  \dodoi{10.1038/nature14500}

\bibitem[{{Carilli} \& {Walter}(2013)}]{2013ARA&A..51..105C}
{Carilli}, C.~L., \& {Walter}, F. 2013, \araa, 51, 105,
  \dodoi{10.1146/annurev-astro-082812-140953}

\bibitem[{{Casey}(2016)}]{2016ApJ...824...36C}
{Casey}, C.~M. 2016, \apj, 824, 36, \dodoi{10.3847/0004-637X/824/1/36}

\bibitem[{{Casey} {et~al.}(2014){Casey}, {Narayanan}, \&
  {Cooray}}]{2014PhR...541...45C}
{Casey}, C.~M., {Narayanan}, D., \& {Cooray}, A. 2014, \physrep, 541, 45,
  \dodoi{10.1016/j.physrep.2014.02.009}

\bibitem[{{Casey} {et~al.}(2015){Casey}, {Cooray}, {Capak}, {Fu}, {Kovac},
  {Lilly}, {Sanders}, {Scoville}, \& {Treister}}]{2015ApJ...808L..33C}
{Casey}, C.~M., {Cooray}, A., {Capak}, P., {et~al.} 2015, \apjl, 808, L33,
  \dodoi{10.1088/2041-8205/808/2/L33}

\bibitem[{{Chabrier}(2003)}]{2003PASP..115..763C}
{Chabrier}, G. 2003, \pasp, 115, 763, \dodoi{10.1086/376392}

\bibitem[{{Chapman} {et~al.}(2009){Chapman}, {Blain}, {Ibata}, {Ivison},
  {Smail}, \& {Morrison}}]{2009ApJ...691..560C}
{Chapman}, S.~C., {Blain}, A., {Ibata}, R., {et~al.} 2009, \apj, 691, 560,
  \dodoi{10.1088/0004-637X/691/1/560}

\bibitem[{{Chapman} {et~al.}(2005){Chapman}, {Blain}, {Smail}, \&
  {Ivison}}]{2005ApJ...622..772C}
{Chapman}, S.~C., {Blain}, A.~W., {Smail}, I., \& {Ivison}, R.~J. 2005, \apj,
  622, 772, \dodoi{10.1086/428082}

\bibitem[{{Chiang} {et~al.}(2013){Chiang}, {Overzier}, \&
  {Gebhardt}}]{2013ApJ...779..127C}
{Chiang}, Y.-K., {Overzier}, R., \& {Gebhardt}, K. 2013, \apj, 779, 127,
  \dodoi{10.1088/0004-637X/779/2/127}

\bibitem[{{Chiang} {et~al.}(2017){Chiang}, {Overzier}, {Gebhardt}, \&
  {Henriques}}]{2017ApJ...844L..23C}
{Chiang}, Y.-K., {Overzier}, R.~A., {Gebhardt}, K., \& {Henriques}, B. 2017,
  \apjl, 844, L23, \dodoi{10.3847/2041-8213/aa7e7b}

\bibitem[{{Chiang} {et~al.}(2015){Chiang}, {Overzier}, {Gebhardt},
  {Finkelstein}, {Chiang}, {Hill}, {Blanc}, {Drory}, {Chonis}, {Zeimann},
  {Hagen}, {Schneider}, {Jogee}, {Ciardullo}, \&
  {Gronwall}}]{2015ApJ...808...37C}
{Chiang}, Y.-K., {Overzier}, R.~A., {Gebhardt}, K., {et~al.} 2015, \apj, 808,
  37, \dodoi{10.1088/0004-637X/808/1/37}

\bibitem[{{Cimatti} {et~al.}(2008){Cimatti}, {Cassata}, {Pozzetti}, {Kurk},
  {Mignoli}, {Renzini}, {Daddi}, {Bolzonella}, {Brusa}, {Rodighiero},
  {Dickinson}, {Franceschini}, {Zamorani}, {Berta}, {Rosati}, \&
  {Halliday}}]{2008A&A...482...21C}
{Cimatti}, A., {Cassata}, P., {Pozzetti}, L., {et~al.} 2008, \aap, 482, 21,
  \dodoi{10.1051/0004-6361:20078739}

\bibitem[{{Clements} {et~al.}(2014){Clements}, {Braglia}, {Hyde},
  {P{\'e}rez-Fournon}, {Bock}, {Cava}, {Chapman}, {Conley}, {Cooray}, {Farrah},
  {Gonz{\'a}lez Solares}, {Marchetti}, {Marsden}, {Oliver}, {Roseboom},
  {Schulz}, {Smith}, {Vaccari}, {Vieira}, {Viero}, {Wang}, {Wardlow}, {Zemcov},
  \& {de Zotti}}]{2014MNRAS.439.1193C}
{Clements}, D.~L., {Braglia}, F.~G., {Hyde}, A.~K., {et~al.} 2014, \mnras, 439,
  1193, \dodoi{10.1093/mnras/stt2253}

\bibitem[{{Clements} {et~al.}(2016){Clements}, {Braglia}, {Petitpas},
  {Greenslade}, {Cooray}, {Valiante}, {De Zotti}, {O'Halloran}, {Holdship},
  {Morris}, {P{\'e}rez-Fournon}, {Herranz}, {Riechers}, {Baes}, {Bremer},
  {Bourne}, {Dannerbauer}, {Dariush}, {Dunne}, {Eales}, {Fritz},
  {Gonzalez-Nuevo}, {Hopwood}, {Ibar}, {Ivison}, {Leeuw}, {Maddox},
  {Micha{\l}owski}, {Negrello}, {Omont}, {Oteo}, {Serjeant}, {Valtchanov},
  {Vieira}, {Wardlow}, \& {van der Werf}}]{2016MNRAS.461.1719C}
{Clements}, D.~L., {Braglia}, F., {Petitpas}, G., {et~al.} 2016, \mnras, 461,
  1719, \dodoi{10.1093/mnras/stw1224}

\bibitem[{{Coogan} {et~al.}(2018){Coogan}, {Daddi}, {Sargent}, {Strazzullo},
  {Valentino}, {Gobat}, {Magdis}, {Bethermin}, {Pannella}, {Onodera}, {Liu},
  {Cimatti}, {Dannerbauer}, {Carollo}, {Renzini}, \&
  {Tremou}}]{2018MNRAS.479..703C}
{Coogan}, R.~T., {Daddi}, E., {Sargent}, M.~T., {et~al.} 2018, \mnras, 479,
  703, \dodoi{10.1093/mnras/sty1446}

\bibitem[{{Cowley} {et~al.}(2015){Cowley}, {Lacey}, {Baugh}, \&
  {Cole}}]{2015MNRAS.446.1784C}
{Cowley}, W.~I., {Lacey}, C.~G., {Baugh}, C.~M., \& {Cole}, S. 2015, \mnras,
  446, 1784, \dodoi{10.1093/mnras/stu2179}

\bibitem[{{Cucciati} {et~al.}(2018){Cucciati}, {Lemaux}, {Zamorani}, {Le
  F{\`e}vre}, {Tasca}, {Hathi}, {Lee}, {Bardelli}, {Cassata}, {Garilli}, {Le
  Brun}, {Maccagni}, {Pentericci}, {Thomas}, {Vanzella}, {Zucca}, {Lubin},
  {Amorin}, {Cassar{\`a}}, {Cimatti}, {Talia}, {Vergani}, {Koekemoer}, {Pforr},
  \& {Salvato}}]{2018A&A...619A..49C}
{Cucciati}, O., {Lemaux}, B.~C., {Zamorani}, G., {et~al.} 2018, \aap, 619, A49,
  \dodoi{10.1051/0004-6361/201833655}

\bibitem[{{Daddi} {et~al.}(2007){Daddi}, {Dickinson}, {Morrison}, {Chary},
  {Cimatti}, {Elbaz}, {Frayer}, {Renzini}, {Pope}, {Alexander}, {Bauer},
  {Giavalisco}, {Huynh}, {Kurk}, \& {Mignoli}}]{2007ApJ...670..156D}
{Daddi}, E., {Dickinson}, M., {Morrison}, G., {et~al.} 2007, \apj, 670, 156,
  \dodoi{10.1086/521818}

\bibitem[{{Daddi} {et~al.}(2009){Daddi}, {Dannerbauer}, {Stern}, {Dickinson},
  {Morrison}, {Elbaz}, {Giavalisco}, {Mancini}, {Pope}, \&
  {Spinrad}}]{2009ApJ...694.1517D}
{Daddi}, E., {Dannerbauer}, H., {Stern}, D., {et~al.} 2009, \apj, 694, 1517,
  \dodoi{10.1088/0004-637X/694/2/1517}

\bibitem[{{Daddi} {et~al.}(2010{\natexlab{a}}){Daddi}, {Bournaud}, {Walter},
  {Dannerbauer}, {Carilli}, {Dickinson}, {Elbaz}, {Morrison}, {Riechers},
  {Onodera}, {Salmi}, {Krips}, \& {Stern}}]{2010ApJ...713..686D}
{Daddi}, E., {Bournaud}, F., {Walter}, F., {et~al.} 2010{\natexlab{a}}, \apj,
  713, 686, \dodoi{10.1088/0004-637X/713/1/686}

\bibitem[{{Daddi} {et~al.}(2010{\natexlab{b}}){Daddi}, {Elbaz}, {Walter},
  {Bournaud}, {Salmi}, {Carilli}, {Dannerbauer}, {Dickinson}, {Monaco}, \&
  {Riechers}}]{2010ApJ...714L.118D}
{Daddi}, E., {Elbaz}, D., {Walter}, F., {et~al.} 2010{\natexlab{b}}, \apjl,
  714, L118, \dodoi{10.1088/2041-8205/714/1/L118}

\bibitem[{{Daddi} {et~al.}(2015){Daddi}, {Dannerbauer}, {Liu}, {Aravena},
  {Bournaud}, {Walter}, {Riechers}, {Magdis}, {Sargent}, {B{\'e}thermin},
  {Carilli}, {Cibinel}, {Dickinson}, {Elbaz}, {Gao}, {Gobat}, {Hodge}, \&
  {Krips}}]{2015A&A...577A..46D}
{Daddi}, E., {Dannerbauer}, H., {Liu}, D., {et~al.} 2015, \aap, 577, A46,
  \dodoi{10.1051/0004-6361/201425043}

\bibitem[{{Daddi} {et~al.}(2017){Daddi}, {Jin}, {Strazzullo}, {Sargent},
  {Wang}, {Ferrari}, {Schinnerer}, {Smol{\v c}i{\'c}}, {Calabr{\'o}}, {Coogan},
  {Delhaize}, {Delvecchio}, {Elbaz}, {Gobat}, {Gu}, {Liu}, {Novak}, \&
  {Valentino}}]{2017ApJ...846L..31D}
{Daddi}, E., {Jin}, S., {Strazzullo}, V., {et~al.} 2017, \apjl, 846, L31,
  \dodoi{10.3847/2041-8213/aa8808}

\bibitem[{{Dame} {et~al.}(2001){Dame}, {Hartmann}, \&
  {Thaddeus}}]{2001ApJ...547..792D}
{Dame}, T.~M., {Hartmann}, D., \& {Thaddeus}, P. 2001, \apj, 547, 792,
  \dodoi{10.1086/318388}

\bibitem[{{Dannerbauer} {et~al.}(2014){Dannerbauer}, {Kurk}, {De Breuck},
  {Wylezalek}, {Santos}, {Koyama}, {Seymour}, {Tanaka}, {Hatch}, {Altieri},
  {Coia}, {Galametz}, {Kodama}, {Miley}, {R{\"o}ttgering}, {Sanchez-Portal},
  {Valtchanov}, {Venemans}, \& {Ziegler}}]{2014A&A...570A..55D}
{Dannerbauer}, H., {Kurk}, J.~D., {De Breuck}, C., {et~al.} 2014, \aap, 570,
  A55, \dodoi{10.1051/0004-6361/201423771}

\bibitem[{{Dannerbauer} {et~al.}(2017){Dannerbauer}, {Lehnert}, {Emonts},
  {Ziegler}, {Altieri}, {De Breuck}, {Hatch}, {Kodama}, {Koyama}, {Kurk},
  {Matiz}, {Miley}, {Narayanan}, {Norris}, {Overzier}, {R{\"o}ttgering},
  {Sargent}, {Seymour}, {Tanaka}, {Valtchanov}, \&
  {Wylezalek}}]{2017A&A...608A..48D}
{Dannerbauer}, H., {Lehnert}, M.~D., {Emonts}, B., {et~al.} 2017, \aap, 608,
  A48, \dodoi{10.1051/0004-6361/201730449}

\bibitem[{{Diener} {et~al.}(2013){Diener}, {Lilly}, {Knobel}, {Zamorani},
  {Lemson}, {Kampczyk}, {Scoville}, {Carollo}, {Contini}, {Kneib}, {Le Fevre},
  {Mainieri}, {Renzini}, {Scodeggio}, {Bardelli}, {Bolzonella}, {Bongiorno},
  {Caputi}, {Cucciati}, {de la Torre}, {de Ravel}, {Franzetti}, {Garilli},
  {Iovino}, {Kova{\v c}}, {Lamareille}, {Le Borgne}, {Le Brun}, {Maier},
  {Mignoli}, {Pello}, {Peng}, {Perez Montero}, {Presotto}, {Silverman},
  {Tanaka}, {Tasca}, {Tresse}, {Vergani}, {Zucca}, {Bordoloi}, {Cappi},
  {Cimatti}, {Coppa}, {Koekemoer}, {L{\'o}pez-Sanjuan}, {McCracken}, {Moresco},
  {Nair}, {Pozzetti}, \& {Welikala}}]{2013ApJ...765..109D}
{Diener}, C., {Lilly}, S.~J., {Knobel}, C., {et~al.} 2013, \apj, 765, 109,
  \dodoi{10.1088/0004-637X/765/2/109}

\bibitem[{{Diener} {et~al.}(2015){Diener}, {Lilly}, {Ledoux}, {Zamorani},
  {Bolzonella}, {Murphy}, {Capak}, {Ilbert}, \&
  {McCracken}}]{2015ApJ...802...31D}
{Diener}, C., {Lilly}, S.~J., {Ledoux}, C., {et~al.} 2015, \apj, 802, 31,
  \dodoi{10.1088/0004-637X/802/1/31}

\bibitem[{{Donevski} {et~al.}(2018){Donevski}, {Buat}, {Boone}, {Pappalardo},
  {Bethermin}, {Schreiber}, {Mazyed}, {Alvarez-Marquez}, \&
  {Duivenvoorden}}]{2018A&A...614A..33D}
{Donevski}, D., {Buat}, V., {Boone}, F., {et~al.} 2018, \aap, 614, A33,
  \dodoi{10.1051/0004-6361/201731888}

\bibitem[{{Dowell} {et~al.}(2014){Dowell}, {Conley}, {Glenn}, {Arumugam},
  {Asboth}, {Aussel}, {Bertoldi}, {B{\'e}thermin}, {Bock}, {Boselli}, {Bridge},
  {Buat}, {Burgarella}, {Cabrera-Lavers}, {Casey}, {Chapman}, {Clements},
  {Conversi}, {Cooray}, {Dannerbauer}, {De Bernardis}, {Ellsworth-Bowers},
  {Farrah}, {Franceschini}, {Griffin}, {Gurwell}, {Halpern}, {Hatziminaoglou},
  {Heinis}, {Ibar}, {Ivison}, {Laporte}, {Marchetti},
  {Mart{\'{\i}}nez-Navajas}, {Marsden}, {Morrison}, {Nguyen}, {O'Halloran},
  {Oliver}, {Omont}, {Page}, {Papageorgiou}, {Pearson}, {Petitpas},
  {P{\'e}rez-Fournon}, {Pohlen}, {Riechers}, {Rigopoulou}, {Roseboom},
  {Rowan-Robinson}, {Sayers}, {Schulz}, {Scott}, {Seymour}, {Shupe}, {Smith},
  {Streblyanska}, {Symeonidis}, {Vaccari}, {Valtchanov}, {Vieira}, {Viero},
  {Wang}, {Wardlow}, {Xu}, \& {Zemcov}}]{2014ApJ...780...75D}
{Dowell}, C.~D., {Conley}, A., {Glenn}, J., {et~al.} 2014, \apj, 780, 75,
  \dodoi{10.1088/0004-637X/780/1/75}

\bibitem[{{Downes} \& {Solomon}(1998)}]{1998ApJ...507..615D}
{Downes}, D., \& {Solomon}, P.~M. 1998, \apj, 507, 615, \dodoi{10.1086/306339}

\bibitem[{{Draine} \& {Li}(2007)}]{2007ApJ...657..810D}
{Draine}, B.~T., \& {Li}, A. 2007, \apj, 657, 810, \dodoi{10.1086/511055}

\bibitem[{{Draine} {et~al.}(2007){Draine}, {Dale}, {Bendo}, {Gordon}, {Smith},
  {Armus}, {Engelbracht}, {Helou}, {Kennicutt}, {Li}, {Roussel}, {Walter},
  {Calzetti}, {Moustakas}, {Murphy}, {Rieke}, {Bot}, {Hollenbach}, {Sheth}, \&
  {Teplitz}}]{2007ApJ...663..866D}
{Draine}, B.~T., {Dale}, D.~A., {Bendo}, G., {et~al.} 2007, \apj, 663, 866,
  \dodoi{10.1086/518306}

\bibitem[{{Eales} {et~al.}(2010){Eales}, {Dunne}, {Clements}, {Cooray}, {De
  Zotti}, {Dye}, {Ivison}, {Jarvis}, {Lagache}, {Maddox}, {Negrello},
  {Serjeant}, {Thompson}, {Van Kampen}, {Amblard}, {Andreani}, {Baes},
  {Beelen}, {Bendo}, {Benford}, {Bertoldi}, {Bock}, {Bonfield}, {Boselli},
  {Bridge}, {Buat}, {Burgarella}, {Carlberg}, {Cava}, {Chanial}, {Charlot},
  {Christopher}, {Coles}, {Cortese}, {Dariush}, {da Cunha}, {Dalton}, {Danese},
  {Dannerbauer}, {Driver}, {Dunlop}, {Fan}, {Farrah}, {Frayer}, {Frenk},
  {Geach}, {Gardner}, {Gomez}, {Gonz{\'a}lez-Nuevo}, {Gonz{\'a}lez-Solares},
  {Griffin}, {Hardcastle}, {Hatziminaoglou}, {Herranz}, {Hughes}, {Ibar},
  {Jeong}, {Lacey}, {Lapi}, {Lawrence}, {Lee}, {Leeuw}, {Liske},
  {L{\'o}pez-Caniego}, {M{\"u}ller}, {Nandra}, {Panuzzo}, {Papageorgiou},
  {Patanchon}, {Peacock}, {Pearson}, {Phillipps}, {Pohlen}, {Popescu},
  {Rawlings}, {Rigby}, {Rigopoulou}, {Robotham}, {Rodighiero}, {Sansom},
  {Schulz}, {Scott}, {Smith}, {Sibthorpe}, {Smail}, {Stevens}, {Sutherland},
  {Takeuchi}, {Tedds}, {Temi}, {Tuffs}, {Trichas}, {Vaccari}, {Valtchanov},
  {van der Werf}, {Verma}, {Vieria}, {Vlahakis}, \&
  {White}}]{2010PASP..122..499E}
{Eales}, S., {Dunne}, L., {Clements}, D., {et~al.} 2010, \pasp, 122, 499,
  \dodoi{10.1086/653086}

\bibitem[{{Elbaz} {et~al.}(2007){Elbaz}, {Daddi}, {Le Borgne}, {Dickinson},
  {Alexander}, {Chary}, {Starck}, {Brandt}, {Kitzbichler}, {MacDonald},
  {Nonino}, {Popesso}, {Stern}, \& {Vanzella}}]{2007A&A...468...33E}
{Elbaz}, D., {Daddi}, E., {Le Borgne}, D., {et~al.} 2007, \aap, 468, 33,
  \dodoi{10.1051/0004-6361:20077525}

\bibitem[{{Faisst} {et~al.}(2017){Faisst}, {Capak}, {Yan}, {Pavesi},
  {Riechers}, {Bari{\v s}i{\'c}}, {Cooke}, {Kartaltepe}, \&
  {Masters}}]{2017ApJ...847...21F}
{Faisst}, A.~L., {Capak}, P.~L., {Yan}, L., {et~al.} 2017, \apj, 847, 21,
  \dodoi{10.3847/1538-4357/aa886c}

\bibitem[{{Fu} {et~al.}(2013){Fu}, {Cooray}, {Feruglio}, {Ivison}, {Riechers},
  {Gurwell}, {Bussmann}, {Harris}, {Altieri}, {Aussel}, {Baker}, {Bock},
  {Boylan-Kolchin}, {Bridge}, {Calanog}, {Casey}, {Cava}, {Chapman},
  {Clements}, {Conley}, {Cox}, {Farrah}, {Frayer}, {Hopwood}, {Jia}, {Magdis},
  {Marsden}, {Mart{\'{\i}}nez-Navajas}, {Negrello}, {Neri}, {Oliver}, {Omont},
  {Page}, {P{\'e}rez-Fournon}, {Schulz}, {Scott}, {Smith}, {Vaccari},
  {Valtchanov}, {Vieira}, {Viero}, {Wang}, {Wardlow}, \&
  {Zemcov}}]{2013Natur.498..338F}
{Fu}, H., {Cooray}, A., {Feruglio}, C., {et~al.} 2013, \nat, 498, 338,
  \dodoi{10.1038/nature12184}

\bibitem[{{Geach} {et~al.}(2011){Geach}, {Smail}, {Moran}, {MacArthur},
  {Lagos}, \& {Edge}}]{2011ApJ...730L..19G}
{Geach}, J.~E., {Smail}, I., {Moran}, S.~M., {et~al.} 2011, \apjl, 730, L19,
  \dodoi{10.1088/2041-8205/730/2/L19}

\bibitem[{{Genzel} {et~al.}(2010){Genzel}, {Tacconi}, {Gracia-Carpio},
  {Sternberg}, {Cooper}, {Shapiro}, {Bolatto}, {Bouch{\'e}}, {Bournaud},
  {Burkert}, {Combes}, {Comerford}, {Cox}, {Davis}, {Schreiber},
  {Garcia-Burillo}, {Lutz}, {Naab}, {Neri}, {Omont}, {Shapley}, \&
  {Weiner}}]{2010MNRAS.407.2091G}
{Genzel}, R., {Tacconi}, L.~J., {Gracia-Carpio}, J., {et~al.} 2010, \mnras,
  407, 2091, \dodoi{10.1111/j.1365-2966.2010.16969.x}

\bibitem[{{Genzel} {et~al.}(2012){Genzel}, {Tacconi}, {Combes}, {Bolatto},
  {Neri}, {Sternberg}, {Cooper}, {Bouch{\'e}}, {Bournaud}, {Burkert},
  {Comerford}, {Cox}, {Davis}, {F{\"o}rster Schreiber}, {Garcia-Burillo},
  {Gracia-Carpio}, {Lutz}, {Naab}, {Newman}, {Saintonge}, {Shapiro}, {Shapley},
  \& {Weiner}}]{2012ApJ...746...69G}
{Genzel}, R., {Tacconi}, L.~J., {Combes}, F., {et~al.} 2012, \apj, 746, 69,
  \dodoi{10.1088/0004-637X/746/1/69}

\bibitem[{{Genzel} {et~al.}(2015){Genzel}, {Tacconi}, {Lutz}, {Saintonge},
  {Berta}, {Magnelli}, {Combes}, {Garc{\'{\i}}a-Burillo}, {Neri}, {Bolatto},
  {Contini}, {Lilly}, {Boissier}, {Boone}, {Bouch{\'e}}, {Bournaud}, {Burkert},
  {Carollo}, {Colina}, {Cooper}, {Cox}, {Feruglio}, {F{\"o}rster Schreiber},
  {Freundlich}, {Gracia-Carpio}, {Juneau}, {Kovac}, {Lippa}, {Naab}, {Salome},
  {Renzini}, {Sternberg}, {Walter}, {Weiner}, {Weiss}, \&
  {Wuyts}}]{2015ApJ...800...20G}
{Genzel}, R., {Tacconi}, L.~J., {Lutz}, D., {et~al.} 2015, \apj, 800, 20,
  \dodoi{10.1088/0004-637X/800/1/20}

\bibitem[{{Gobat} {et~al.}(2011){Gobat}, {Daddi}, {Onodera}, {Finoguenov},
  {Renzini}, {Arimoto}, {Bouwens}, {Brusa}, {Chary}, {Cimatti}, {Dickinson},
  {Kong}, \& {Mignoli}}]{2011A&A...526A.133G}
{Gobat}, R., {Daddi}, E., {Onodera}, M., {et~al.} 2011, \aap, 526, A133,
  \dodoi{10.1051/0004-6361/201016084}

\bibitem[{{G{\'o}mez-Guijarro} {et~al.}(2018){G{\'o}mez-Guijarro}, {Toft},
  {Karim}, {Magnelli}, {Magdis}, {Jim{\'e}nez-Andrade}, {Capak}, {Fraternali},
  {Fujimoto}, {Riechers}, {Schinnerer}, {Smol{\v c}i{\'c}}, {Aravena},
  {Bertoldi}, {Cortzen}, {Hasinger}, {Hu}, {Jones}, {Koekemoer}, {Lee},
  {McCracken}, {Micha{\l}owski}, {Navarrete}, {Povi{\'c}}, {Puglisi},
  {Romano-D{\'{\i}}az}, {Sheth}, {Silverman}, {Staguhn}, {Steinhardt},
  {Stockmann}, {Tanaka}, {Valentino}, {van Kampen}, \&
  {Zirm}}]{2018ApJ...856..121G}
{G{\'o}mez-Guijarro}, C., {Toft}, S., {Karim}, A., {et~al.} 2018, \apj, 856,
  121, \dodoi{10.3847/1538-4357/aab206}

\bibitem[{{Greenslade} {et~al.}(2018){Greenslade}, {Clements}, {Cheng}, {De
  Zotti}, {Scott}, {Valiante}, {Eales}, {Bremer}, {Dannerbauer}, {Birkinshaw},
  {Farrah}, {Harrison}, {Micha{\l}owski}, {Valtchanov}, {Oteo}, {Baes},
  {Cooray}, {Negrello}, {Wang}, {van der Werf}, {Dunne}, \&
  {Dye}}]{2018MNRAS.476.3336G}
{Greenslade}, J., {Clements}, D.~L., {Cheng}, T., {et~al.} 2018, \mnras, 476,
  3336, \dodoi{10.1093/mnras/sty023}

\bibitem[{{Griffin} {et~al.}(2010){Griffin}, {Abergel}, {Abreu}, {Ade},
  {Andr{\'e}}, {Augueres}, {Babbedge}, {Bae}, {Baillie}, {Baluteau}, {Barlow},
  {Bendo}, {Benielli}, {Bock}, {Bonhomme}, {Brisbin}, {Brockley-Blatt},
  {Caldwell}, {Cara}, {Castro-Rodriguez}, {Cerulli}, {Chanial}, {Chen},
  {Clark}, {Clements}, {Clerc}, {Coker}, {Communal}, {Conversi}, {Cox},
  {Crumb}, {Cunningham}, {Daly}, {Davis}, {de Antoni}, {Delderfield}, {Devin},
  {di Giorgio}, {Didschuns}, {Dohlen}, {Donati}, {Dowell}, {Dowell}, {Duband},
  {Dumaye}, {Emery}, {Ferlet}, {Ferrand}, {Fontignie}, {Fox}, {Franceschini},
  {Frerking}, {Fulton}, {Garcia}, {Gastaud}, {Gear}, {Glenn}, {Goizel},
  {Griffin}, {Grundy}, {Guest}, {Guillemet}, {Hargrave}, {Harwit}, {Hastings},
  {Hatziminaoglou}, {Herman}, {Hinde}, {Hristov}, {Huang}, {Imhof}, {Isaak},
  {Israelsson}, {Ivison}, {Jennings}, {Kiernan}, {King}, {Lange}, {Latter},
  {Laurent}, {Laurent}, {Leeks}, {Lellouch}, {Levenson}, {Li}, {Li},
  {Lilienthal}, {Lim}, {Liu}, {Lu}, {Madden}, {Mainetti}, {Marliani}, {McKay},
  {Mercier}, {Molinari}, {Morris}, {Moseley}, {Mulder}, {Mur}, {Naylor},
  {Nguyen}, {O'Halloran}, {Oliver}, {Olofsson}, {Olofsson}, {Orfei}, {Page},
  {Pain}, {Panuzzo}, {Papageorgiou}, {Parks}, {Parr-Burman}, {Pearce},
  {Pearson}, {P{\'e}rez-Fournon}, {Pinsard}, {Pisano}, {Podosek}, {Pohlen},
  {Polehampton}, {Pouliquen}, {Rigopoulou}, {Rizzo}, {Roseboom}, {Roussel},
  {Rowan-Robinson}, {Rownd}, {Saraceno}, {Sauvage}, {Savage}, {Savini},
  {Sawyer}, {Scharmberg}, {Schmitt}, {Schneider}, {Schulz}, {Schwartz},
  {Shafer}, {Shupe}, {Sibthorpe}, {Sidher}, {Smith}, {Smith}, {Smith},
  {Spencer}, {Stobie}, {Sudiwala}, {Sukhatme}, {Surace}, {Stevens}, {Swinyard},
  {Trichas}, {Tourette}, {Triou}, {Tseng}, {Tucker}, {Turner}, {Vaccari},
  {Valtchanov}, {Vigroux}, {Virique}, {Voellmer}, {Walker}, {Ward}, {Waskett},
  {Weilert}, {Wesson}, {White}, {Whitehouse}, {Wilson}, {Winter}, {Woodcraft},
  {Wright}, {Xu}, {Zavagno}, {Zemcov}, {Zhang}, \&
  {Zonca}}]{2010A&A...518L...3G}
{Griffin}, M.~J., {Abergel}, A., {Abreu}, A., {et~al.} 2010, \aap, 518, L3,
  \dodoi{10.1051/0004-6361/201014519}

\bibitem[{{Groves} {et~al.}(2015){Groves}, {Schinnerer}, {Leroy}, {Galametz},
  {Walter}, {Bolatto}, {Hunt}, {Dale}, {Calzetti}, {Croxall}, \&
  {Kennicutt}}]{2015ApJ...799...96G}
{Groves}, B.~A., {Schinnerer}, E., {Leroy}, A., {et~al.} 2015, \apj, 799, 96,
  \dodoi{10.1088/0004-637X/799/1/96}

\bibitem[{{Hayashi} {et~al.}(2018){Hayashi}, {Tadaki}, {Kodama}, {Kohno},
  {Yamaguchi}, {Hatsukade}, {Koyama}, {Shimakawa}, {Tamura}, \&
  {Suzuki}}]{2018ApJ...856..118H}
{Hayashi}, M., {Tadaki}, K.-i., {Kodama}, T., {et~al.} 2018, \apj, 856, 118,
  \dodoi{10.3847/1538-4357/aab3e7}

\bibitem[{{Hayward} {et~al.}(2013){Hayward}, {Behroozi}, {Somerville},
  {Primack}, {Moreno}, \& {Wechsler}}]{2013MNRAS.434.2572H}
{Hayward}, C.~C., {Behroozi}, P.~S., {Somerville}, R.~S., {et~al.} 2013,
  \mnras, 434, 2572, \dodoi{10.1093/mnras/stt1202}

\bibitem[{{Hayward} {et~al.}(2018){Hayward}, {Chapman}, {Steidel}, {Golob},
  {Casey}, {Smith}, {Zitrin}, {Blain}, {Bremer}, {Chen}, {Coppin}, {Farrah},
  {Ibar}, {Micha{\l}owski}, {Sawicki}, {Scott}, {van der Werf}, {Fazio},
  {Geach}, {Gurwell}, {Petitpas}, \& {Wilner}}]{2018MNRAS.476.2278H}
{Hayward}, C.~C., {Chapman}, S.~C., {Steidel}, C.~C., {et~al.} 2018, \mnras,
  476, 2278, \dodoi{10.1093/mnras/sty304}

\bibitem[{{Hodge} {et~al.}(2013){Hodge}, {Karim}, {Smail}, {Swinbank},
  {Walter}, {Biggs}, {Ivison}, {Weiss}, {Alexander}, {Bertoldi}, {Brandt},
  {Chapman}, {Coppin}, {Cox}, {Danielson}, {Dannerbauer}, {De Breuck},
  {Decarli}, {Edge}, {Greve}, {Knudsen}, {Menten}, {Rix}, {Schinnerer},
  {Simpson}, {Wardlow}, \& {van der Werf}}]{2013ApJ...768...91H}
{Hodge}, J.~A., {Karim}, A., {Smail}, I., {et~al.} 2013, \apj, 768, 91,
  \dodoi{10.1088/0004-637X/768/1/91}

\bibitem[{Hunter(2007)}]{Hunter:2007}
Hunter, J.~D. 2007, Computing In Science \& Engineering, 9, 90,
  \dodoi{10.1109/MCSE.2007.55}

\bibitem[{{Ilbert} {et~al.}(2006){Ilbert}, {Arnouts}, {McCracken},
  {Bolzonella}, {Bertin}, {Le F{\`e}vre}, {Mellier}, {Zamorani}, {Pell{\`o}},
  {Iovino}, {Tresse}, {Le Brun}, {Bottini}, {Garilli}, {Maccagni}, {Picat},
  {Scaramella}, {Scodeggio}, {Vettolani}, {Zanichelli}, {Adami}, {Bardelli},
  {Cappi}, {Charlot}, {Ciliegi}, {Contini}, {Cucciati}, {Foucaud}, {Franzetti},
  {Gavignaud}, {Guzzo}, {Marano}, {Marinoni}, {Mazure}, {Meneux}, {Merighi},
  {Paltani}, {Pollo}, {Pozzetti}, {Radovich}, {Zucca}, {Bondi}, {Bongiorno},
  {Busarello}, {de La Torre}, {Gregorini}, {Lamareille}, {Mathez}, {Merluzzi},
  {Ripepi}, {Rizzo}, \& {Vergani}}]{2006A&A...457..841I}
{Ilbert}, O., {Arnouts}, S., {McCracken}, H.~J., {et~al.} 2006, \aap, 457, 841,
  \dodoi{10.1051/0004-6361:20065138}

\bibitem[{{Ivison} {et~al.}(2013){Ivison}, {Swinbank}, {Smail}, {Harris},
  {Bussmann}, {Cooray}, {Cox}, {Fu}, {Kov{\'a}cs}, {Krips}, {Narayanan},
  {Negrello}, {Neri}, {Pe{\~n}arrubia}, {Richard}, {Riechers}, {Rowlands},
  {Staguhn}, {Targett}, {Amber}, {Baker}, {Bourne}, {Bertoldi}, {Bremer},
  {Calanog}, {Clements}, {Dannerbauer}, {Dariush}, {De Zotti}, {Dunne},
  {Eales}, {Farrah}, {Fleuren}, {Franceschini}, {Geach}, {George}, {Helly},
  {Hopwood}, {Ibar}, {Jarvis}, {Kneib}, {Maddox}, {Omont}, {Scott}, {Serjeant},
  {Smith}, {Thompson}, {Valiante}, {Valtchanov}, {Vieira}, \& {van der
  Werf}}]{2013ApJ...772..137I}
{Ivison}, R.~J., {Swinbank}, A.~M., {Smail}, I., {et~al.} 2013, \apj, 772, 137,
  \dodoi{10.1088/0004-637X/772/2/137}

\bibitem[{{Jarvis} {et~al.}(2013){Jarvis}, {Bonfield}, {Bruce}, {Geach},
  {McAlpine}, {McLure}, {Gonz{\'a}lez-Solares}, {Irwin}, {Lewis}, {Yoldas},
  {Andreon}, {Cross}, {Emerson}, {Dalton}, {Dunlop}, {Hodgkin}, {Le},
  {Karouzos}, {Meisenheimer}, {Oliver}, {Rawlings}, {Simpson}, {Smail},
  {Smith}, {Sullivan}, {Sutherland}, {White}, \& {Zwart}}]{2013MNRAS.428.1281J}
{Jarvis}, M.~J., {Bonfield}, D.~G., {Bruce}, V.~A., {et~al.} 2013, \mnras, 428,
  1281, \dodoi{10.1093/mnras/sts118}

\bibitem[{{Kennicutt}(1998)}]{kennicutt98}
{Kennicutt}, Jr., R.~C. 1998, \araa, 36, 189,
  \dodoi{10.1146/annurev.astro.36.1.189}

\bibitem[{{Kurk} {et~al.}(2000){Kurk}, {R{\"o}ttgering}, {Pentericci}, {Miley},
  {van Breugel}, {Carilli}, {Ford}, {Heckman}, {McCarthy}, \&
  {Moorwood}}]{2000A&A...358L...1K}
{Kurk}, J.~D., {R{\"o}ttgering}, H.~J.~A., {Pentericci}, L., {et~al.} 2000,
  \aap, 358, L1

\bibitem[{{Lee} {et~al.}(2016){Lee}, {Hennawi}, {White}, {Prochaska},
  {Font-Ribera}, {Schlegel}, {Rich}, {Suzuki}, {Stark}, {Le F{\`e}vre},
  {Nugent}, {Salvato}, \& {Zamorani}}]{2016ApJ...817..160L}
{Lee}, K.-G., {Hennawi}, J.~F., {White}, M., {et~al.} 2016, \apj, 817, 160,
  \dodoi{10.3847/0004-637X/817/2/160}

\bibitem[{{Lee} {et~al.}(2017){Lee}, {Tanaka}, {Kawabe}, {Kohno}, {Kodama},
  {Kajisawa}, {Yun}, {Nakanishi}, {Iono}, {Tamura}, {Hatsukade}, {Umehata},
  {Saito}, {Izumi}, {Aretxaga}, {Tadaki}, {Zeballos}, {Ikarashi}, {Wilson},
  {Hughes}, \& {Ivison}}]{2017ApJ...842...55L}
{Lee}, M.~M., {Tanaka}, I., {Kawabe}, R., {et~al.} 2017, \apj, 842, 55,
  \dodoi{10.3847/1538-4357/aa74c2}

\bibitem[{{Leroy} {et~al.}(2011){Leroy}, {Bolatto}, {Gordon}, {Sandstrom},
  {Gratier}, {Rosolowsky}, {Engelbracht}, {Mizuno}, {Corbelli}, {Fukui}, \&
  {Kawamura}}]{2011ApJ...737...12L}
{Leroy}, A.~K., {Bolatto}, A., {Gordon}, K., {et~al.} 2011, \apj, 737, 12,
  \dodoi{10.1088/0004-637X/737/1/12}

\bibitem[{{Lonsdale} {et~al.}(2003){Lonsdale}, {Smith}, {Rowan-Robinson},
  {Surace}, {Shupe}, {Xu}, {Oliver}, {Padgett}, {Fang}, {Conrow},
  {Franceschini}, {Gautier}, {Griffin}, {Hacking}, {Masci}, {Morrison},
  {O'Linger}, {Owen}, {P{\'e}rez-Fournon}, {Pierre}, {Puetter}, {Stacey},
  {Castro}, {Polletta}, {Farrah}, {Jarrett}, {Frayer}, {Siana}, {Babbedge},
  {Dye}, {Fox}, {Gonzalez-Solares}, {Salaman}, {Berta}, {Condon}, {Dole}, \&
  {Serjeant}}]{2003PASP..115..897L}
{Lonsdale}, C.~J., {Smith}, H.~E., {Rowan-Robinson}, M., {et~al.} 2003, \pasp,
  115, 897, \dodoi{10.1086/376850}

\bibitem[{{Magdis} {et~al.}(2012){Magdis}, {Daddi}, {B{\'e}thermin}, {Sargent},
  {Elbaz}, {Pannella}, {Dickinson}, {Dannerbauer}, {da Cunha}, {Walter},
  {Rigopoulou}, {Charmandaris}, {Hwang}, \& {Kartaltepe}}]{2012ApJ...760....6M}
{Magdis}, G.~E., {Daddi}, E., {B{\'e}thermin}, M., {et~al.} 2012, \apj, 760, 6,
  \dodoi{10.1088/0004-637X/760/1/6}

\bibitem[{{Magdis} {et~al.}(2017){Magdis}, {Rigopoulou}, {Daddi}, {Bethermin},
  {Feruglio}, {Sargent}, {Dannerbauer}, {Dickinson}, {Elbaz}, {Gomez Guijarro},
  {Huang}, {Toft}, \& {Valentino}}]{2017A&A...603A..93M}
{Magdis}, G.~E., {Rigopoulou}, D., {Daddi}, E., {et~al.} 2017, \aap, 603, A93,
  \dodoi{10.1051/0004-6361/201731037}

\bibitem[{{Magnelli} {et~al.}(2011){Magnelli}, {Elbaz}, {Chary}, {Dickinson},
  {Le Borgne}, {Frayer}, \& {Willmer}}]{2011A&A...528A..35M}
{Magnelli}, B., {Elbaz}, D., {Chary}, R.~R., {et~al.} 2011, \aap, 528, A35,
  \dodoi{10.1051/0004-6361/200913941}

\bibitem[{{Martinache} {et~al.}(2018){Martinache}, {Rettura}, {Dole},
  {Lehnert}, {Frye}, {Altieri}, {Beelen}, {B{\'e}thermin}, {Le Floc'h},
  {Giard}, {Hurier}, {Lagache}, {Montier}, {Omont}, {Pointecouteau},
  {Polletta}, {Puget}, {Scott}, {Soucail}, \& {Welikala}}]{2018A&A...620A.198M}
{Martinache}, C., {Rettura}, A., {Dole}, H., {et~al.} 2018, \aap, 620, A198,
  \dodoi{10.1051/0004-6361/201833198}

\bibitem[{{Mauduit} {et~al.}(2012){Mauduit}, {Lacy}, {Farrah}, {Surace},
  {Jarvis}, {Oliver}, {Maraston}, {Vaccari}, {Marchetti}, {Zeimann},
  {Gonz{\'a}les-Solares}, {Pforr}, {Petric}, {Henriques}, {Thomas}, {Afonso},
  {Rettura}, {Wilson}, {Falder}, {Geach}, {Huynh}, {Norris}, {Seymour},
  {Richards}, {Stanford}, {Alexander}, {Becker}, {Best}, {Bizzocchi},
  {Bonfield}, {Castro}, {Cava}, {Chapman}, {Christopher}, {Clements}, {Covone},
  {Dubois}, {Dunlop}, {Dyke}, {Edge}, {Ferguson}, {Foucaud}, {Franceschini},
  {Gal}, {Grant}, {Grossi}, {Hatziminaoglou}, {Hickey}, {Hodge}, {Huang},
  {Ivison}, {Kim}, {LeFevre}, {Lehnert}, {Lonsdale}, {Lubin}, {McLure},
  {Messias}, {Mart{\'{\i}}nez-Sansigre}, {Mortier}, {Nielsen}, {Ouchi},
  {Parish}, {Perez-Fournon}, {Pierre}, {Rawlings}, {Readhead}, {Ridgway},
  {Rigopoulou}, {Romer}, {Rosebloom}, {Rottgering}, {Rowan-Robinson}, {Sajina},
  {Simpson}, {Smail}, {Squires}, {Stevens}, {Taylor}, {Trichas}, {Urrutia},
  {van Kampen}, {Verma}, \& {Xu}}]{2012PASP..124..714M}
{Mauduit}, J.-C., {Lacy}, M., {Farrah}, D., {et~al.} 2012, \pasp, 124, 714,
  \dodoi{10.1086/666945}

\bibitem[{{McMullin} {et~al.}(2007){McMullin}, {Waters}, {Schiebel}, {Young},
  \& {Golap}}]{2007ASPC..376..127M}
{McMullin}, J.~P., {Waters}, B., {Schiebel}, D., {Young}, W., \& {Golap}, K.
  2007, in Astronomical Society of the Pacific Conference Series, Vol. 376,
  Astronomical Data Analysis Software and Systems XVI, ed. R.~A. {Shaw},
  F.~{Hill}, \& D.~J. {Bell}, 127

\bibitem[{{Miller} {et~al.}(2018){Miller}, {Chapman}, {Aravena}, {Ashby},
  {Hayward}, {Vieira}, {Wei{\ss}}, {Babul}, {B{\'e}thermin}, {Bradford},
  {Brodwin}, {Carlstrom}, {Chen}, {Cunningham}, {De Breuck}, {Gonzalez},
  {Greve}, {Harnett}, {Hezaveh}, {Lacaille}, {Litke}, {Ma}, {Malkan},
  {Marrone}, {Morningstar}, {Murphy}, {Narayanan}, {Pass}, {Perry}, {Phadke},
  {Rennehan}, {Rotermund}, {Simpson}, {Spilker}, {Sreevani}, {Stark},
  {Strandet}, \& {Strom}}]{2018Natur.556..469M}
{Miller}, T.~B., {Chapman}, S.~C., {Aravena}, M., {et~al.} 2018, \nat, 556,
  469, \dodoi{10.1038/s41586-018-0025-2}

\bibitem[{{Mu{\~n}oz Arancibia} {et~al.}(2015){Mu{\~n}oz Arancibia},
  {Navarrete}, {Padilla}, {Cora}, {Gawiser}, {Kurczynski}, \&
  {Ruiz}}]{2015MNRAS.446.2291M}
{Mu{\~n}oz Arancibia}, A.~M., {Navarrete}, F.~P., {Padilla}, N.~D., {et~al.}
  2015, \mnras, 446, 2291, \dodoi{10.1093/mnras/stu2237}

\bibitem[{{Muldrew} {et~al.}(2015){Muldrew}, {Hatch}, \&
  {Cooke}}]{2015MNRAS.452.2528M}
{Muldrew}, S.~I., {Hatch}, N.~A., \& {Cooke}, E.~A. 2015, \mnras, 452, 2528,
  \dodoi{10.1093/mnras/stv1449}

\bibitem[{{Narayanan} {et~al.}(2018){Narayanan}, {Dav{\'e}}, {Johnson},
  {Thompson}, {Conroy}, \& {Geach}}]{2018MNRAS.474.1718N}
{Narayanan}, D., {Dav{\'e}}, R., {Johnson}, B.~D., {et~al.} 2018, \mnras, 474,
  1718, \dodoi{10.1093/mnras/stx2860}

\bibitem[{{Negrello} {et~al.}(2010){Negrello}, {Hopwood}, {De Zotti}, {Cooray},
  {Verma}, {Bock}, {Frayer}, {Gurwell}, {Omont}, {Neri}, {Dannerbauer},
  {Leeuw}, {Barton}, {Cooke}, {Kim}, {da Cunha}, {Rodighiero}, {Cox},
  {Bonfield}, {Jarvis}, {Serjeant}, {Ivison}, {Dye}, {Aretxaga}, {Hughes},
  {Ibar}, {Bertoldi}, {Valtchanov}, {Eales}, {Dunne}, {Driver}, {Auld},
  {Buttiglione}, {Cava}, {Grady}, {Clements}, {Dariush}, {Fritz}, {Hill},
  {Hornbeck}, {Kelvin}, {Lagache}, {Lopez-Caniego}, {Gonzalez-Nuevo}, {Maddox},
  {Pascale}, {Pohlen}, {Rigby}, {Robotham}, {Simpson}, {Smith}, {Temi},
  {Thompson}, {Woodgate}, {York}, {Aguirre}, {Beelen}, {Blain}, {Baker},
  {Birkinshaw}, {Blundell}, {Bradford}, {Burgarella}, {Danese}, {Dunlop},
  {Fleuren}, {Glenn}, {Harris}, {Kamenetzky}, {Lupu}, {Maddalena}, {Madore},
  {Maloney}, {Matsuhara}, {Micha{\l}owski}, {Murphy}, {Naylor}, {Nguyen},
  {Popescu}, {Rawlings}, {Rigopoulou}, {Scott}, {Scott}, {Seibert}, {Smail},
  {Tuffs}, {Vieira}, {van der Werf}, \& {Zmuidzinas}}]{2010Sci...330..800N}
{Negrello}, M., {Hopwood}, R., {De Zotti}, G., {et~al.} 2010, Science, 330,
  800, \dodoi{10.1126/science.1193420}

\bibitem[{{Noble} {et~al.}(2017){Noble}, {McDonald}, {Muzzin}, {Nantais},
  {Rudnick}, {van Kampen}, {Webb}, {Wilson}, {Yee}, {Boone}, {Cooper},
  {DeGroot}, {Delahaye}, {Demarco}, {Foltz}, {Hayden}, {Lidman},
  {Manilla-Robles}, \& {Perlmutter}}]{2017ApJ...842L..21N}
{Noble}, A.~G., {McDonald}, M., {Muzzin}, A., {et~al.} 2017, \apjl, 842, L21,
  \dodoi{10.3847/2041-8213/aa77f3}

\bibitem[{{Noeske} {et~al.}(2007){Noeske}, {Weiner}, {Faber}, {Papovich},
  {Koo}, {Somerville}, {Bundy}, {Conselice}, {Newman}, {Schiminovich}, {Le
  Floc'h}, {Coil}, {Rieke}, {Lotz}, {Primack}, {Barmby}, {Cooper}, {Davis},
  {Ellis}, {Fazio}, {Guhathakurta}, {Huang}, {Kassin}, {Martin}, {Phillips},
  {Rich}, {Small}, {Willmer}, \& {Wilson}}]{2007ApJ...660L..43N}
{Noeske}, K.~G., {Weiner}, B.~J., {Faber}, S.~M., {et~al.} 2007, \apjl, 660,
  L43, \dodoi{10.1086/517926}

\bibitem[{{Obreschkow} {et~al.}(2009){Obreschkow}, {Croton}, {De Lucia},
  {Khochfar}, \& {Rawlings}}]{2009ApJ...698.1467O}
{Obreschkow}, D., {Croton}, D., {De Lucia}, G., {Khochfar}, S., \& {Rawlings},
  S. 2009, \apj, 698, 1467, \dodoi{10.1088/0004-637X/698/2/1467}

\bibitem[{{Oliver} {et~al.}(2012){Oliver}, {Bock}, {Altieri}, {Amblard},
  {Arumugam}, {Aussel}, {Babbedge}, {Beelen}, {B{\'e}thermin}, {Blain},
  {Boselli}, {Bridge}, {Brisbin}, {Buat}, {Burgarella},
  {Castro-Rodr{\'{\i}}guez}, {Cava}, {Chanial}, {Cirasuolo}, {Clements},
  {Conley}, {Conversi}, {Cooray}, {Dowell}, {Dubois}, {Dwek}, {Dye}, {Eales},
  {Elbaz}, {Farrah}, {Feltre}, {Ferrero}, {Fiolet}, {Fox}, {Franceschini},
  {Gear}, {Giovannoli}, {Glenn}, {Gong}, {Gonz{\'a}lez Solares}, {Griffin},
  {Halpern}, {Harwit}, {Hatziminaoglou}, {Heinis}, {Hurley}, {Hwang}, {Hyde},
  {Ibar}, {Ilbert}, {Isaak}, {Ivison}, {Lagache}, {Le Floc'h}, {Levenson},
  {Faro}, {Lu}, {Madden}, {Maffei}, {Magdis}, {Mainetti}, {Marchetti},
  {Marsden}, {Marshall}, {Mortier}, {Nguyen}, {O'Halloran}, {Omont}, {Page},
  {Panuzzo}, {Papageorgiou}, {Patel}, {Pearson}, {P{\'e}rez-Fournon}, {Pohlen},
  {Rawlings}, {Raymond}, {Rigopoulou}, {Riguccini}, {Rizzo}, {Rodighiero},
  {Roseboom}, {Rowan-Robinson}, {S{\'a}nchez Portal}, {Schulz}, {Scott},
  {Seymour}, {Shupe}, {Smith}, {Stevens}, {Symeonidis}, {Trichas}, {Tugwell},
  {Vaccari}, {Valtchanov}, {Vieira}, {Viero}, {Vigroux}, {Wang}, {Ward},
  {Wardlow}, {Wright}, {Xu}, \& {Zemcov}}]{2012MNRAS.424.1614O}
{Oliver}, S.~J., {Bock}, J., {Altieri}, B., {et~al.} 2012, \mnras, 424, 1614,
  \dodoi{10.1111/j.1365-2966.2012.20912.x}

\bibitem[{{Oteo} {et~al.}(2018){Oteo}, {Ivison}, {Dunne}, {Manilla-Robles},
  {Maddox}, {Lewis}, {de Zotti}, {Bremer}, {Clements}, {Cooray}, {Dannerbauer},
  {Eales}, {Greenslade}, {Omont}, {Perez-Fourn{\'o}n}, {Riechers}, {Scott},
  {van der Werf}, {Weiss}, \& {Zhang}}]{2018ApJ...856...72O}
{Oteo}, I., {Ivison}, R.~J., {Dunne}, L., {et~al.} 2018, \apj, 856, 72,
  \dodoi{10.3847/1538-4357/aaa1f1}

\bibitem[{{Overzier}(2016)}]{2016A&ARv..24...14O}
{Overzier}, R.~A. 2016, \aapr, 24, 14, \dodoi{10.1007/s00159-016-0100-3}

\bibitem[{{Papadopoulos} {et~al.}(2012{\natexlab{a}}){Papadopoulos}, {van der
  Werf}, {Xilouris}, {Isaak}, \& {Gao}}]{2012ApJ...751...10P}
{Papadopoulos}, P.~P., {van der Werf}, P., {Xilouris}, E., {Isaak}, K.~G., \&
  {Gao}, Y. 2012{\natexlab{a}}, \apj, 751, 10,
  \dodoi{10.1088/0004-637X/751/1/10}

\bibitem[{{Papadopoulos} {et~al.}(2012{\natexlab{b}}){Papadopoulos}, {van der
  Werf}, {Xilouris}, {Isaak}, {Gao}, \& {M{\"u}hle}}]{2012MNRAS.426.2601P}
{Papadopoulos}, P.~P., {van der Werf}, P.~P., {Xilouris}, E.~M., {et~al.}
  2012{\natexlab{b}}, \mnras, 426, 2601,
  \dodoi{10.1111/j.1365-2966.2012.21001.x}

\bibitem[{{Pavesi} {et~al.}(2016){Pavesi}, {Riechers}, {Capak}, {Carilli},
  {Sharon}, {Stacey}, {Karim}, {Scoville}, \& {Smol{\v
  c}i{\'c}}}]{2016ApJ...832..151P}
{Pavesi}, R., {Riechers}, D.~A., {Capak}, P.~L., {et~al.} 2016, \apj, 832, 151,
  \dodoi{10.3847/0004-637X/832/2/151}

\bibitem[{{Pavesi} {et~al.}(2018{\natexlab{a}}){Pavesi}, {Riechers}, {Sharon},
  {Smol{\v c}i{\'c}}, {Faisst}, {Schinnerer}, {Carilli}, {Capak}, {Scoville},
  \& {Stacey}}]{2018ApJ...861...43P}
{Pavesi}, R., {Riechers}, D.~A., {Sharon}, C.~E., {et~al.} 2018{\natexlab{a}},
  \apj, 861, 43, \dodoi{10.3847/1538-4357/aac6b6}

\bibitem[{{Pavesi} {et~al.}(2018{\natexlab{b}}){Pavesi}, {Sharon}, {Riechers},
  {Hodge}, {Decarli}, {Walter}, {Carilli}, {Daddi}, {Smail}, {Dickinson},
  {Ivison}, {Sargent}, {da Cunha}, {Aravena}, {Darling}, {Smol{\v c}i{\'c}},
  {Scoville}, {Capak}, \& {Wagg}}]{2018ApJ...864...49P}
{Pavesi}, R., {Sharon}, C.~E., {Riechers}, D.~A., {et~al.} 2018{\natexlab{b}},
  \apj, 864, 49, \dodoi{10.3847/1538-4357/aacb79}

\bibitem[{{Peng} {et~al.}(2002){Peng}, {Ho}, {Impey}, \&
  {Rix}}]{2002AJ....124..266P}
{Peng}, C.~Y., {Ho}, L.~C., {Impey}, C.~D., \& {Rix}, H.-W. 2002, \aj, 124,
  266, \dodoi{10.1086/340952}

\bibitem[{{Pettini} \& {Pagel}(2004)}]{2004MNRAS.348L..59P}
{Pettini}, M., \& {Pagel}, B.~E.~J. 2004, \mnras, 348, L59,
  \dodoi{10.1111/j.1365-2966.2004.07591.x}

\bibitem[{{Planck Collaboration} {et~al.}(2016){Planck Collaboration}, {Ade},
  {Aghanim}, {Arnaud}, {Aumont}, {Baccigalupi}, {Banday}, {Barreiro},
  {Bartolo}, {Battaner}, {Benabed}, {Benoit-L{\'e}vy}, {Bernard}, {Bersanelli},
  {Bielewicz}, {Bonaldi}, {Bonavera}, {Bond}, {Borrill}, {Bouchet},
  {Boulanger}, {Burigana}, {Butler}, {Calabrese}, {Catalano}, {Chiang},
  {Christensen}, {Clements}, {Colombo}, {Couchot}, {Coulais}, {Crill}, {Curto},
  {Cuttaia}, {Danese}, {Davies}, {Davis}, {de Bernardis}, {de Rosa}, {de
  Zotti}, {Delabrouille}, {Dickinson}, {Diego}, {Dole}, {Dor{\'e}}, {Douspis},
  {Ducout}, {Dupac}, {Elsner}, {En{\ss}lin}, {Eriksen}, {Falgarone}, {Finelli},
  {Flores-Cacho}, {Frailis}, {Fraisse}, {Franceschi}, {Galeotta}, {Galli},
  {Ganga}, {Giard}, {Giraud-H{\'e}raud}, {Gjerl{\o}w}, {Gonz{\'a}lez-Nuevo},
  {G{\'o}rski}, {Gregorio}, {Gruppuso}, {Gudmundsson}, {Hansen}, {Harrison},
  {Helou}, {Hern{\'a}ndez-Monteagudo}, {Herranz}, {Hildebrandt}, {Hivon},
  {Hobson}, {Hornstrup}, {Hovest}, {Huffenberger}, {Hurier}, {Jaffe}, {Jaffe},
  {Keih{\"a}nen}, {Keskitalo}, {Kisner}, {Kneissl}, {Knoche}, {Kunz},
  {Kurki-Suonio}, {Lagache}, {Lamarre}, {Lasenby}, {Lattanzi}, {Lawrence},
  {Leonardi}, {Levrier}, {Liguori}, {Lilje}, {Linden-V{\o}rnle},
  {L{\'o}pez-Caniego}, {Lubin}, {Mac{\'{\i}}as-P{\'e}rez}, {Maffei}, {Maggio},
  {Maino}, {Mandolesi}, {Mangilli}, {Maris}, {Martin},
  {Mart{\'{\i}}nez-Gonz{\'a}lez}, {Masi}, {Matarrese}, {Melchiorri},
  {Mennella}, {Migliaccio}, {Mitra}, {Miville-Desch{\^e}nes}, {Moneti},
  {Montier}, {Morgante}, {Mortlock}, {Munshi}, {Murphy}, {Nati}, {Natoli},
  {Nesvadba}, {Noviello}, {Novikov}, {Novikov}, {Oxborrow}, {Pagano}, {Pajot},
  {Paoletti}, {Partridge}, {Pasian}, {Pearson}, {Perdereau}, {Perotto},
  {Pettorino}, {Piacentini}, {Piat}, {Plaszczynski}, {Pointecouteau},
  {Polenta}, {Pratt}, {Prunet}, {Puget}, {Rachen}, {Reinecke}, {Remazeilles},
  {Renault}, {Renzi}, {Ristorcelli}, {Rocha}, {Rosset}, {Rossetti}, {Roudier},
  {Rubi{\~n}o-Mart{\'{\i}}n}, {Rusholme}, {Sandri}, {Santos}, {Savelainen},
  {Savini}, {Scott}, {Spencer}, {Stolyarov}, {Stompor}, {Sudiwala}, {Sunyaev},
  {Suur-Uski}, {Sygnet}, {Tauber}, {Terenzi}, {Toffolatti}, {Tomasi},
  {Tristram}, {Tucci}, {T{\"u}rler}, {Umana}, {Valenziano}, {Valiviita}, {Van
  Tent}, {Vielva}, {Villa}, {Wade}, {Wandelt}, {Wehus}, {Welikala}, {Yvon},
  {Zacchei}, \& {Zonca}}]{2016A&A...596A.100P}
{Planck Collaboration}, {Ade}, P.~A.~R., {Aghanim}, N., {et~al.} 2016, \aap,
  596, A100, \dodoi{10.1051/0004-6361/201527206}

\bibitem[{{Popping} {et~al.}(2017){Popping}, {Puglisi}, \&
  {Norman}}]{2017MNRAS.472.2315P}
{Popping}, G., {Puglisi}, A., \& {Norman}, C.~A. 2017, \mnras, 472, 2315,
  \dodoi{10.1093/mnras/stx2202}

\bibitem[{{Ricciardelli} {et~al.}(2010){Ricciardelli}, {Trujillo}, {Buitrago},
  \& {Conselice}}]{2010MNRAS.406..230R}
{Ricciardelli}, E., {Trujillo}, I., {Buitrago}, F., \& {Conselice}, C.~J. 2010,
  \mnras, 406, 230, \dodoi{10.1111/j.1365-2966.2010.16693.x}

\bibitem[{{Riechers} {et~al.}(2010){Riechers}, {Capak}, {Carilli}, {Cox},
  {Neri}, {Scoville}, {Schinnerer}, {Bertoldi}, \& {Yan}}]{2010ApJ...720L.131R}
{Riechers}, D.~A., {Capak}, P.~L., {Carilli}, C.~L., {et~al.} 2010, \apjl, 720,
  L131, \dodoi{10.1088/2041-8205/720/2/L131}

\bibitem[{{Riechers} {et~al.}(2013){Riechers}, {Bradford}, {Clements},
  {Dowell}, {P{\'e}rez-Fournon}, {Ivison}, {Bridge}, {Conley}, {Fu}, {Vieira},
  {Wardlow}, {Calanog}, {Cooray}, {Hurley}, {Neri}, {Kamenetzky}, {Aguirre},
  {Altieri}, {Arumugam}, {Benford}, {B{\'e}thermin}, {Bock}, {Burgarella},
  {Cabrera-Lavers}, {Chapman}, {Cox}, {Dunlop}, {Earle}, {Farrah}, {Ferrero},
  {Franceschini}, {Gavazzi}, {Glenn}, {Solares}, {Gurwell}, {Halpern},
  {Hatziminaoglou}, {Hyde}, {Ibar}, {Kov{\'a}cs}, {Krips}, {Lupu}, {Maloney},
  {Martinez-Navajas}, {Matsuhara}, {Murphy}, {Naylor}, {Nguyen}, {Oliver},
  {Omont}, {Page}, {Petitpas}, {Rangwala}, {Roseboom}, {Scott}, {Smith},
  {Staguhn}, {Streblyanska}, {Thomson}, {Valtchanov}, {Viero}, {Wang},
  {Zemcov}, \& {Zmuidzinas}}]{2013Natur.496..329R}
{Riechers}, D.~A., {Bradford}, C.~M., {Clements}, D.~L., {et~al.} 2013, \nat,
  496, 329, \dodoi{10.1038/nature12050}

\bibitem[{{Riechers} {et~al.}(2014){Riechers}, {Carilli}, {Capak}, {Scoville},
  {Smol{\v c}i{\'c}}, {Schinnerer}, {Yun}, {Cox}, {Bertoldi}, {Karim}, \&
  {Yan}}]{2014ApJ...796...84R}
{Riechers}, D.~A., {Carilli}, C.~L., {Capak}, P.~L., {et~al.} 2014, \apj, 796,
  84, \dodoi{10.1088/0004-637X/796/2/84}

\bibitem[{{Robitaille} \& {Bressert}(2012)}]{2012ascl.soft08017R}
{Robitaille}, T., \& {Bressert}, E. 2012, {APLpy: Astronomical Plotting Library
  in Python}, Astrophysics Source Code Library.
\newblock \doeprint{1208.017}

\bibitem[{{Rudnick} {et~al.}(2017){Rudnick}, {Hodge}, {Walter}, {Momcheva},
  {Tran}, {Papovich}, {da Cunha}, {Decarli}, {Saintonge}, {Willmer}, {Lotz}, \&
  {Lentati}}]{2017ApJ...849...27R}
{Rudnick}, G., {Hodge}, J., {Walter}, F., {et~al.} 2017, \apj, 849, 27,
  \dodoi{10.3847/1538-4357/aa87b2}

\bibitem[{{Safarzadeh} {et~al.}(2017){Safarzadeh}, {Hayward}, \&
  {Ferguson}}]{2017ApJ...840...15S}
{Safarzadeh}, M., {Hayward}, C.~C., \& {Ferguson}, H.~C. 2017, \apj, 840, 15,
  \dodoi{10.3847/1538-4357/aa6c5b}

\bibitem[{{Sargent} {et~al.}(2014){Sargent}, {Daddi}, {B{\'e}thermin},
  {Aussel}, {Magdis}, {Hwang}, {Juneau}, {Elbaz}, \& {da
  Cunha}}]{2014ApJ...793...19S}
{Sargent}, M.~T., {Daddi}, E., {B{\'e}thermin}, M., {et~al.} 2014, \apj, 793,
  19, \dodoi{10.1088/0004-637X/793/1/19}

\bibitem[{{Schinnerer} {et~al.}(2016){Schinnerer}, {Groves}, {Sargent},
  {Karim}, {Oesch}, {Magnelli}, {LeFevre}, {Tasca}, {Civano}, {Cassata}, \&
  {Smol{\v{c}}i{\'c}}}]{2016ApJ...833..112S}
{Schinnerer}, E., {Groves}, B., {Sargent}, M.~T., {et~al.} 2016, \apj, 833,
  112, \dodoi{10.3847/1538-4357/833/1/112}

\bibitem[{{Schmidt}(1959)}]{1959ApJ...129..243S}
{Schmidt}, M. 1959, \apj, 129, 243, \dodoi{10.1086/146614}

\bibitem[{{Schreiber} {et~al.}(2015){Schreiber}, {Pannella}, {Elbaz},
  {B{\'e}thermin}, {Inami}, {Dickinson}, {Magnelli}, {Wang}, {Aussel}, {Daddi},
  {Juneau}, {Shu}, {Sargent}, {Buat}, {Faber}, {Ferguson}, {Giavalisco},
  {Koekemoer}, {Magdis}, {Morrison}, {Papovich}, {Santini}, \&
  {Scott}}]{2015A&A...575A..74S}
{Schreiber}, C., {Pannella}, M., {Elbaz}, D., {et~al.} 2015, \aap, 575, A74,
  \dodoi{10.1051/0004-6361/201425017}

\bibitem[{{Scoville} {et~al.}(2014){Scoville}, {Aussel}, {Sheth}, {Scott},
  {Sanders}, {Ivison}, {Pope}, {Capak}, {Vanden Bout}, {Manohar}, {Kartaltepe},
  {Robertson}, \& {Lilly}}]{2014ApJ...783...84S}
{Scoville}, N., {Aussel}, H., {Sheth}, K., {et~al.} 2014, \apj, 783, 84,
  \dodoi{10.1088/0004-637X/783/2/84}

\bibitem[{{Scoville} {et~al.}(2017){Scoville}, {Lee}, {Vanden Bout},
  {Diaz-Santos}, {Sanders}, {Darvish}, {Bongiorno}, {Casey}, {Murchikova},
  {Koda}, {Capak}, {Vlahakis}, {Ilbert}, {Sheth}, {Morokuma-Matsui}, {Ivison},
  {Aussel}, {Laigle}, {McCracken}, {Armus}, {Pope}, {Toft}, \&
  {Masters}}]{2017ApJ...837..150S}
{Scoville}, N., {Lee}, N., {Vanden Bout}, P., {et~al.} 2017, \apj, 837, 150,
  \dodoi{10.3847/1538-4357/aa61a0}

\bibitem[{{Silva} {et~al.}(2015){Silva}, {Sajina}, {Lonsdale}, \&
  {Lacy}}]{2015ApJ...806L..25S}
{Silva}, A., {Sajina}, A., {Lonsdale}, C., \& {Lacy}, M. 2015, \apjl, 806, L25,
  \dodoi{10.1088/2041-8205/806/2/L25}

\bibitem[{{Smol{\v c}i{\'c}} {et~al.}(2012){Smol{\v c}i{\'c}}, {Navarrete},
  {Aravena}, {Ilbert}, {Yun}, {Sheth}, {Salvato}, {McCracken}, {Diener},
  {Aretxaga}, {Riechers}, {Finoguenov}, {Bertoldi}, {Capak}, {Hughes}, {Karim},
  {Schinnerer}, {Scoville}, \& {Wilson}}]{2012ApJS..200...10S}
{Smol{\v c}i{\'c}}, V., {Navarrete}, F., {Aravena}, M., {et~al.} 2012, \apjs,
  200, 10, \dodoi{10.1088/0067-0049/200/1/10}

\bibitem[{{Solomon} {et~al.}(1992){Solomon}, {Downes}, \&
  {Radford}}]{1992ApJ...398L..29S}
{Solomon}, P.~M., {Downes}, D., \& {Radford}, S.~J.~E. 1992, \apjl, 398, L29,
  \dodoi{10.1086/186569}

\bibitem[{{Solomon} {et~al.}(1997){Solomon}, {Downes}, {Radford}, \&
  {Barrett}}]{1997ApJ...478..144S}
{Solomon}, P.~M., {Downes}, D., {Radford}, S.~J.~E., \& {Barrett}, J.~W. 1997,
  \apj, 478, 144, \dodoi{10.1086/303765}

\bibitem[{{Spitler} {et~al.}(2012){Spitler}, {Labb{\'e}}, {Glazebrook},
  {Persson}, {Monson}, {Papovich}, {Tran}, {Poole}, {Quadri}, {van Dokkum},
  {Kelson}, {Kacprzak}, {McCarthy}, {Murphy}, {Straatman}, \&
  {Tilvi}}]{2012ApJ...748L..21S}
{Spitler}, L.~R., {Labb{\'e}}, I., {Glazebrook}, K., {et~al.} 2012, \apjl, 748,
  L21, \dodoi{10.1088/2041-8205/748/2/L21}

\bibitem[{{Steidel} {et~al.}(1998){Steidel}, {Adelberger}, {Dickinson},
  {Giavalisco}, {Pettini}, \& {Kellogg}}]{1998ApJ...492..428S}
{Steidel}, C.~C., {Adelberger}, K.~L., {Dickinson}, M., {et~al.} 1998, \apj,
  492, 428, \dodoi{10.1086/305073}

\bibitem[{{Strong} \& {Mattox}(1996)}]{1996A&A...308L..21S}
{Strong}, A.~W., \& {Mattox}, J.~R. 1996, \aap, 308, L21

\bibitem[{{Tacconi} {et~al.}(2008){Tacconi}, {Genzel}, {Smail}, {Neri},
  {Chapman}, {Ivison}, {Blain}, {Cox}, {Omont}, {Bertoldi}, {Greve},
  {F{\"o}rster Schreiber}, {Genel}, {Lutz}, {Swinbank}, {Shapley}, {Erb},
  {Cimatti}, {Daddi}, \& {Baker}}]{2008ApJ...680..246T}
{Tacconi}, L.~J., {Genzel}, R., {Smail}, I., {et~al.} 2008, \apj, 680, 246,
  \dodoi{10.1086/587168}

\bibitem[{{Tacconi} {et~al.}(2010){Tacconi}, {Genzel}, {Neri}, {Cox}, {Cooper},
  {Shapiro}, {Bolatto}, {Bouch{\'e}}, {Bournaud}, {Burkert}, {Combes},
  {Comerford}, {Davis}, {Schreiber}, {Garcia-Burillo}, {Gracia-Carpio}, {Lutz},
  {Naab}, {Omont}, {Shapley}, {Sternberg}, \& {Weiner}}]{2010Natur.463..781T}
{Tacconi}, L.~J., {Genzel}, R., {Neri}, R., {et~al.} 2010, \nat, 463, 781,
  \dodoi{10.1038/nature08773}

\bibitem[{{Tacconi} {et~al.}(2018){Tacconi}, {Genzel}, {Saintonge}, {Combes},
  {Garc{\'{\i}}a-Burillo}, {Neri}, {Bolatto}, {Contini}, {F{\"o}rster
  Schreiber}, {Lilly}, {Lutz}, {Wuyts}, {Accurso}, {Boissier}, {Boone},
  {Bouch{\'e}}, {Bournaud}, {Burkert}, {Carollo}, {Cooper}, {Cox}, {Feruglio},
  {Freundlich}, {Herrera-Camus}, {Juneau}, {Lippa}, {Naab}, {Renzini},
  {Salome}, {Sternberg}, {Tadaki}, {{\"U}bler}, {Walter}, {Weiner}, \&
  {Weiss}}]{2018ApJ...853..179T}
{Tacconi}, L.~J., {Genzel}, R., {Saintonge}, A., {et~al.} 2018, \apj, 853, 179,
  \dodoi{10.3847/1538-4357/aaa4b4}

\bibitem[{{Toft} {et~al.}(2014){Toft}, {Smol{\v c}i{\'c}}, {Magnelli}, {Karim},
  {Zirm}, {Michalowski}, {Capak}, {Sheth}, {Schawinski}, {Krogager}, {Wuyts},
  {Sanders}, {Man}, {Lutz}, {Staguhn}, {Berta}, {Mccracken}, {Krpan}, \&
  {Riechers}}]{2014ApJ...782...68T}
{Toft}, S., {Smol{\v c}i{\'c}}, V., {Magnelli}, B., {et~al.} 2014, \apj, 782,
  68, \dodoi{10.1088/0004-637X/782/2/68}

\bibitem[{{Umehata} {et~al.}(2015){Umehata}, {Tamura}, {Kohno}, {Ivison},
  {Alexander}, {Geach}, {Hatsukade}, {Hughes}, {Ikarashi}, {Kato}, {Izumi},
  {Kawabe}, {Kubo}, {Lee}, {Lehmer}, {Makiya}, {Matsuda}, {Nakanishi}, {Saito},
  {Smail}, {Yamada}, {Yamaguchi}, \& {Yun}}]{2015ApJ...815L...8U}
{Umehata}, H., {Tamura}, Y., {Kohno}, K., {et~al.} 2015, \apjl, 815, L8,
  \dodoi{10.1088/2041-8205/815/1/L8}

\bibitem[{{Valentino} {et~al.}(2015){Valentino}, {Daddi}, {Strazzullo},
  {Gobat}, {Onodera}, {Bournaud}, {Juneau}, {Renzini}, {Arimoto}, {Carollo}, \&
  {Zanella}}]{2015ApJ...801..132V}
{Valentino}, F., {Daddi}, E., {Strazzullo}, V., {et~al.} 2015, \apj, 801, 132,
  \dodoi{10.1088/0004-637X/801/2/132}

\bibitem[{{Valentino} {et~al.}(2016){Valentino}, {Daddi}, {Finoguenov},
  {Strazzullo}, {Le Brun}, {Vignali}, {Bournaud}, {Dickinson}, {Renzini},
  {B{\'e}thermin}, {Zanella}, {Gobat}, {Cimatti}, {Elbaz}, {Onodera},
  {Pannella}, {Sargent}, {Arimoto}, {Carollo}, \&
  {Starck}}]{2016ApJ...829...53V}
{Valentino}, F., {Daddi}, E., {Finoguenov}, A., {et~al.} 2016, \apj, 829, 53,
  \dodoi{10.3847/0004-637X/829/1/53}

\bibitem[{{Walter} {et~al.}(2012){Walter}, {Decarli}, {Carilli}, {Bertoldi},
  {Cox}, {da Cunha}, {Daddi}, {Dickinson}, {Downes}, {Elbaz}, {Ellis}, {Hodge},
  {Neri}, {Riechers}, {Weiss}, {Bell}, {Dannerbauer}, {Krips}, {Krumholz},
  {Lentati}, {Maiolino}, {Menten}, {Rix}, {Robertson}, {Spinrad}, {Stark}, \&
  {Stern}}]{2012Natur.486..233W}
{Walter}, F., {Decarli}, R., {Carilli}, C., {et~al.} 2012, \nat, 486, 233,
  \dodoi{10.1038/nature11073}

\bibitem[{{Wang} {et~al.}(2016){Wang}, {Elbaz}, {Daddi}, {Finoguenov}, {Liu},
  {Schreiber}, {Mart{\'{\i}}n}, {Strazzullo}, {Valentino}, {van der Burg},
  {Zanella}, {Ciesla}, {Gobat}, {Le Brun}, {Pannella}, {Sargent}, {Shu}, {Tan},
  {Cappelluti}, \& {Li}}]{2016ApJ...828...56W}
{Wang}, T., {Elbaz}, D., {Daddi}, E., {et~al.} 2016, \apj, 828, 56,
  \dodoi{10.3847/0004-637X/828/1/56}

\bibitem[{{Wang} {et~al.}(2018){Wang}, {Elbaz}, {Daddi}, {Liu}, {Kodama},
  {Tanaka}, {Schreiber}, {Zanella}, {Valentino}, {Sargent}, {Kohno}, {Xiao},
  {Pannella}, {Ciesla}, {Gobat}, \& {Koyama}}]{2018ApJ...867L..29W}
---. 2018, \apjl, 867, L29, \dodoi{10.3847/2041-8213/aaeb2c}

\bibitem[{{Wardlow} {et~al.}(2013){Wardlow}, {Cooray}, {De Bernardis},
  {Amblard}, {Arumugam}, {Aussel}, {Baker}, {B{\'e}thermin}, {Blundell},
  {Bock}, {Boselli}, {Bridge}, {Buat}, {Burgarella}, {Bussmann},
  {Cabrera-Lavers}, {Calanog}, {Carpenter}, {Casey}, {Castro-Rodr{\'{\i}}guez},
  {Cava}, {Chanial}, {Chapin}, {Chapman}, {Clements}, {Conley}, {Cox},
  {Dowell}, {Dye}, {Eales}, {Farrah}, {Ferrero}, {Franceschini}, {Frayer},
  {Frazer}, {Fu}, {Gavazzi}, {Glenn}, {Gonz{\'a}lez Solares}, {Griffin},
  {Gurwell}, {Harris}, {Hatziminaoglou}, {Hopwood}, {Hyde}, {Ibar}, {Ivison},
  {Kim}, {Lagache}, {Levenson}, {Marchetti}, {Marsden}, {Martinez-Navajas},
  {Negrello}, {Neri}, {Nguyen}, {O'Halloran}, {Oliver}, {Omont}, {Page},
  {Panuzzo}, {Papageorgiou}, {Pearson}, {P{\'e}rez-Fournon}, {Pohlen},
  {Riechers}, {Rigopoulou}, {Roseboom}, {Rowan-Robinson}, {Schulz}, {Scott},
  {Scoville}, {Seymour}, {Shupe}, {Smith}, {Streblyanska}, {Strom},
  {Symeonidis}, {Trichas}, {Vaccari}, {Vieira}, {Viero}, {Wang}, {Xu}, {Yan},
  \& {Zemcov}}]{2013ApJ...762...59W}
{Wardlow}, J.~L., {Cooray}, A., {De Bernardis}, F., {et~al.} 2013, \apj, 762,
  59, \dodoi{10.1088/0004-637X/762/1/59}

\bibitem[{{Wardlow} {et~al.}(2018){Wardlow}, {Simpson}, {Smail}, {Swinbank},
  {Blain}, {Brandt}, {Chapman}, {Chen}, {Cooke}, {Dannerbauer}, {Gullberg},
  {Hodge}, {Ivison}, {Knudsen}, {Scott}, {Thomson}, {Wei{\ss}}, \& {van der
  Werf}}]{2018MNRAS.479.3879W}
{Wardlow}, J.~L., {Simpson}, J.~M., {Smail}, I., {et~al.} 2018, \mnras, 479,
  3879, \dodoi{10.1093/mnras/sty1526}

\bibitem[{{Yuan} {et~al.}(2014){Yuan}, {Nanayakkara}, {Kacprzak}, {Tran},
  {Glazebrook}, {Kewley}, {Spitler}, {Poole}, {Labb{\'e}}, {Straatman}, \&
  {Tomczak}}]{2014ApJ...795L..20Y}
{Yuan}, T., {Nanayakkara}, T., {Kacprzak}, G.~G., {et~al.} 2014, \apjl, 795,
  L20, \dodoi{10.1088/2041-8205/795/1/L20}

\end{thebibliography}

\end{document}